# Geophysical Observations of the 24 September 2023 OSIRIS-REx Sample Return Capsule Re-Entry

Elizabeth A. Silber[1,*], Daniel C. Bowman[2], Chris G. Carr[3], David P. Eisenberg[4], Brian R. Elbing[5], Benjamin Fernando[6], Milton A. Garcés[7,8], Robert Haaser[9], Siddharth Krishnamoorthy[10], Charles A. Langston[11], Yasuhiro Nishikawa[12], Jeremy Webster[3], Jacob F. Anderson[13], Stephen Arrowsmith[14], Sonia Bazargan[11], Luke Beardslee[15], Brant Beck[4], Jordan W. Bishop[3], Philip Blom[3], Grant Bracht[4], David L. Chichester[16], Anthony Christe[17], Jacob Clarke[14], Kenneth Cummins[18], James Cutts[10], Lisa Danielson[3], Carly Donahue[3], Kenneth Eack[9], Michael Fleigle[2], Douglas Fox[5], Ashish Goel[10], David Green[19], Yuta Hasumi[12], Chris Hayward[14], Dan Hicks[20], Jay Hix[16], Stephen Horton[11], Emalee Hough[5], David P. Huber[21], Madeline A. Hunt[13], Jennifer Inman[22], S. M. Ariful Islam[11], Jacob Izraelevitz[10], Jamey D. Jacob[5], James Johnson[16], Real J. KC[5], Attila Komjathy[10], Eric Lam[23], Justin LaPierre[24], Kevin Lewis[6], Richard D. Lewis[25], Patrick Liu[23], Léo Martire[10], Meaghan McCleary[22], Elisa A. McGhee[26], Ipsita Mitra[11], Amitabh Nag[9], Luis Ocampo Giraldo[16], Karen Pearson[27], Mathieu Plaisir[18], Sarah K. Popenhagen[7], Hamid Rassoul[18], Miro Ronac Giannone[1,14], Mirza Samnani[10], Nicholas Schmerr[28], Kate Spillman[5], Girish Srinivas[4], Samuel K. Takazawa[7], Alex Tempert[18], Reagan Turley[28], Cory Van Beek[4], Loïc Viens[3], Owen A. Walsh[13], Nathan Weinstein[4], Robert White[29], Brian Williams[7], Trevor C. Wilson[5], Shirin Wyckoff[7], Masa-yuki Yamamoto[12], Zachary Yap[5], Tyler Yoshiyama[7], Cleat Zeiler[29]

[1]Sandia National Laboratories, Geophysics, Albuquerque, NM, 87123, USA
[2]Sandia National Laboratories, Geophysical Detection Programs, Albuquerque, NM, 87123, USA
[3]Los Alamos National Laboratory, Earth and Environmental Sciences Division, PO Box 1663, Los Alamos, NM, 87544, USA
[4]TDA Research Inc, 12345 W 52nd Avenue, Wheat Ridge, CO, 80033, USA
[5]Oklahoma State University, Mechanical and Aerospace Engineering, Stillwater, OK, 74078, USA
[6]Johns Hopkins University, Department of Earth and Planetary Sciences, Baltimore, MD, 21218, USA
[7]University of Hawaii, Infrasound Lab, 73970 Makako Bay Drive 205, Kailua-Kona, HI, 96740, USA
[8]University of Hawaii, Applied Research Lab, 73970 Makako Bay Drive 117, Kailua-Kona, HI, 96740, USA
[9]Los Alamos National Laboratory, ISR-2, P.O. Box 1663, MS-B241, Los Alamos, NM, 87545, USA
[10]NASA Jet Propulsion Laboratory, California Institute of Technology, 4800 Oak Grove Drive, Pasadena, CA, 91109, USA
[11]University of Memphis, Center for Earthquake Research and Information, 3876 Central Ave., Ste. 1, Memphis, TN, 38152, USA
[12]Kochi University of Technology, School of Systems Engineering, 185 Miyanokuchi, Tosayamada, Kami, Kochi, 7828502, Japan
[13]Boise State University, Geosciences, 1910 University Dr, Boise, ID, 83725, USA
[14]Southern Methodist University, Earth Sciences, 3225 Daniel Avenue, Dallas, TX, 75205
[15]Silixa Ltd, 3102 Broadway St., Suite E, Missoula, MT, 59808, USA
[16]Idaho National Laboratory, Nuclear Nonproliferation Division, PO Box 1625, MS2212, Idaho Falls, ID, 83415, USA
[17]RedVox, Inc., 73970 Makako Bay Drive 116, Kailua-Kona, HI, 96740
[18]Florida Institute of Technology, Geospace Physics Laboratory, 150 W. University Blvd., Melbourne, FL, 32901, USA
[19]AWE Blacknest, Brimpton Common, Reading, Berkshire, RG7 4RS, UK
[20]United States Department of Defense, KBR Consultant, Las Cruces, NM, USA

[21]University of Texas-El Paso, Earth, Environmental, and Resource Sciences, 500 W University, El Paso, TX, 79902, USA
[22]NASA Langley Research Center, SCIFLI, Hampton, Virginia, 23681, USA
[23]Air Force Research Laboratory, Sensing Effects & Analysis, 2241 Avionics Circle, Wright-Patterson AFB, Ohio, 45433, USA
[24]Sandia National Laboratories, Atmospheric Science, Albuquerque, NM, 87123, USA
[25]Defense Threat Reduction Agency (DTRA), 8725 John J. Kingman Rd, Stop 6201, Ft. Belvoir, VA, 22060, USA
[26]Colorado State University, Department of Geosciences, 1482 Campus Delivery, Ft. Collins, CO, 80523, USA
[27]Independent Researcher, Bowie, MD, 20715, USA
[28]University of Maryland, Dept. of Geology, College Park, MD, 20742, USA
[29]Nevada National Security Sites, Science and Technology, 2621 Losee Rd, North Las Vegas, NV, 89030, USA

[*]Corresponding Author e-mail: esilbe@sandia.gov

**Abstract**

Sample Return Capsules (SRCs) entering Earth's atmosphere at hypervelocity from interplanetary space are a valuable resource for studying meteor phenomena. The 24 September 2023 arrival of the OSIRIS-REx (Origins, Spectral Interpretation, Resource Identification, and Security-Regolith Explorer) SRC provided an unprecedented chance for geophysical observations of a well-characterized source with known parameters, including timing and trajectory. A collaborative effort involving researchers from 16 institutions executed a carefully planned geophysical observational campaign at strategically chosen locations, deploying over 400 ground-based sensors encompassing infrasound, seismic, distributed acoustic sensing (DAS), and GPS technologies. Additionally, balloons equipped with infrasound sensors were launched to capture signals at higher altitudes. This campaign (the largest of its kind so far) yielded a wealth of invaluable data anticipated to fuel scientific inquiry for years to come. The success of the observational campaign is evidenced by the near-universal detection of signals across instruments, both proximal and distal. This paper presents a comprehensive overview of the collective scientific effort, field deployment, and preliminary findings. The early findings have the potential to inform future space missions and terrestrial campaigns, contributing to our understanding of meteoroid interactions with planetary atmospheres. Furthermore, the dataset collected during this campaign will improve entry and propagation models as well as augment the study of atmospheric dynamics and shock phenomena generated by meteoroids and similar sources.

**Keywords:** meteoroid, meteor, fireball, bolide, infrasound, seismic, DAS, planetary atmosphere

## 1. Introduction

Geophysical sensing of objects entering planetary atmospheres and surfaces is of immense importance for understanding impact-induced physical processes on Earth and beyond. Interplanetary space is teeming with meteoroids, asteroids, and comets (e.g., Belton, 2004; Chapman, 2008), and is sometimes even visited by objects originating from beyond our solar system, such as 1I/'Oumuamua (Meech et al., 2017). While the rate of large, extremely energetic and planet altering impacts has largely dissipated over time since the Late Heavy Bombardment, smaller impacts continue to happen on Earth and elsewhere. For example, the lunar surface is frequently impacted by objects large enough to produce light emissions visible from Earth (e.g., Avdellidou and Vaubaillon, 2019; Ortiz et al., 2006; Ortiz et al., 2015). Moreover, bright flashes seen in the atmosphere of Venus have been attributed to meteoroids (Blaske et al., 2023). On Mars, a freshly formed impact craters have been found (Daubar et al., 2023; Posiolova et al., 2022), and possible seismic and acoustic wave signatures from meteoroid entries were detected by NASA's InSight lander (Garcia et al., 2022).

Unfortunately, it is profoundly difficult to predict impacts of meter-sized and larger objects with sufficient temporal and spatial accuracy and with long enough advance notice to allow comprehensive observational campaign planning. Thus, it is nearly impossible and prohibitively costly to mount a comprehensive observational campaign using a full range of sensing modalities. Therefore, virtually all observations are incidental – instruments either passively "wait" for an event to happen over a certain region (e.g., Devillepoix et al., 2020) or they make a detection as a byproduct of a completely different observational mission (e.g., Jenniskens et al., 2018). While small meteoroids are numerous, objects in a meter-size range are significantly more scarce and thus profoundly more difficult to capture with a multitude of instruments. Even if detailed observations take place, source characterization does not come without its own challenges. Impeding factors include incomplete ground truth, inability to directly measure and sample the object, lack of comparable events (no two natural objects are alike), limitations in models and theoretical considerations, and other uncertainties (Silber, 2024).

Therefore, it is imperative to make use of well-characterized artificial objects that can serve as natural meteoroid/asteroid analogues (e.g., ReVelle et al., 2005). Ideal candidates are space mission sample return capsules (SRCs) that re-enter from interplanetary space and thus achieve speeds that match those of (slow) natural meteors (> 11 km/s). Their speed is also relatively close to the mean speed of natural asteroid entries (25 – 30 km/s) (Janches et al., 2006). Only five sample return missions have re-entered from interplanetary space since the end of the Apollo era:

Genesis (ReVelle et al., 2005), Stardust (ReVelle and Edwards, 2006), Hayabusa 1 (Ishihara et al., 2012), Hayabusa 2 (Sansom et al., 2022), and OSIRIS-REx (Origins, Spectral Interpretation, Resource Identification, and Security-Regolith Explorer) (Fernando et al., 2024; Silber et al., 2023a). All five were detected by dedicated geophysical instruments (infrasound or/and seismic) (see Silber et al., 2023a for details).

To understand the relevance and application of artificial objects, such as SRCs, towards the study of larger meteoroid dynamics in the planetary atmospheres, we start with a brief overview of meteor phenomena.

Approximately $10^5$ tons per year of extraterrestrial material enter the Earth’s atmosphere, ranging in size from dust particles to meters (Plane, 2012). Most particles peak in diameters around 2 x $10^{-4}$ m (Kalashnikova et al., 2000; Plane, 2012), with only an extremely small fraction corresponding to meter-sized and larger objects  (Drolshagen et al., 2017; Moorhead et al., 2017). Typical entry speeds are 11.2 – 72.8 km/s for objects originating in the Solar System (Ceplecha et al., 1998). Speeds greater than ~73 km/s correspond to objects visiting from interstellar space, although some exceptions around that velocity have been noted (Peña-Asensio et al., 2024). Asteroids (≥1m in diameter) and meteoroids (<1 m in diameter), through their collisions with local atmosphere and subsequent ablation, produce a light phenomenon known as a meteor or a shooting star (Ceplecha et al., 1998). Very bright meteors are known as fireballs (brighter than Venus, magnitude -4) and bolides (brighter than magnitude -14 (Belton, 2004)), and exceptionally bright events (exceeding magnitude -20) as superbolides (Ceplecha et al., 1998).

Of particular interest to the scientific and planetary defense communities are the asteroids and a subset of sufficiently large and fast meteoroids that produce shock waves upon entering the upper regions of the atmosphere (Bronshten, 1983; Ceplecha et al., 1998; Silber et al., 2018; Tsikulin, 1970). Specifically, the shock waves can lead to formation of secondary physical phenomena, from low frequency acoustic waves and seismic shaking (e.g., Arrowsmith et al., 2008a; Arrowsmith et al., 2007; Caudron et al., 2016; Ceplecha et al., 1998; Evers and Haak, 2003; Ishihara, 2004; Pilger et al., 2020; ReVelle, 1974; Silber and Brown, 2019) to ionospheric disturbances (e.g., Luo et al., 2020; Perevalova et al., 2015; Yang et al., 2014). When recorded by geophysical instruments, the signatures of these phenomena can be analyzed to infer physical properties and characteristics of the emitting source (e.g., ReVelle, 1976). Smaller meteoroids with diameters 0.1 – 10 cm, while still capable of generating shock waves, are not of interest in this study as in most cases these completely ablate at altitudes between ~70 and 100 km (Ceplecha et al., 1998; Silber and Brown, 2014).

Large objects can penetrate deep into the atmosphere, depositing a tremendous amount of energy at mid and low altitudes (typically below 50 km) and sometimes, their fragments may reach the surface as meteorites. A recent example is the Chelyabinsk superbolide, whose arrival caught the scientific community by surprise. The Chelyabinsk impactor was ~18 m in diameter and it deposited energy of approximately 500 kt of TNT equivalent (1 TNT = 1.484 x $10^{12}$ J), leaving a wake of destruction beneath its path (Brown et al., 2013; Popova et al., 2013).

SRCs can serve as reasonable analogues for meter-sized objects that are generally studied using a variety of sensing modalities, from ground-based (e.g., optical (e.g., Devillepoix et al., 2020), radar (e.g., Janches et al., 2006), infrasound (e.g., Silber and Brown, 2014), seismic (e.g., Edwards et al., 2008)) to space-based instruments (e.g., US government sensors (e.g., Brown et al., 2002), Geostationary Lightning Mapper (GLM) (e.g., Jenniskens et al., 2018)). In this paper, we place an emphasis on geophysical observations that include infrasound (ground-based and airborne), acoustic (audible), seismic, Distributed Acoustic Sensing (DAS), and Global Positioning System (GPS). We will outline the function of these in Section 2.

We present multi-modal observations of the OSIRIS-REx re-entry, the largest geophysical observational campaign of a controlled re-entry ever conducted. Multi-modal, large scale observational campaigns of re-entry and similar phenomena with well-known ground truth require careful planning, coordination, and execution. There is also only one chance to get it right – the object's re-entry cannot be delayed or modified to meet the observation campaign's needs. Given that this was an enormous undertaking that involved many scientists from over a dozen institutions, we felt that it was pertinent to consolidate our efforts into a single publication that will provide a complete contextual picture of the campaign and serve as a scientific reference for data types and sources, study replication, and for building upon this work by others. Furthermore, campaigns like this one provide an unparalleled learning opportunity for future "one shot" terrestrial and space exploration missions.

This paper is organized as follows: in Section 2, we give a brief background on geophysical sensing modalities, and in Section 3 we outline a primer on meteor generated shock waves and how they can be detected by geophysical instruments. The OSIRIS-REx re-entry is presented in Section 4. In Section 5 we describe the institutional involvement and the geographical context, and in Section 6, the field deployment effort and various instruments used. In Section 7, we present the preliminary results, and in Section 8 we outline our conclusions and path forward.

## 2. A Brief Primer on Geophysical Sensing Modalities

Infrasound is defined as sound waves below the limit of human hearing (<20 Hz). Infrasound sensing finds widespread utility in monitoring natural phenomena such as volcanic eruptions (e.g., Matoza et al., 2019), earthquakes (e.g., Arrowsmith et al., 2011) and meteorological events (Stopa et al., 2012). Additionally, it serves as a critical tool for detecting and characterizing anthropogenic activities, including explosions (e.g., Arrowsmith et al., 2008b; Mutschlecner and Whitaker, 2006; Obenberger et al., 2022) and rockets (e.g., Balachandran and Donn, 1971; Pilger et al., 2021). Infrasound monitoring also supports efforts in nuclear test ban verification (Brachet et al., 2010). The most typical instruments include ground-based sensors. These can be permanent or temporary installations. The latter are useful for short-term observational campaigns. In recent years, there has been an emergence of balloon-borne infrasound sensing (Bowman and Albert, 2018; Silber et al., 2023b) which has opened new avenues for detection and characterization of ground and elevated infrasound sources, and for validation of theoretical predictions (Albert et al., 2023). Balloon-borne infrasound has been proposed as a feasible mode of exploration for planets with thick atmospheres where other sensing modalities are either not possible or are exceedingly more costly (Krishnamoorthy and Bowman, 2023). Balloons also offer a unique vantage point, away from heavy tropospheric noise, and in a presumably quieter region of the atmosphere (Krishnamoorthy et al., 2020). There have been successful detections of high-altitude and ground-based phenomena on balloons, including rocket launches, atmospheric explosions, chemical explosions, storms, and gravity waves (Albert et al., 2023; Bowman and Lees, 2018). While bolide detection by a balloon-borne infrasound sensor has never been confirmed, it is expected that these platforms would readily detect a bolide should one occur in the vicinity.

Much of seismic analysis involves observing and modeling the wavefields generated from sources interior to or on the surface of the Earth. Earthquakes, volcanic disturbances, chemical and nuclear explosions, and artificial energy sources such as vibration-producing trucks or even hand-held hammer blows can provide seismic wavefields that can be modeled to determine the physical characteristics of the Earth over scales from meters to 10,000 km. Impulsive atmospheric sources such as explosions (e.g., Matoza et al., 2022), bolide sonic booms (D'Auria et al., 2006; Langston, 2004; Le Pichon, 2002), or even thunder (Lin and Langston, 2009a; Lin and Langston, 2009b), can be interesting in their own right as well as providing for new wavefields for investigating Earth structure using records from seismometers.

Distributed Acoustic Sensing (DAS) systems are a rapidly emerging technology that provide spatially dense (~1–10 m), extensive (10s of km), and high-fidelity seismic measurements by sensing with fiber-optic cables (e.g., Hartog, 2017). Previous studies have demonstrated that DAS can capture a variety of seismo-acoustic signals including seismic waves from earthquakes and explosions (e.g., Fang et al., 2020; Lindsey et al., 2017), and meteorites (Vera Rodriguez et al., 2023). While seismometers and infrasound sensors have been employed to measure signals from spacecraft re-entry events prior to the OSIRIS-REx sample return capsule (e.g., Edwards et al., 2007; ReVelle and Edwards, 2006), the OSIRIS-REx event is the first instance of DAS deployment to record a re-entry. Similarly, while DAS has not yet been deployed in extraterrestrial settings, data returned from seismometers and/or infrasound sensors have provided information about the seismic activity and structure of the Moon and Mars (e.g., Giardini et al., 2020; Lognonné et al., 2020; Nunn et al., 2020) and the atmosphere of Mars (e.g., Banfield et al., 2020; Ortiz et al., 2022).

Energetic and explosive events near the Earth are well-known to produce acoustic (compression) waves that propagate upward and outward into the atmosphere. When these acoustic waves propagate upward into the ionosphere, they couple with ionospheric plasma, producing electron density fluctuations (Forbes and Roble, 1990; Miller et al., 1986). Wave periods generated by these events range from 2-16 minutes, with ground-level speeds between 300-400 m/s. Speeds dramatically increase with altitude above the mesosphere due to changes in density and increases in background thermosphere temperatures. At altitudes near peak electron density (250-400 km), speeds can range from 700-900 m/s. In all, it takes approximately 8 to 10 minutes for the wave generated at the Earth's surface to propagate to these altitudes.

These waves create electron density perturbation signals that are probed remotely, such as using electromagnetic instruments. Impulsive events like meteors (e.g., Luo et al., 2020; Perevalova et al., 2015; Yang et al., 2014), volcanic eruptions (e.g., Shimada et al., 1990), earthquakes (e.g., Otsuka et al., 2006), tsunamis (e.g., Blewitt et al., 2009), rocket launches (e.g., Afraimovich et al., 2001), and ground explosions (e.g., Fitzgerald, 1997) have been examined using these methods. Here, we are interested in examining the effects of the shock wave generated by the hypersonic OSIRIS-REx SRC re-entry, and how it might affect the ionosphere and signals used to probe it. We employ GPS L-band signals which are frequently used to probe the ionosphere, but signatures often suffer a 1 to 30 minute delay before they impact the ionosphere and can be detected.

## 3. Meteor Generated Shock Waves

In this section, we offer a concise overview of meteor-generated shock waves and their correlation with acoustic and, on occasion, seismic waves (e.g., Edwards et al., 2008; Silber and Brown, 2019). We also briefly outline the similarities between shock waves generated by natural objects and their artificial analogues (Figure 1).

In principle, the mechanisms governing meteor-generated shock waves are broadly relevant to artificial hypersonic analogues (and vice versa), including space mission re-entries, rendering them valuable proxies for investigating meteor phenomena (ReVelle et al., 2005; Silber et al., 2023a).

When a meteoroid (or an asteroid) reaches the continuum flow regime as it descends through the Earth's exponentially denser atmosphere at hypervelocity, it generates a shock wave (Krehl, 2011; Silber et al., 2018 and references therein). Meteoroids travel at extremely high Mach numbers, from 35 to 240. Mach number ($M$) is the ratio of the meteoroid speed and the local speed of sound. At such speeds, the Mach cone angle (outlining the adiabatically expanding ablational flowfield) is small enough that the shock front can be approximated as a cylinder, therefore forming the so-called cylindrical line source (Plooster, 1970; ReVelle, 1976). The shock travels ballistically, or perpendicularly to the meteoroid flight trajectory, and energy is deposited into the surrounding atmosphere as a function of path length. It is important to note that the fundamental difference in shock waves between the large meteoroids/asteroids and artificial analogues is the fact that the shock waves generated by natural objects are ablationally amplified, while objects such as SRCs have a very limited ablation rate. Another key difference is the significantly higher magnitude and intensity of the hyperthermally driven chemical reactions in the ablationally amplified meteor/asteroid flowfield, which ultimately affects the strength of a shock wave. Moreover, meteoroids and asteroids frequently experience fragmentation, which can occur in the form of continuous fragmentation, gross fragmentation, or some combination of the two (e.g., Silber, 2024; Trigo-Rodríguez et al., 2021). In such cases, the shock wave is no longer generated by the object's hypersonic passage alone, but it also includes the shock with quasi-spherical or spherical geometry (also known as point source). The SRCs do not include such a point source component allowing the hypersonic shock wave to be studied without interference from such signals.

The blast radius ($R_0$) represents the volume of a region containing superheated adiabatically expanding plasma in the flow-field immediately behind the shock front, where highly non-linear processes take place. The mathematical expression is: $R_0 = (E_0/p_0)^{0.5}$, where $E_0$ is the energy per unit length and $p_0$ is the ambient pressure. Assuming no fragmentation, the blast radius can be approximated as $R_0 \sim Md_m$. where $M$ is the Mach number and $d_m$ is the meteoroid diameter. It is generally accepted that beyond approximately $10R_0$, the shock decays to a weak shock regime, and at some point, to a linear acoustic wave (Plooster, 1970; ReVelle, 1976; Tsikulin, 1970). A comprehensive review of meteor generated shock waves can be found in Silber et al. (2018), and a review on meteor-generated infrasound in Silber and Brown (2019).

Shock waves generated by hypersonic passage of meteoroids (and other impulsive sources) ultimately decay to infrasound, which has the remarkable ability to propagate over vast distances with minimal attenuation (Evans et al., 1972). At a very close range from the source, the acoustic wave might have both inaudible and audible components. A familiar example of the audible component would be a sonic boom generated by a jet when it breaks the sound barrier. The 1908 Tunguska airburst was the first documented bolide-generated infrasound (Whipple, 1930), followed by a dozen or so events during the 1960s and the 1970s (Revelle, 1997; Silber et al., 2009). Since the inception of the Comprehensive Nuclear-Test-Ban Treaty (CTBT) in the mid-1990s (Brachet et al., 2010), infrasound has gained momentum as a vital sensing modality used towards global detection of large bolides (e.g., Pilger et al., 2015; Pilger et al., 2020).

In some instances, infrasound might have enough energy to induce seismic waves, known as air-coupled or atmospheric seismic waves, which can be detected by seismometers and other seismic monitoring instruments (Edwards et al., 2008). The characteristics of these seismic waves, such as their amplitude, frequency, and arrival time, can provide valuable information about the source. It's important to note that while bolide sonic booms can induce seismic waves, the seismic signals produced are typically much weaker and shorter-lived compared to those generated by earthquakes or other large-scale seismic events (e.g., impacts). There are three modes of coupling: (1) direct coupling (the incident acoustic wave induces ground motion); (2) precursory (generated by the infrasound wave impacting the ground at specific incidence angles that allow resonant coupling to subsurface seismic propagation modes that then travel independently to the recording station); and (3) impact (surface and body waves generated when a fragment of a meteoroid hits the surface) (see Cumming, 1989; Edwards et al., 2008).

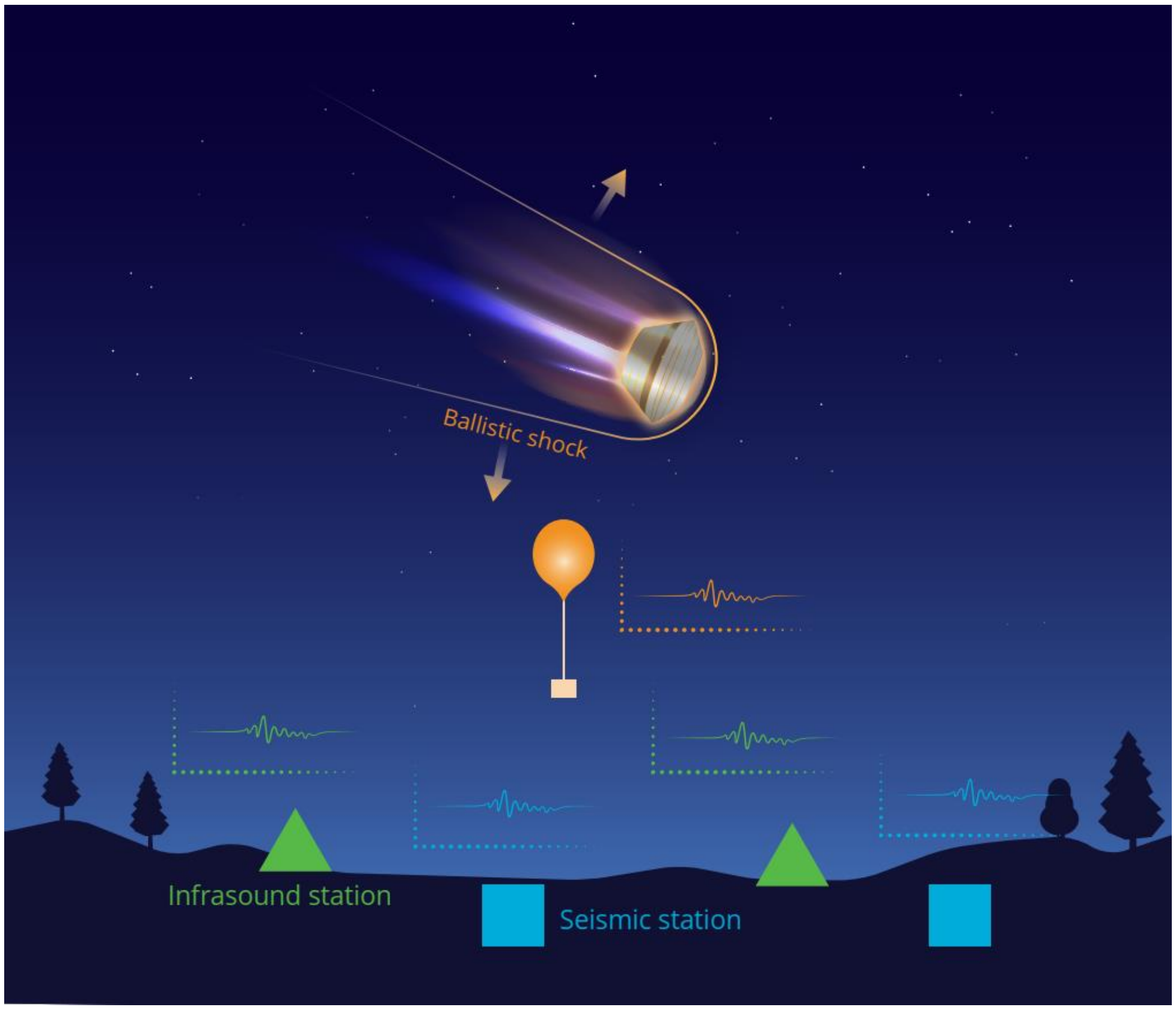

Figure 1: Diagram showing the shock wave generated by a sample return capsule (analogue to meteoroids) as it travels through the atmosphere. Ground-based infrasound and seismic instruments and airborne infrasound instruments could detect the shock waves depending on circumstances. Diagram not to scale. Diagram does not include all possible sensing modalities that might be used in geophysical observations of re-entry.

## 4. OSIRIS-REx and Geophysical Observation Considerations

Controlled and well-characterized re-entries from interplanetary space at velocities exceeding 11 km/s are exceptionally rare. Only five such re-entries took place since the end of the Apollo era, with the most recent one being OSIRIS-REx in September 2023 (Silber et al., 2023a). Prior to that, Genesis landed in 2004 (ReVelle et al., 2005), Stardust in 2008 (ReVelle and Edwards, 2006), Hayabusa 1 in 2010 (Fujita et al., 2011; Watanabe et al., 2011; Yamamoto et al., 2011), and Hayabusa 2 in 2020 (Nishikawa et al., 2022; Sansom et al., 2022; Sarli and Tsuda, 2017; Yamada and Yoshihara, 2022). The physical parameters for the five SRCs are listed in Silber et al. (2023a). Both Genesis and Stardust landed in the USA, and their signals were detected by instruments installed at West Wendover, UT airport (ReVelle et al., 2005; ReVelle and Edwards, 2006). Genesis was detected via infrasound, and the acoustic signatures generated by the latter three were recorded by both infrasound and seismic instruments. A review describing seismo-acoustic detections of these four re-entries is given by Silber et al. (2023a), and we therefore keep the discussion to a minimum.

The OSIRIS-REx space mission was designed to collect samples of the near-Earth asteroid Bennu and bring those particles to Earth for analysis (Ajluni et al., 2015; Beshore et al., 2015; Lauretta et al., 2017). This was the first asteroid sample collection mission for the USA. On September 24, 2023, the SRC separated from the main craft and entered the atmosphere at a very shallow angle (nearly horizontal). The SRC is 81 cm wide and has a mass of ~46 kg. The atmospheric interface was at an altitude of 125 km above a point close to San Francisco, CA (14:42 UTC, 8:42 MDT). After a few minutes flight, it safely touched down on the Department of Defense's Utah Test and Training Range (UTTR) at a speed of 5 m/s (18 km/h), slightly faster than originally anticipated (Gran, 2023). The re-entry consisted of several flight phases, including hypersonic, transonic, and dark flight (see Silber et al., 2023a for further details). The drogue parachute was supposed to open at an altitude of ~30.4 km to slow down and stabilize the SRC before the main parachute sequence but it failed due to faulty wiring (Francis et al., 2024). When the main parachute opened at 2.74 km altitude, the drogue was also released but because it was already cut loose, it flew off. The main parachute managed to sufficiently slow the SRC down, facilitating soft landing at 14:52 UTC, a minute earlier than planned (Francis et al., 2024; Gran, 2023).

The SRC return of the OSIRIS-REx mission offered a rare opportunity to record both the incident atmospheric pressure field of the incoming Mach cone at the ground surface and induced seismic motions near the Earth's surface at a known location in time and space. Factors that are important in deducing the effects of the acoustic-seismic interaction at the ground surface include obtaining basic knowledge about the incident wavefield such as its horizontal slowness and azimuth of approach (Lin and Langston, 2009a). It is testament to the accuracy of NASA orbital dynamics that the trajectory and arrival time of the returning capsule could be controlled for a landing in western Utah. However, the exact behavior of the expanding Mach cone and how it interacts with a particular place on the ground depends not only on the precise path geometry but atmospheric winds that can cause lateral variations in sound speed distorting the acoustic wavefront on its descent from the upper atmosphere. In principle, the horizontal phase velocity of the downgoing acoustic wave could vary from infinite velocity (vertical incidence) to approximately 0.33 km/s for near-horizontal wave propagation. Conversion of the atmospheric acoustic wave into propagating P and S body waves or Rayleigh waves will strongly depend on the local wave slowness. Directional attributes, such as particle motion, will also depend on wave azimuth of approach.

The nominal trajectory, based on the entry, descent, and landing (EDL) simulations (Francis et al., 2024) was provided by NASA (M. Moreau, personal communication); the ground track is shown in Figure 2. The color represents the altitude. The peak heating, dynamic pressure, and Mach number as a function of altitude are also shown in Figure 2. We note that any trajectory-related values (geographical coordinates, altitudes, peak heating, dynamic pressure, and Mach number) presented here correspond to the nominal trajectory (Francis et al., 2024) that was released by NASA before the OSIRIS-REx re-entry, and it therefore may not accurately represent the real-time re-entry trajectory. However, it is expected that the two would be in a very close agreement.

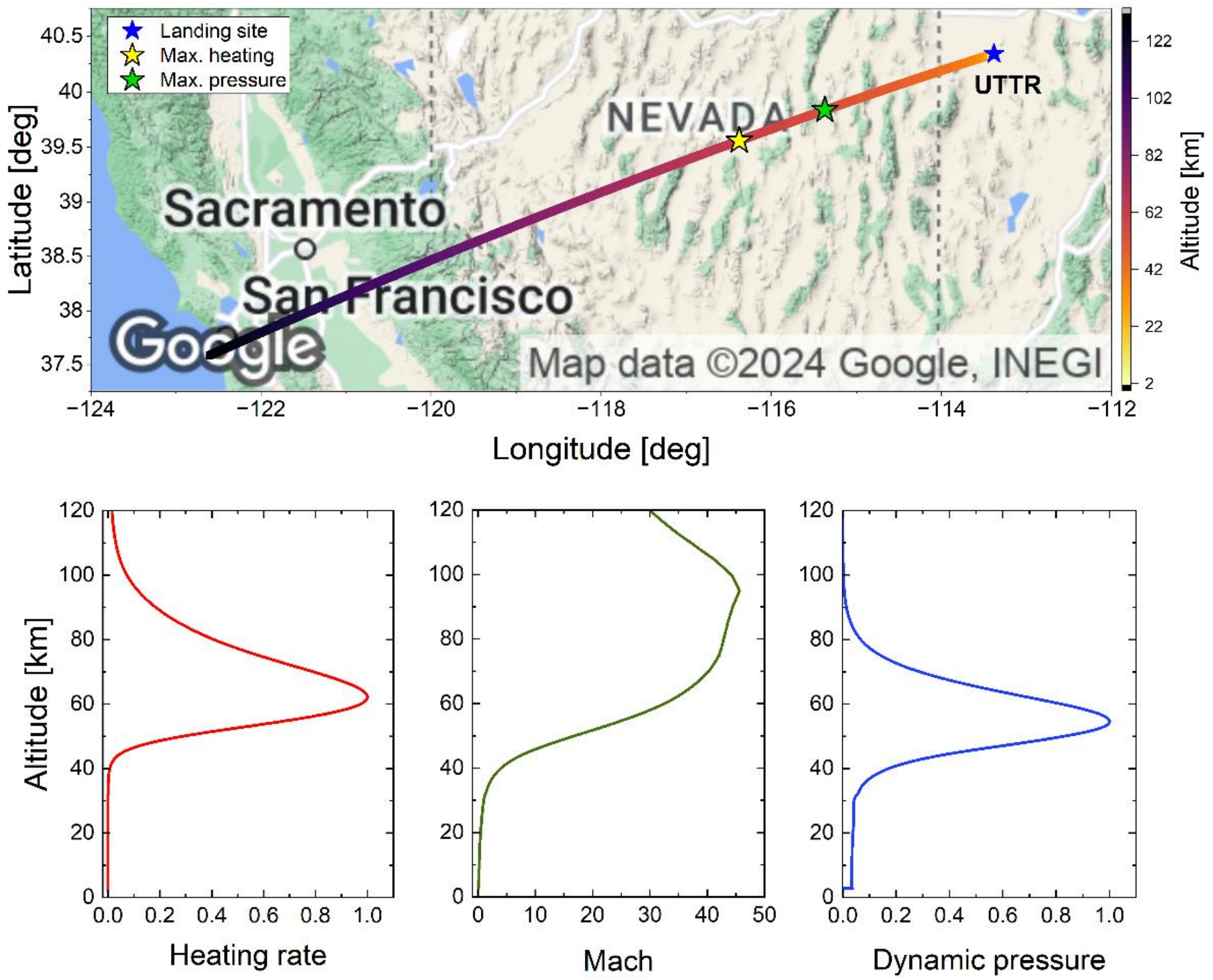

Figure 2: (Top panel) OSIRIS-REx ground track. Yellow and green stars show the points of maximum heating and maximum dynamic pressure, respectively. The landing site is indicated with a blue star. (Bottom panels) The normalized heating rate, Mach number, and normalized dynamic pressure as a function of altitude are shown in the three panels below the map. Trajectory data are courtesy of NASA.

One significant distinction between re-entry phenomena and conventional controlled experiments, such as static sources (e.g., chemical explosions) or sources with constrained lateral movement at mid- and low altitudes (e.g., rocket launches), is the extensive geographical area (Figure 2) covered and the multitude of potential observation points available to capture the dynamic changes occurring along the trajectory. Consequently, selecting the most relevant regions of interest becomes pivotal to ensure the acquisition of high-fidelity data. Although permanent infrasound and seismic stations are established throughout the USA, the immediate vicinity beneath and adjacent to the trajectory lacks sufficient geophysical instrumentation to reliably capture signals with a high degree of certainty. While incidental detections remain plausible, there exists uncertainty regarding the likelihood of signal recording. This uncertainty stems from the fact that the altitude of the SRC could be too elevated or the SRC ground track would be at a considerable distance from the existing permanent instrument installations. Given the infrequency of re-entry events from interplanetary space, relying solely on distant instruments to gather data entails substantial risk. Therefore, systematic planning of a dedicated observational campaign is imperative.

Our geophysical observational campaign was carefully planned to maximize the scientific output and ensure the highest chance of success while considering various intrinsic and extrinsic factors. These include, but are not limited, to the following:

i. Signal detection and data collection locations (e.g., identify locations through research and analysis where signals related to the re-entry are most likely to be detected; consider factors such as geological features, topography, and historical data on similar events to pinpoint optimal observation sites and prioritize locations that are accessible for instrumentation setup and maintenance; evaluate the potential for multiple observation points to capture different aspects of the re-entry and enhance data coverage).

ii. Cost considerations (e.g., develop a budget that accounts for all expenses associated with the campaign, including personnel salaries, fieldwork costs, equipment procurement and maintenance, logistics, and administrative expenses; allocate resources strategically to ensure the campaign's financial viability and explore cost-saving measures such as optimizing logistical arrangements to minimize expenses without compromising scientific objectives).

iii. Instrument synergy and deployment (e.g., assess the compatibility and capabilities of different instrumentation options to ensure they complement each other and provide comprehensive data coverage; consider factors such as instrument reliability, data transmission capabilities, and power requirements when selecting deployment locations and configurations; implement contingency plans and redundancy measures to mitigate the risk of instrument failure and ensure continuous data collection throughout the campaign).

iv. Environmental and infrastructure impact (e.g., conduct environmental risk assessments to identify potential safety hazards associated with fieldwork activities, such as exposure to extreme temperatures, rough terrain, or wildlife encounters; engage with relevant stakeholders, regulatory bodies, and environmental agencies to obtain permits and approvals for fieldwork activities; implement mitigation measures to minimize environmental disruption, such as site restoration efforts and adherence to best practices for minimizing habitat disturbance; coordinate with landowners, facility managers, and infrastructure operators to ensure minimal interference with existing infrastructure and facilities).

v. Personnel allocation (e.g., define roles and responsibilities for scientific and technical staff, students, and collaborators involved in the campaign and provide specialized training to personnel on wilderness safety, navigation techniques, and emergency first aid to enhance their ability to respond to safety incidents in remote locations; foster a collaborative and inclusive work environment that encourages communication, teamwork, and knowledge sharing among team members; prioritize the inclusion of personnel with a cardiopulmonary resuscitation (CPR) certification and wilderness knowledge in field teams to oversee safety protocols and emergency response procedures; establish clear lines of communication and designated safety officers within field teams to facilitate coordination and decision-making in emergency situations).

vi. Timeline and coordination (e.g., schedule safety briefings and training sessions to reinforce safety protocols and address any emerging safety concerns; establish communication protocols for reporting safety incidents or concerns, including designated channels for contacting emergency services and obtaining assistance; incorporate safety checkpoints into the campaign timeline to review safety procedures, assess risks, and adjust plans as necessary to ensure the ongoing safety of all personnel involved).

With respect to the most optimal observational point, the region beneath the peak heating point would be of primary interest since the energy deposition would be the greatest, and the signals the strongest. The peak heating was projected to take place over Eureka, NV, a small town along US Route 50, known as “the loneliest road in America”. Therefore, this region was deemed to be ideal for emplacement of geophysical instruments to best capture the signals from OSIRIS-REx.

At the coordinates corresponding to the peak heating (39.5585 N, -116.3852 E, altitude 62.1 km), the dynamic pressure was predicted to be 69% of the maximum. The SRC was estimated to be travelling at Mach 34.8. At this speed the Mach cone angle is only 1.65 degrees. Thus, for the purposes of modeling and signal prediction, the shock can be approximated as a cylindrical line source (Figure 1), travelling ballistically relative to the path of the object (Silber & Brown 2019; ReVelle 1976). Consequently, emplacing some number of instruments roughly perpendicular to the trajectory would theoretically capture the shock decay as a function of distance. Moreover, emplacing instruments in several locations beneath and roughly parallel to the trajectory, would theoretically capture signals generated at different parts of the trail and aid in studying the signal characteristics as a function of altitude and other factors (e.g., velocity, atmospheric specifications).

Another region of scientific interest from the observational standpoint would be the trajectory segment related to the SRC deceleration and the flight regime change from hypersonic to supersonic and finally subsonic. The lower altitude of the SRC would ideally provide ample opportunity for the signals to be detected by geophysical instruments. The maximum dynamic pressure was predicted to occur at an altitude of 54.5 km (39.8365 N, -115.3717 E). Here, the heating would have decreased to 66% of the maximum. In the case of OSIRIS-REx, this is just beyond the Nevada-Utah state line. For reference, the West Wendover, UT airport is due north.

The highest Mach achieved by the SRC was 45.6, at 95 km altitude (38.5178 N, -119.8486 E). Here, the heating rate was estimated at only 11% of the maximum achieved and dynamic pressure at ~0.5% of the maximum. Based on the available parameters, the onset of the shock wave is estimated to occur at an altitude of approximately ~80 km (or slightly higher). The shock wave, as soon as it forms, would also produce infrasound. It would be scientifically interesting to attempt to capture the shock wave as it forms at these altitudes. However, considering that the SRC is at a high altitude and the energy deposition is much lower than at the peak heating point, such an endeavor would carry a high risk of non-detection. Therefore, our observational campaign focused on the geographic region spanning from slightly west of Eureka, NV (roughly the peak heating point) towards east, in the area relatively close to the landing site (Figure 2).

## 5. Institutional Engagement and Site Selections

### 5.1 Institutional Engagement

Approximately 80 investigators from over a dozen institutions participated in this historical observational campaign. The primary participating institutions were: Sandia National Laboratories (SNL), Los Alamos National Laboratory (LANL), NASA Jet Propulsion Laboratory (JPL), Air Force Research Laboratory (AFRL), Atomic Weapons Establishment (AWE) Blacknest, Boise State University (BSU), Defense Threat Reduction Agency (DTRA), Idaho National Laboratory (INL), Johns Hopkins University (JHU), Kochi University of Technology (KUT), Nevada National Security Site (NNSS), Oklahoma State University (OSU), Southern Methodist University (SMU), TDA Research Inc. (TDA), University of Hawaii (UH), and University of Memphis (UM). For brevity, affiliations of those who were involved and/or contributed through primary institutions (e.g., student exchange, internships, second-level collaboration, and similar means) are stated at the front of the paper but not reiterated here.

To keep focus on the scientific aspect of the campaign, the main text might not always differentiate who did what unless contextually necessary. While each team had their own scientific objectives, the entire multi-institutional group collaborated towards the common goal of gathering high-fidelity data. In the remainder of this section, we first describe the geographical context, followed by ground-based observations, and conclude with balloon-borne observations. We also include the appendices and supplemental materials (SM) with additional pertinent information, which we will refer to throughout the main text.

### 5.2 Geographical Context

#### *5.2.1 West Region: Eureka, NV*

Two primary locations were selected as deployment sites in the region of Eureka, NV: Eureka Municipal Airport (EUE) (39.6039 N, -116.0036 E) and the Newark Valley (centered at 39.6833 N, -115.7217 E). EUE was selected because it was situated almost directly beneath the OSIRIS-REx re-entry path, was access controlled, and had large areas of pavement for equipment layout. Moreover, for balloon deployment, trajectory calculations using weather model outputs from previous years indicated that the balloons would most likely remain close to the re-entry path if launched from the EUE. Finally, there were very few other suitable sites in the area for multiple balloon releases. The town of Eureka graciously allowed us to use the airport. The Newark Valley was selected because it is traversed by Strawberry Road, which is not traffic heavy, passes through large plots of land owned by the Bureau of Land Management (BLM), and most importantly, runs north to south with a section situated directly beneath the nominal re-entry trajectory. This is ideal for configuring a transect with instrument installations perpendicular to the nominal trajectory. Permits or confirmation of casual use compliance were obtained from BLM to install infrasound and seismic sensors, as well as DAS. Bean Flat Rest Area (BFRA) was an additional location (39.4996 N, -116.5095 E), further west and very close to the point of peak heating, that was selected for seismic instrument installations. The list of institutions that deployed in the West Region is shown in Table 1. The map is shown in Figure 3.

#### *5.2.2 East Region: Utah-Nevada*

The East Region included several locations, chosen because of their proximity to the nominal re-entry trajectory and the ease of access. These included the NV-UT state line, West Wendover Airport (ENV) in Utah, and two locations east of the UTTR. The area around the NV-UT state line (centered around 40.201 N, -114.047 E), was selected because of accessibility and the Bureau of Land Management (BLM) landownership. Importantly, on the Utah side just beyond the NV-UT state line (centered around 40.1738 N, -113.9960 E), there is a local road (N Ibapah Road) that runs approximately north-south beneath the nominal re-entry trajectory. On the Nevada side, US Route 93 runs from the nominal trajectory and north up to ENV (the half-way point is approximately at 40.4773 N, -114.1555 E). ENV (40.7280 N, -114.0212 E) was previously utilized in observing the re-entry of Genesis and Stardust, although their nominal trajectories were significantly closer than that of OSIRIS-REx. Preliminary propagation modeling and hypersonic carpet prediction modeling using averaged atmospheric specifications and winds from previous years showed that signals generated by the OSIRIS-REx SRC would be received at ENV. The airport offered a secure large space that would allow sensor setup without any tampering. Two other locations east of the UTTR were selected, one north of Dugway, UT (40.2571 N, -112.7404 E), and the other one was in Clive, UT (40.7089 N, -113.1167 E). Dugway is ~63 km east, and Clive is ~52 km northeast from the nominal landing site. See Table 1 for institutions that deployed in the East Region. The map is shown in Figure 3.

#### *5.2.3 Distal Region*

Distal Region stations consisted of permanent infrasound array stations, which were part of larger networks and wider trial series, as well as dedicated stations that were installed for the purpose of detecting the OSIRIS-REx SRC re-entry. These stations were situated in several areas east (Price, UT), north (Boise, ID) and south (St. George, UT and NNSS, NV) relative to the SRC's nominal trajectory, and at distances ranging from ~250 to ~400 km. The list of institutions that operated these infrasound assets is shown in Table 1.

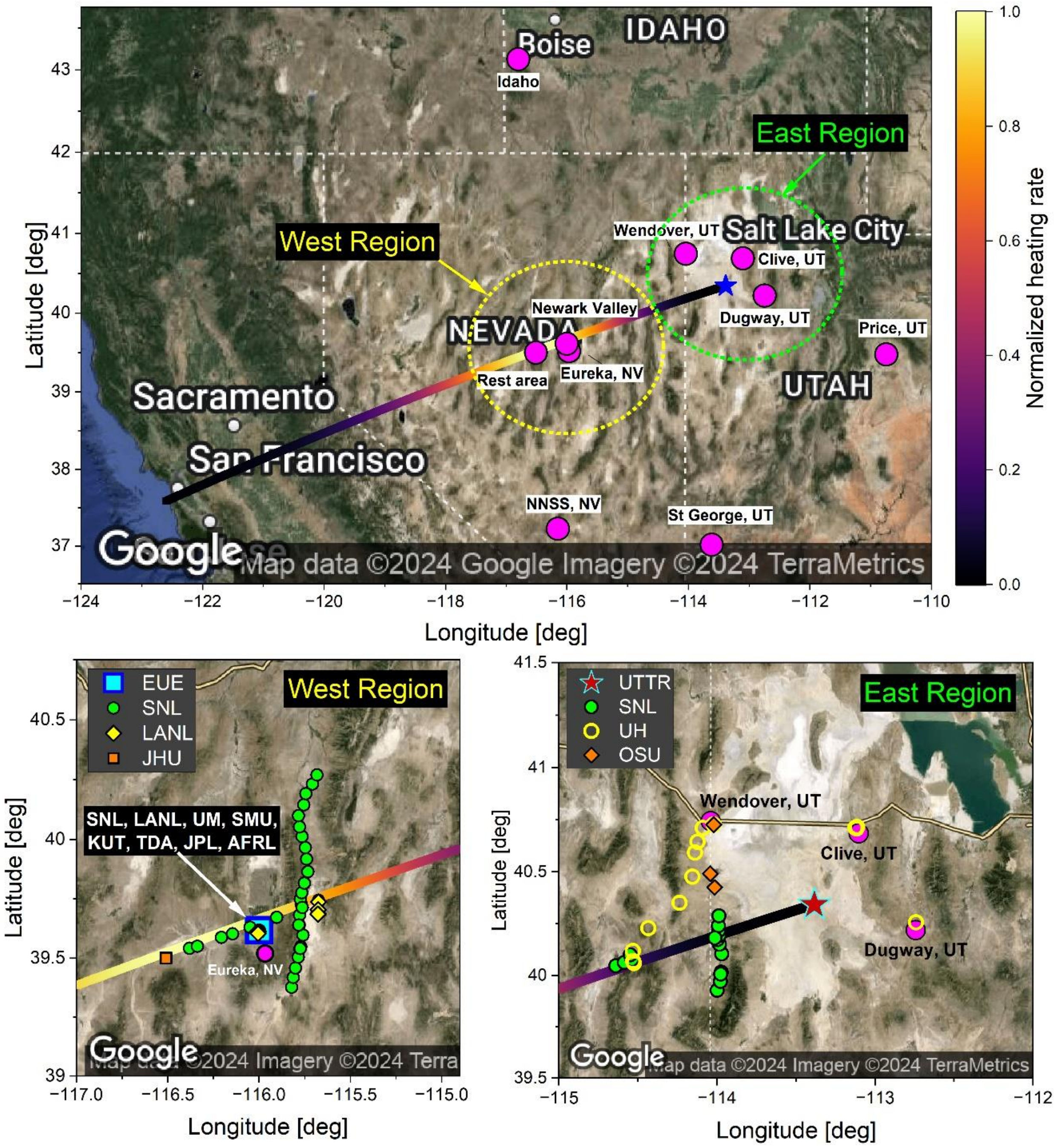

Figure 3: Maps showing deployment regions (top panel). Areas beyond the circled areas are within the Distal Region. Bottom left: Zoomed in West Region. Bottom right: Zoomed in East Region. Pink circles are various landmarks (to be explained further in the text). In the West Region, most institutions deployed at EUE. Installations beyond the airport are plotted separately by institutions for the West and East Regions. Instrument type is further enumerated Appendices B and C.

Table 1: List of institutions that deployed instruments in the various regions.

| West Region | | | East Region | | Distal Region |
|---|---|---|---|---|---|
| Eureka Airport | Newark Valley | Bean Flat Rest Area | West Wendover Airport | NV/UT | All sites |
| SNL, LANL, UM, SMU, KUT, TDA, JPL, AFRL | SNL, LANL | JHU | OSU, UH | UH, AFRL, INL | LANL, AWE, BSU, NNSS |

## 6. Instruments and Deployment

The team deployed over 400 sensors combined among all participating institutions, marking the most instrumented re-entry in history. For clarity, we present instruments and describe field deployments by instrument type (e.g., infrasound, seismic, etc.). Because it is not possible to include the particulars for that many instruments in the main text, we include the detailed list (consisting of the instrument make, sampling rate, the geographical coordinates, and the affiliated institution) in the Tables in the appendices. Our goal was to deploy instruments beneath the trajectory as well as perpendicular to it wherever possible in order to evaluate signal characteristics as a function of the SRC flight along and away from the propagation path (see Figure 3). The numbers and type of instruments each institution deployed are shown in Table 2.

Table 2: Summary of all instruments observing the re-entry of OSIRIS-REx. The detailed list of instruments with geographical coordinates can be found in Appendices.

| Institution | Infrasound (single sensor) | Infrasound (array) | Total # of sensors in arrays | Large N-array (# of sensors) | Audible microphone | Smartphone | Seismic | DAS | GPS | Balloons |
|---|---|---|---|---|---|---|---|---|---|---|
| SNL | 47 | 3 (x4) | 12 | - | - | 2 | 19 | - | - | 6 |
| LANL | 6 | 2 (x4) + 1 (x6) | 14 | - | 1 | 1 | 6 | 2 | 5 | |
| AWE | - | 1 (x4) | 4 | - | - | - | - | - | - | |
| BSU | - | 3 (x4), 1 (x44) | 56 | - | - | - | - | - | - | |
| JHU | - | - | - | - | - | - | 11 | - | - | |
| JPL | - | - | - | - | - | - | - | - | - | 2 |
| KUT | - | 1 (x4) | 4 | - | 5 | - | - | - | - | |
| OSU | 12 | 1 (x4) | 4 | - | - | - | - | - | - | |
| SMU | - | 1 (x4) | 4 | - | - | - | - | - | - | |
| TDA | - | | | 114 | - | - | - | - | - | |
| UH, AFRL, INL | - | - | - | - | - | 33 | - | - | - | |
| UM | - | 2 (x4) | 8 | - | - | - | 20 + 96 | - | - | |
| | **65** | **16** | **106** | **114** | **6** | **36** | **56** | **2** | **5** | **8** |

### 6.1 Ground-based Infrasound and Audible Acoustic

SNL deployed three 4-element arrays and 47 single sensor stations in the West and East Regions. The three infrasound arrays, deployed in the West Region, consisted of analog and digital Hyperion sensors arranged in a triangular formation. Hyperion sensors are manufactured by Hyperion Technology Group, Inc. and are widely used in a variety of infrasound monitoring applications (e.g., Bowman and Albert, 2018). One array was at the EUE, in the northeast corner, and the other two were in Newark Valley along Strawberry Rd. We aimed to arrange the arrays into an “L”, such that two arrays are positioned roughly parallel to the nominal OSIRIS-REx trajectory (~10-12 km shortest path), and one perpendicular to it (~24 km shortest path). The reasoning is that such configuration could potentially capture the signals from different parts of the trail and help evaluate signal properties as it propagates away from the re-entry path. Each array had two colocated seismic nodes. The instruments were powered by marine deep-cycle batteries and solar panels. Timing was tracked through GPS.

In addition to the three arrays, SNL and DTRA also deployed 46 single stations consisting of Gem infrasound loggers (Figure 4A) in various locations (30 in the West Region, and 16 in the East Region). Gems (version 1.01) are self-contained sensor loggers optimized for deployment and maintenance in large numbers (i.e., small, lightweight, low power consumption, cable-free, and fast deployment process) (Figure 4A) (Anderson et al., 2018). They sample at 100 Hz, with a flat response between 0.039-27.1 Hz and root-mean-square self-noise of 1.55 mPa (0.5-2 Hz) and 3.9 mPa (0.1-20 Hz). Gem infrasound loggers have previously been used in campaigns using large numbers of sensors (Anderson et al., 2023; Rosenblatt et al., 2022; Scamfer and Anderson, 2023), campaigns where ease of concealment was essential (Ronan, 2017; Tatum et al., 2023), and airborne infrasound recording requiring a lightweight sensor (Bowman and Albert, 2018; Bowman and Krishnamoorthy, 2021; Krishnamoorthy et al., 2020; Silber et al., 2023b). Twenty-three Gems were installed along Strawberry Rd. to form a transect relative to the nominal trajectory, extending both north and south. Out of these, 11 were to the south relative to the nominal trajectory (~44 km due south, and ~40 km shortest distance), and 12 to the north (~56 km due north, and ~51 km shortest distance). The approximate separation between stations was 5 km. The sensors were powered with batteries and portable solar panels. To reduce the adverse effect of wind noise, Gems were emplaced inside bushes, but the solar panels were left exposed. The remaining 7 Gems were deployed about 3-5 km from and parallel to the nominal ground track, extending from the point of peak heating towards east, just beyond Strawberry Rd. The total end-to-end distance was ~65 km. In the Eastern Region, SNL deployed 15 Gems. Four of these were very close to the nominal trajectory (~1 – 3 km) and in the vicinity of US Route 93 Alternate (NV). Some ~45 km to the east, ten Gems were installed along the NV-UT state line, forming a 41 km long transect (N Ibapah Rd.), with an additional Gem positioned slightly to the west to coincide with the nominal re-entry path. All instruments were installed two days before the OSIRIS-REx re-entry and removed either immediately after the re-entry or the next day.

TDA Research deployed a 115-element Large N-array, collocated with SNL's infrasound array in the northeast corner of EUE, in a 100 x 100 m array. The sensors (Figure 4E), designed and built by TDA Research Inc., were previously tested during a controlled field experiment. This observational campaign was the second fielding, and the first time the sensors were used against a real-life event. TDA's sensors are low cost, and specifically designed to be networked into large and dense arrays with hundreds of sensors and wirelessly stream data to a central computer. The array design is modular and flexible, and its size can be anywhere from 5 sensors up to 500 sensors added in groups of five. These sensors have a sensitivity of <0.1 Pa and a sampling frequency of 200 Hz (that can be increased up to 330 Hz if needed). Each sensor has an on-board battery with a battery lifetime of approximately nine days. They also come equipped with a solar panel that will recharge a day's worth of power in 1-2 hours of sunlight and will fully recharge the sensors in 9-13 hours. They are located using a differential GPS system with accuracy of <1cm and clocks from all sensors in the array are synced to within 1 ms. The sensors weigh 0.703 kg each and are 33 cm tall when fully assembled. TDA's sensors minimize wind noise by sampling at only 1.3 cm off the ground, taking advantage of the ground's boundary layer to reduce effective wind speed by 75%.

There were three teams from LANL, two in the West Region (DAS/seismo-acoustic, and GPS (see Section 6.5)), and one in the Distal Region. The DAS/seismo-acoustic LANL team in the West Region deployed DAS at EUE and Newark Valley, also collocating infrasound (Figure 4B) and seismic sensors. Infrasound sensors (Hyperion IFS-3000) were installed at strategic locations along the DAS fiber (for complete instrumentation details, see Appendix). Additionally, at the Newark Valley site, a PCB microphone sampling at 50 kHz was deployed near the DAS interrogator setup location, and a personal iPhone was filming during the re-entry.

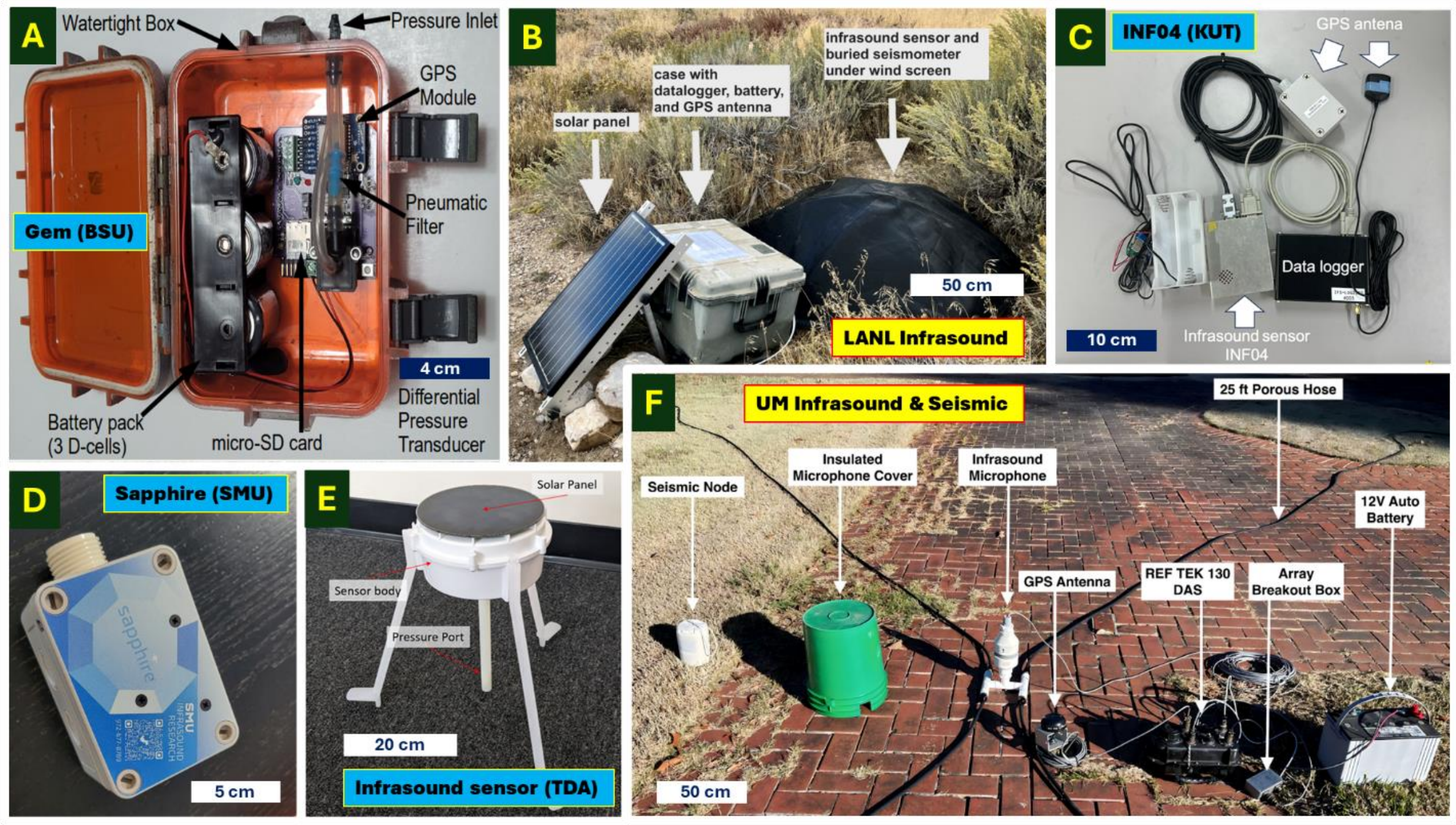

Figure 4: Representative examples of infrasound and seismic sensors and their field installations. (A) Gem Infrasound logger developed by BSU (photo credit: J. Anderson); (B) infrasound installation (LANL) (photo credit: C. Carr); (C) INF04 infrasound sensor (KUT) (photo credit: Y. Nikishawa); (D) Sapphire infrasound sensor (SMU) (photo credit: E. Silber); (E) TDA's infrasound sensor (photo credit: D. Eisenberg); (F) Infrasound and seismic instruments (UM) (photo credit: S. Bazargan).

The Distal region LANL team (alongside AWE, BSU, and NNSS) focused on infrasound data collection at distal infrasound stations. The high speed of arrival and the known time and path made this an ideal opportunity to test LANL's shock wave propagation estimate algorithms (Blom et al., 2024). Three arrays were deployed by LANL, one in Price, UT, one in St. George UT, and one on the north side of the NNSS, NV. These sites were chosen because an early estimate of regional infrasound arrivals indicated that they were likely to occur at these locations. The sensors deployed were Hyperion IFS-3000 infrasound sensors. Digitization was done with RefTek 130 digitizers powered by batteries that were kept charged by solar panels. These three arrays were deployed during the week prior to the re-entry. There were no team activities on the day of the re-entry.

At EUE, KUT deployed five microphones, as well as four infrasound sensors (SAYA INF04), the same kind as those used to observe Hayabusa 1 and Hayabusa 2 SRCs (Figure 4C). Hayabusa 1 was also observed with Chaparral microbarometers (Ishihara et al., 2012; Sansom et al., 2022; Yamamoto et al., 2011). The sampling rate of SAYA INF04 was 100 Hz (using Mathematical Assist Design Lab's INFLOG 2020). These infrasound sensors were arranged to form an elongated north-south triangle array (see Appendix for sensor locations).

UM installed a four-element infrasound array at each of two subarrays of the seismic array (described in the next section) at EUE. Instruments were made by VLF Designs and are flat to pressure between 0.25 to 25 Hz. These were placed on the ground surface within a plastic bucket insulated with 10cm rubber foam. Four porous 7.5m hoses attach to the bottom pressure manifold of the microphone. The four elements were arranged in a triangular configuration with a central element. Elements communicated with a RefTek 130 digitizer via 30 m cables to a breakout box that routed infrasound signals to channels 1-4. Timing was through GPS. The instruments and digital acquisition system were powered by a 12V car battery (Figure 4F). The rationale for having two infrasound arrays was to ensure that we could record the atmospheric pressure signal at a minimum of one position with the other being a backup. The small aperture (~60m) infrasound arrays precluded high resolution for determining the incident wave slowness and azimuth for the expected near-vertical incidence angle for the incoming N-wave. However, having four instruments in each array added an additional redundancy in measuring the incident wave pressure to compare with the test seismometer that was placed at the center of each array.

SMU deployed four lightweight portable Sapphire sensors (Figure 4D) at the EUE. The Sapphire is an infrasound nodal recorder developed at SMU inspired by the Gem infrasound node (Anderson et al., 2018) and similar in design (although developed independently) to the node described by den Ouden et al. (2021). The Sapphire response is reasonably flat above 0.1 seconds with a sensor that is factory calibrated to 0.25% amplitude response, enabling the recorder to apply the factory calibration constant and log pressures in units of pascals thus for most experiments avoiding the need to do instrument corrections. The unit records continuously for about two weeks on four AA batteries, recording at 128 Hz with a GPS controlled clock. Although the Sapphire sensor has a higher self-noise than more expensive infrasound microphones, its low-cost, good calibration, quick deployment, and small size make it convenient for many experiments that expect reasonable signal-to-noise (SNR) in the 1-5 Hz band.

OSU set up their instruments at the ENV, deploying three different models of microbarometers. A 4-sensor infrasonic array was formed with Chaparral Physics Model 64S sensors and deployed at ENV. Each sensor had a nominal sensitivity of 0.08 mV/Pa and a flat response to within 3 dB from 0.01 Hz to 245 Hz. Each sensor was mounted within a weatherproof case (1300 Case, Pelican). A single data acquisition system (PGS-140 4-channel, Pegasus) was used to record the Chaparral Physics sensors at 1000 Hz. The nominal separation between each sensor and the center sensor was 60 m, which produced an aperture of 112 m. OSU also deployed eight model ISSM23 microbarometers manufactured by the Wilson Engineering Research and Development (WERD). These sensors had a nominal frequency range of 0.1 to 200 Hz and had on-board sampling at 400 Hz. One of the sensors malfunctioned during the deployment leaving only seven that successfully recorded the re-entry. Four of the sensors were arranged in a triangle with an aperture of 51 m and centered on the Chaparral Physics array center sensor, with the fourth WERD sensor colocated at the center. The remaining three WERD sensors were positioned around the southwest sensor of the Chaparral Physics array with an aperture of 51 m.

OSU also deployed five Gem infrasound loggers (Anderson et al., 2018). Each sensor was secured within an enclosure that was then secured inside of a small Styrofoam box with the side walls replaced with the windscreens developed in Swaim et al. (2023). Three of the Gem sensors were arranged in a triangle with an aperture of 49 m and the east corner of the triangle colocated with the center Chaparral Physics sensor (and the one WERD sensor). It should also be noted that the UH also colocated a RedVox sensor at this central location, and UH colocated several of their sensors with OSU sensors. The remaining two Gem sensors were located between ENV and the OSIRIS-REx SRC flight path to the south of the airport. These Gem sensors were 26 km and 34 km south of the ENV deployment.

UH, along with AFRL and INL, deployed smart phones with RedVox app which utilizes the phone's built-in microphone (Garcés et al., 2022) in a variety of locations, including the West, East, and Distal Regions. Deployment locations in the East Region included Clive, UT, Dugway, UT, and around US Route 93 Alternate, NV.

The BSU infrasound campaign used Gem infrasound sensors (version 1.01). The instruments used belong to BSU and were deployed as part of a temporary network active between July-October 2023. The southwest Idaho infrasound network operated by BSU was deployed in Reynolds Creek Experimental Watershed (Seyfried et al., 2018) from July-October 2023 with objectives of recording a prescribed wildland fire and regional signals in addition to the OSIRIS-REx entry. The large array ("TOP") included 44 sensors approximately in a rectangle with overall dimensions of 210 m x 120 m; its large size was intended to facilitate detecting weak signals while providing precise backazimuths in beamforming operations. Additionally, three four-element small arrays (JDNB, JDSA, JDSB), were deployed within 1.5 km of TOP, helping to increase the spatial extent of the overall network. The smaller dimensions of these arrays are due to being constrained to small protected zones within the anticipated prescribed burn area. When possible, sensors were placed in or under shrubs in order to mitigate wind noise in these treeless sites.

### 6.2 Seismic

SNL deployed two seismic nodes colocated with each infrasound array. The nodes were buried at EUE because we had permission to dig holes. However, land permits at the other two arrays did not allow for digging and the nodes were placed on the surface. An additional 12 seismic nodes were deployed at EUE, distributed across a large area. As mentioned in the previous section, LANL installed 6 seismometers colocated with infrasound sensors (see Appendix for further details).

At EUE, UM also set up seismic sensors. A 1 km aperture, phased seismic array was sized to fit in the northern part of county land associated with the EUE (see Appendix C). The EUE was chosen because of its proximity to the ground track of the incoming capsule trajectory (within 10 km) and because Eureka County allowed seismometers to be buried. The phased array is relatively unusual and based on the "Golay 3x6" geometry (F. Followill, personal communication 2006). It consists of 6 tripartite subarrays arranged in a surprisingly open geometry following design principles of minimizing the number of array elements while maximizing the array spatial bandwidth (Followill et al., 1997). This can be seen in the co-array diagram that shows uniform sampling in space (Figure S16 in Appendix A). The broadband array response (Nawab et al., 1985) shows a highly focused beam that can resolve the slowness of the expected infrasound signal to 0.02 s/km. Instrumentation for the seismic array elements consisted of Magseis-Fairfield three-component nodal seismometers (see Appendix C). These seismometers have a low-frequency corner near 5 Hz, have self-contained GPS timing, and power and data storage for 30 day deployments. Installation consists of digging a 20 cm deep hole with a posthole shovel such that the top of the seismometer is within 5 cm of the surface in order to maintain GPS lock. Seismometers were oriented with respect to the North using a magnetic compass. Instruments were installed on September 21 and 22 at locations determined using a handheld GPS receiver. Thus, location accuracy is estimated to be within 3m of the target locations. Note that 20 seismic sensors served to detect OSIRIS-REx. The refraction profiles included 48 vertical component geophones and 48 horizontal component geophones but these were not used to detect the SRC signals.

At the BFRA site, JHU installed Fairfield ZLand 3-component nodes equipped with GPS timing and inbuilt power supply. These are deployed from a handheld terminal and placed in a small hole in the ground. Recovery also uses this terminal. The sampling rate is 2000 Hz. Instruments were placed into an 11-station array, in a cross-shape with the long axis aligned parallel to the OSIRIS-REx trajectory. The field site was left unattended, and instruments were collected back in the afternoon after the re-entry.

### 6.3 DAS with Co-located Seismo-acoustic Sensors

LANL deployed single mode optical fiber at two sites: EUE and in Newark Valley. Fiber was laid on the ground (deployment photos are in Appendix A). An AP Sensing instrument (N5225B-R100) was used at the airport to probe 4.5 km of fiber. The AP Sensing DAS had a sampling frequency of 500 Hz, a gauge length of 5 m, and a channel spacing of 1.23 m. In Newark Valley, a Silixa iDASv2 (Version 2.4.1.111) and an Alcatel OptoDAS were connected to 7.5 km of single mode optical fiber. The iDAS used a sampling frequency of 500 Hz, a gauge length of 10 m, and a channel spacing of 2.0 m. The OptoDAS used a sampling frequency of 10 kHz, with a gauge length of 5.1 m, and a channel spacing of 1.02 m. All DAS units operated intermittently for testing purposes in addition to during the re-entry, unlike the seismometers and infrasound sensors that operated continuously. LANL installed six of each colocated seismometers (Geospace HS-1 3 Component) and infrasound sensors (Hyperion IFS-3000) at strategic locations along the DAS fiber (for complete instrumentation details, see Appendix). The seismometers and infrasound sensors recorded at 200 Hz with each seismometer-infrasound sensor pair connected to a RefTek 130 datalogger with timing information provided by Garmin GPS 16x-HVS antennas. All instruments were deployed specifically for the purposes of capturing the re-entry. Instruments were installed over several days prior to the re-entry, and removed by the evening of 24 September 2023 (local time).

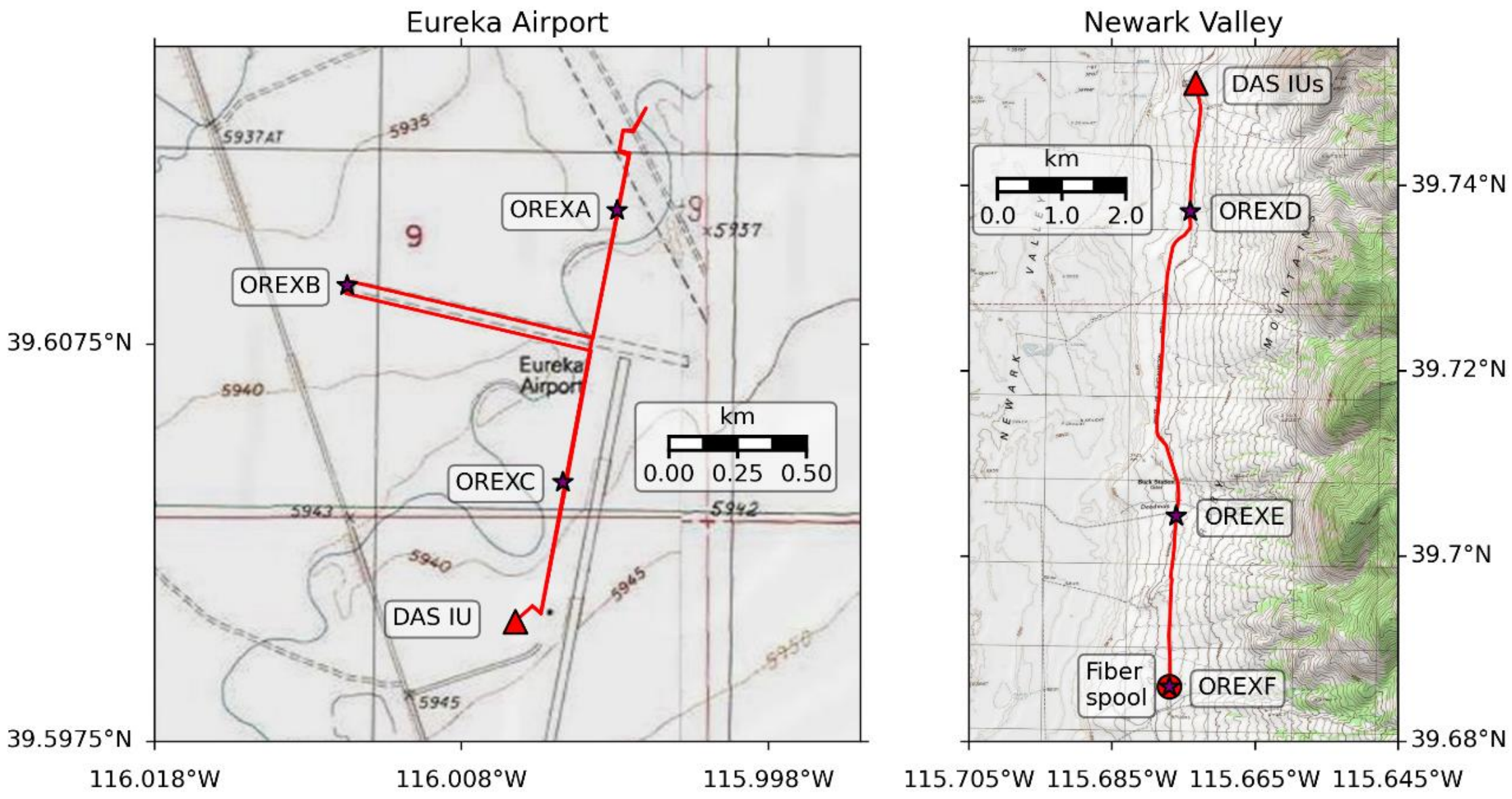

Figure 5: Map of DAS and colocated seismometers and infrasound instruments deployed at two sites: (left) the Eureka Airport and (right) in Newark Valley. OREXA, B, C, D, E, and F are colocated seismometer/infrasound pairs. DAS IU(s) indicates the DAS interrogator unit(s). At the Eureka airport, fiber was placed running from the DAS IU towards OREXC, and continuing along the edge of the main taxi runway before bending 90 degrees towards the west (towards OREXB) along the cross runway, then returning to the main taxiway and continuing north towards OREXA. The fiber returned from the turnaround point to the north of OREXA and continued directly towards OREXC without returning to OREXB. In Newark Valley, the fiber ran from the DAS IU south past OREXD, OREXE, and OREXF. A spool with the remaining fiber was placed near OREXF.

### 6.4 Balloons

SNL and JPL deployed balloons carrying sensor payloads. There are several types of balloons capable of bearing infrasound payloads. All of them depend on relatively low winds at the launch site, and some have additional restrictions, such as requiring sunlight to fly. To increase the odds that at least one balloon would be successfully deployed during the OSIRIS-REx overflight, SNL and JPL deployed two helium zero-pressure balloons, two helium meteorological balloons, two 7 m diameter heliotrope solar hot-air balloons towed aloft using helium meteorological balloons, and two ‘cloudskimmer’ 3.5 m heliotrope solar-hot air balloons. Each balloon carried a parachute to slow the payload during descent.  They also carried a flight termination system that ended the deployment when the balloon crossed a pre-programmed geofence and/or after a certain amount of time had elapsed. Some balloons utilized Balloon Ascent Technologies Bounder and others used a High Altitude Science Stratocut termination system. The balloons were tracked using StratoTrack Automatic Package Retrieval System (APRS) radios that transmitted during flight and a SPOT TRACE unit that reported the payload position after landing using the Globalstar satellite network. The location of this equipment on the balloon is shown in Figure 6.

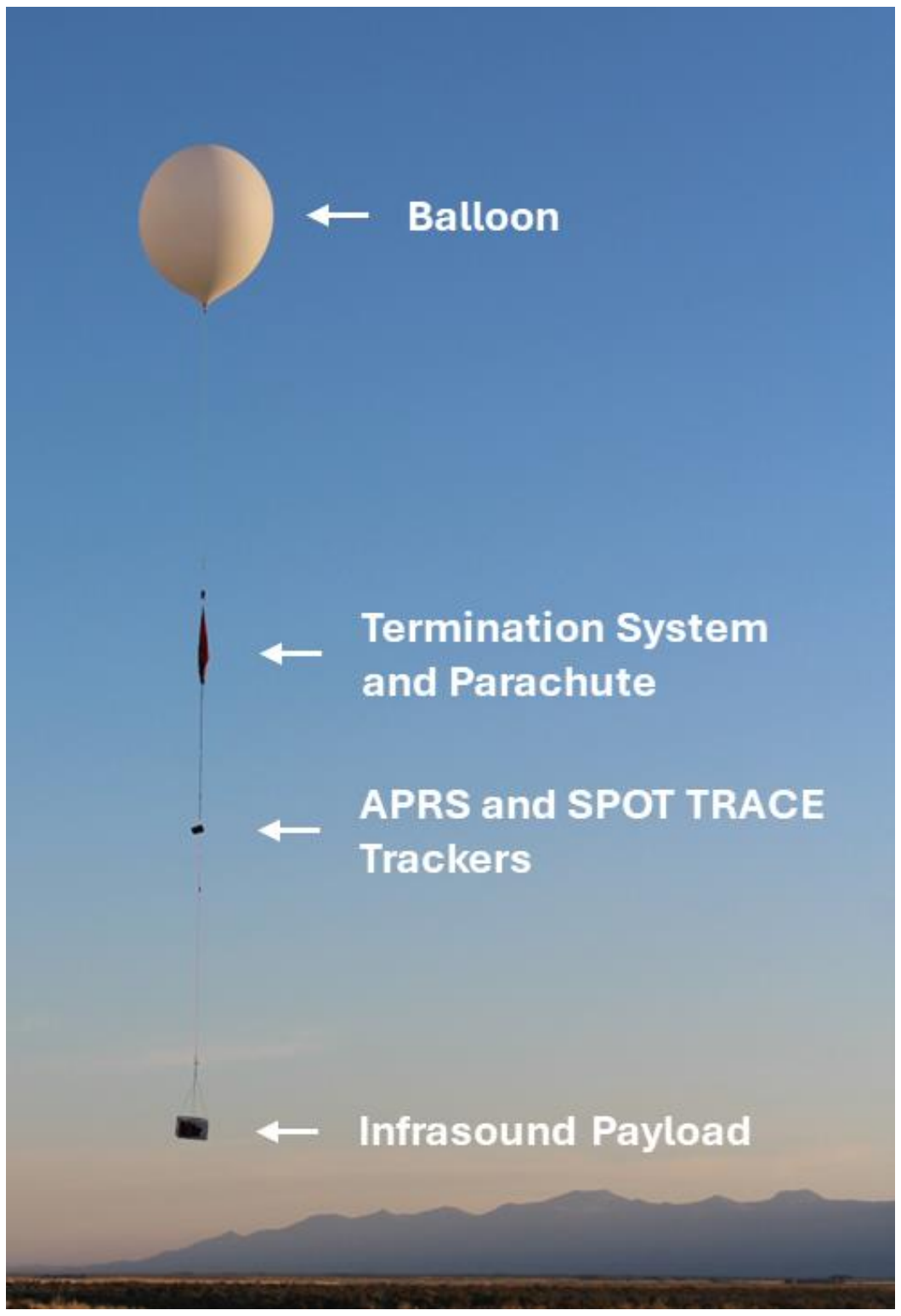

Figure 6: Example infrasound balloon configuration from the OSIRIS-REx deployment. This is the low-altitude weather balloon just after release. Photo credit: R. Lewis.

Helium zero pressure balloons climb until they reach their neutral buoyancy altitude. Two zero pressure balloons with a maximum capacity of 4300 cu ft (121.8 $m^3$) were fielded, each targeting a different altitude in the lower stratosphere. The sensor packages consisted of Paroscientific Digiquartz 15A-IS microbarometers and an InertialSense µINS inertial measurement with a Raspberry Pi flight computer and custom-built interface board (similar to Brissaud et al., 2021). The Paroscientific microbarometers recorded at 158 Hz with an internal five-stage anti-alias filter set at 25.1 Hz, providing a pre-digitized output. The inertialSense IMU was sampled at 15 Hz. Each balloon had two sensor packages separated by a 33 m tether. The inertialSense IMU also provided timing information from the Global Navigation Satellite Systems (GNSS) network to the Paroscientific barometers through the custom interface board for timestamps accurate to 1 microsecond for time-of-flight analysis.

Meteorological balloons climb continuously until they either burst or the flight is terminated by other means. Despite relatively high levels of wind noise during ascent, Popenhagen et al. (2023)

found that distant acoustic events could be recorded on such platforms. Thus, we deployed two meteorological balloons with infrasound payloads: a Kaymont 3000 g envelope released early with a rapid ascent rate and a Kaymont 600 g envelope released later with a very slow ascent rate. The rapid ascent balloon was meant to capture the OSIRIS-REx signal in the lower/middle stratosphere, and the slow ascent balloon was meant to capture the signal within a few acoustic wavelengths of the ground. Each balloon carried two InfraBSU microbarometers (Marcillo et al., 2012) and a Camas condenser microphone (Slad and Merchant, 2021) digitized on a DiGOS DATA-CUBE recorder digitizing at 400 Hz. One of the two infraBSU microbarometers was polarity reversed, which assists in discriminating between true pressure signals and spurious non-pressure fluctuations like those caused by electronic interference (Bowman et al., 2019).

Heliotrope solar hot-air balloons rise into the lower stratosphere, where they maintain a constant altitude until sunset or the flight is terminated by other means (Bowman et al, 2020). They are less expensive and can be launched at a more rapid cadence than zero pressure balloons but have more restrictive launch conditions. This constant altitude greatly reduces wind noise on infrasound microbarometers, permitting much fainter signals to be recorded. However, the OSIRIS-REx overflight occurred too close to dawn, meaning the heliotropes would not be able to reach their neutral buoyancy in time to record the acoustic signal. Therefore, two 7 m diameter heliotropes were towed aloft by meteorological balloons and then released into sunlight, achieving level flight before the OSIRIS-REx event. After dawn, two 3.5 m 'cloudskimmer' heliotropes were launched from the ground in the hopes that their slow ascent rate would reduce wind noise to acceptable levels. One 7 m heliotrope carried a single Gem microbarometer (Anderson et al., 2018), and the other carried two Gem microbarometers separated by a 30.5 m long tether. The Cloudskimmer balloons each carried an Android cellphone running the RedVox infrasound recording app (Garcés et al., 2022; Popenhagen et al., 2023). The RedVox phones recorded pressure data at 800 Hz.

Starting about a week before the re-entry, trajectory calculations were performed at least daily in order to refine our launch times and ascent rates. The area of operation was challenging, with restricted airspace on three sides. Furthermore, the OSIRIS-REx landing site was to the east of Eureka Airport – exactly where the balloons were expected to go. Therefore, we set termination geofences to prevent balloons from drifting into restricted airspace, including the OSIRIS-REx landing site.

Balloons were launched in two batches: an initial salvo meant to capture signals at high altitude (>20 km) and the second intended for very low altitude recordings (within several acoustic wavelengths of the ground). The first batch included the 7 m heliotrope balloons towed aloft using helium-filled meteorological balloons, one meteorological balloon with an infrasound payload, and the two helium zero pressure balloons. The first balloon was launched at 11:54 UTC, and all but one had been launched by 12:12 UTC. The final zero pressure balloon was launched at 13:12 UTC. The second batch consisted of a very slowly ascending meteorological balloon and the two cloudskimmer heliotropes. The meteorological balloon was released at 14:00 UTC and the last cloudskimmer was released at 14:13 UTC.

Despite gusty winds in the town of Eureka, conditions were calm at the airport due to a strong temperature inversion that had set up overnight. Winds began to pick up around 14:05 UTC, resulting in the decision to add helium to the cloudskimmers to help them get off the ground faster. Because of the rising wind and the imminent arrival of the OSIRIS-REx SRC, we opted not to launch the spare heliotrope balloon. Instead, we left the still-recording spare payloads at the launch site and proceeded to the pilots' lounge. After the capsule overflight, the balloon flights were automatically terminated. Payloads from the low altitude meteorological balloon and both cloudskimmers were recovered on the same day. The remainder of the payloads were recovered the following day.

**6.5 GPS**

LANL's GPS team deployed in the West region, with the aim of measuring and characterizing ionospheric Total Electron Content (TEC) signatures (via GPS L-band measurements) as well as atmospheric current and electric field signatures. Analyses involve modeling and measuring signature speed and period, geolocating likely sources, and estimating source strength using the LANL/GPS Rex-five stations with controlled/compact placement and large data-rate collections to probe small-scale ionospheric effects. The most significant caveat of this method, of course, is that ionospheric TEC signatures with the highest signal-to-noise ratio (SNR) are known to be tens of minutes delayed from any source in the troposphere. The LANL/GPS team deployed five GPS/GNSS Ground Stations at EUE along the runways: four along one taxiway (stations Rex-2, Rex-3, Rex-4 and Rex-5), and one at the end of its orthogonal dirt runway (station Rex-1). The geographical coordinates are listed in the appendix. Septentrio PolaRX5s GPS receivers were used in all stations with Veraphase 6000/High-Precision Full GNSS Spectrum Antennas. Standard solar panels and voltage regulators were used to supply power. Power expectations were designed to ensure that consistent power was maintained.

When properly calibrated, GPS Ground stations measure the group and phase dispersion of L-band signals (1-2 GHz) to all GNSS satellites simultaneously. These measured quantities determine the integrated electron density along the line-of-sight or Total Electron Content (TEC) to each satellite. Nearly all of that dispersion occurs in the ionosphere, and mostly near the altitude of peak ionization (250-400 km). As a result, these lines of sight, mapped though the ionosphere, can be used to scan for small changes and to characterize and locate atmosphere-impacting events perturbing the upper atmosphere.

The GPS stations, atmospheric current, and electric field measuring instrumentation were set up a few days prior to the OSIRIS-REx SRC re-entry to ensure all units were working properly and calibrated. The GPS stations were deployed less than 400 m apart from one another to allow detection of small-scale ionospheric disturbances. The most widely separated stations were 2000 m apart, nearly spanning the EUE main taxiway. In addition, the data-measuring interval was set to 50 ms to enhance the resolution of small time-scale measurements of ionospheric disturbances. The OSIRIS-REx SRC re-entry occurred nearly an hour after dawn under clear skies.

## 7. Preliminary Results

Here we present a snapshot of the preliminary data and our initial results. Nearly all instruments located near the nominal trajectory (i.e., direct arrivals) successfully detected the signals produced by the re-entry of OSIRIS-REx. Additionally, numerous stations in the Distal Region also captured the signals. Illustrative examples of detections are shown throughout this section to demonstrate the remarkable success of this largest ever geophysical observational campaign of a re-entry. We note that some results are omitted from this paper because detailed analyses are underway by various teams and will be disseminated as separate studies in due course.

### 7.1 Witness Reports

During the anticipated time window during which the OSIRIS-REx SRC overflight was expected to take place, the teams in the West Region exercised the so-called silent observation time. At EUE, some team members were stationed at the airport entrance to prevent inadvertent vehicular intrusion into the observation zone. The highway adjacent to EUE is commonly busy, and with steady tractor-trailer traffic. We asked the local police if they could assist in temporarily closing the highway during the overflight, but the request needed to be escalated with the Nevada State police. Despite lacking direct confirmation regarding the feasibility of our request, the EUE team observed a notable absence of vehicular activity for several minutes prior to re-entry, indicative of a road closure. The area was nearly windless and very quiet, and the only dominant audible noise came from birds and a rooster. Regrettably, approximately one minute before the anticipated OSIRIS-REx overflight, we noted a gradual resumption of vehicular activity. By the time traffic returned to full speed, the OSIRIS-REx SRC had already passed but not before some traffic noise started to become apparent. At EUE, the experiences in audible perception of a possible sound generated by the OSIRIS-REx re-entry varied. Some people perceived it as a single soft “thump”, some as a double “thump”, while others heard nothing. Notably, the observed audible signatures (or lack of thereof) exhibited strong dependence on the locality where the witnesses were present at the time. The sound may have gone unnoticed if individuals were engaged in casual conversation.

In Newark Valley, during the silent observation time, team members observed several airplanes, bird noise, and various wind noises. One team member out of six likely saw the OSIRIS-REx SRC. Four observers in Newark Valley heard a double boom at 14:45:52 UTC, other observers recorded the time to the minute as 14:45 UTC. Newark Valley observers perceived the sound as coming from the east (two observers), southwest (one observer), northwest (one observer). The four Newark Valley observers agreed the sound was distinct and unmistakable given the quiet conditions but could have been missed if a loud conversation had been happening. A clip from the video recorded by Carr’s personal phone is included in a .tar.gz package. The animation clip begins at 14:45:40 UTC and ends at 14:46:00 UTC. A double boom is audible about 11 seconds into the clip, corresponding to 14:45:51 UTC. The sample return capsule is not visible in the recording. A more detailed account of the LANL team’s visual and audible observations in Newark Valley can be found in Appendix A. There were no visual or audible observations from the BFRA site at the time of re-entry.

In the East Region, both the OSU and UH teams heard the sound. The entire OSU team viewed the re-entry from their lodging location (40.7347 N, -114.0805 E). One audible boom was heard at 14:46 UTC that was not loud but easily noticeable. The JSU team, situated at 39.2646 N, -116.0269 E at the time of re-entry, heard a clear single “thud” at 14:46:45 UTC. The OSU team had noted that the sound would likely be brushed off as another ancillary source of noise if not expected. It was very calm with little to no wind (i.e., no wind at the observation location as perceived by the team), and mostly clear skies. The team members located in Clive, UT also heard a soft, double “thump” resulting from the sonic boom.

### 7.2 Signal Detections on Ground-based Sensors

All SNL’s arrays and most Gems detected clear signals generated by the OSIRIS-REx SRC re-entry. A plot with the arrivals received at Gem stations south of the nominal re-entry path is shown in Figure 7. The timeseries were filtered with a highpass filter at 10 s. The difference in timing is due to the airwave travelling a longer path to more distant stations. The N-wave indicates a ballistic arrival. Station A05 was closest to the nominal trajectory (~13 km), and A01 the farthest (~40 km). More detailed analysis is needed to determine whether all stations received the signal from a single point of the re-entry path or perhaps from different points along the trajectory. The latter is more plausible when one considers the ballistic nature of the shock wave.

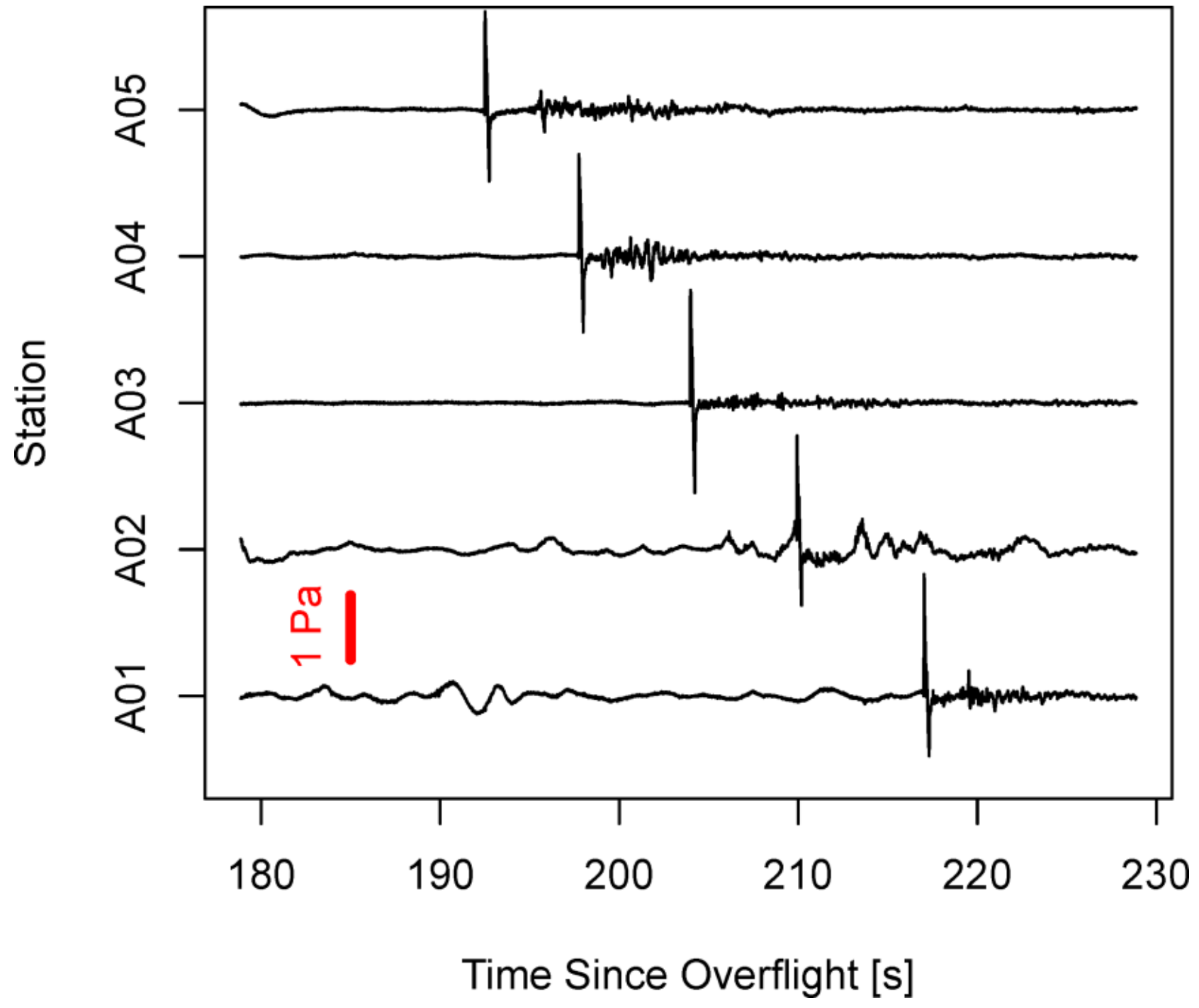

Figure 7: Signals received at Gem stations situated south of the nominal re-entry path in Newark Valley (Strawberry Rd). The timeseries were filtered (highpass, 10 s). The x-axis is in seconds from 14:45:50 UTC.

All the instruments deployed by LANL in the West Region captured the signal from the capsule. All seismometers and infrasound sensors recorded high SNR signals (Figure 8). The Silixa DAS data recorded a move out of the signal as it propagated along the fiber that can be seen without any pre-processing (Figure 8 shows data from fiber wrapped around a spool near OREXF). The AP Sensing (EUE) and OptoDAS (Newark Valley) recorded the signal, but the signal is only visible after data preprocessing. In the examples shown in Figure 8, we manually pick arrivals because the SNR is so large (red circles in Figure 8). For infrasound detection, we choose the corner at the start of the increase in pressure of the incoming N-wave (peak amplitude). For the seismic records, we pick the corresponding corner at the start of the rise towards the first high SNR peak amplitude on the vertical channel. For the DAS detection, we pick the corresponding corner at the start of the rise toward the maximum peak in strain. Detection metadata are in Appendix A, Table S1.

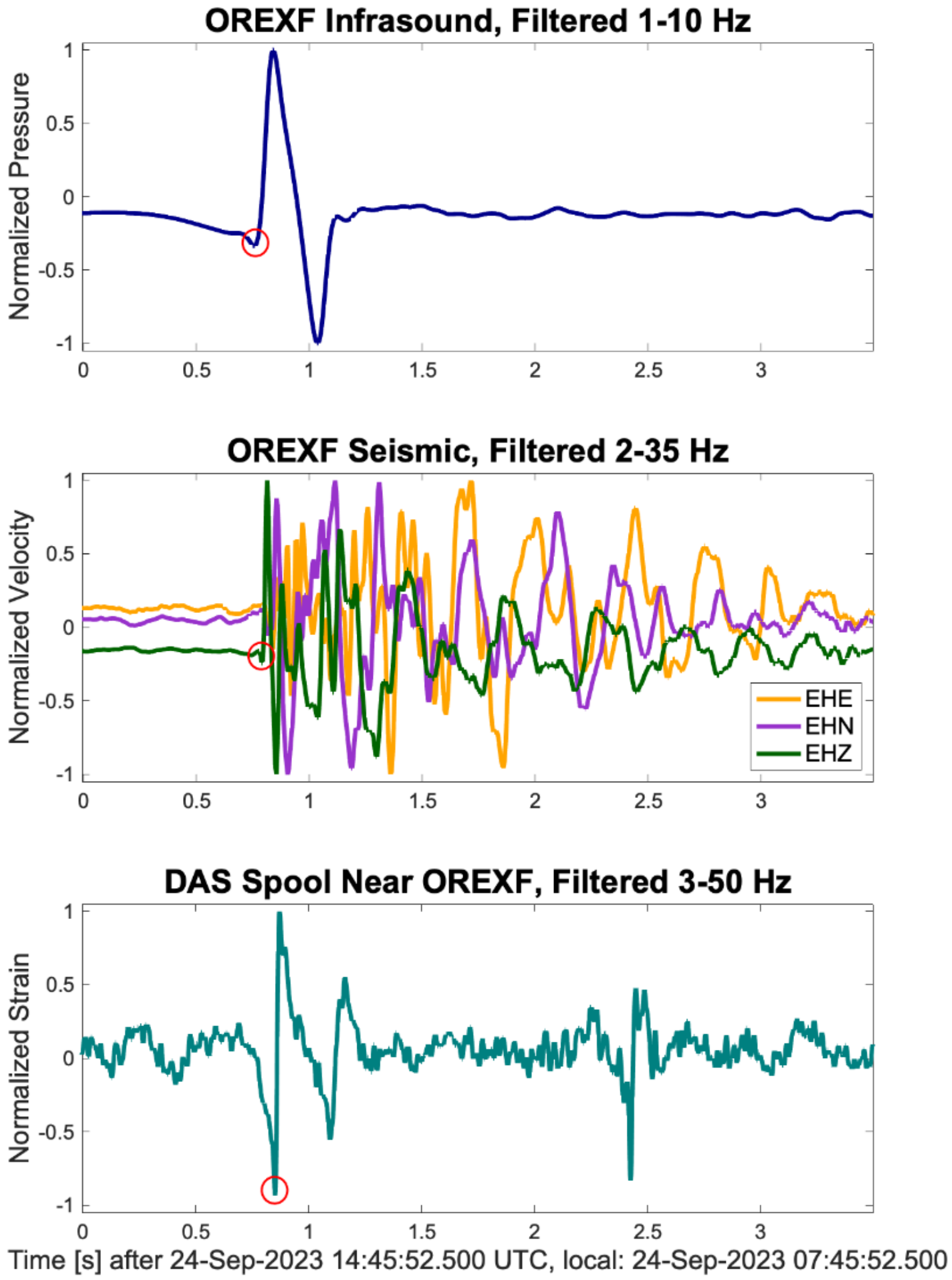

Figure 8: High SNR arrivals on the co-located OREXF (top) infrasound and (middle) 3-channel seismic sensors and (bottom) on fiber wrapped around a spool near OREXF, the sensor is at 7436 m from the interrogator at channel 3718. Data are filtered with a bandpass filter from [1-10] Hz for the infrasound channel, [2-35] Hz for the seismic channels, and [3-50] for the DAS channel. Time picks are shown with open circles.

TDA's Large N-array started collecting data at 19:23 UTC on September 23 and stopped collecting data at 15:05 UTC the next day, shortly after the re-entry. At the time of the re-entry (14:46 UTC), 114 of the 115 sensors were collecting data and detected the N-wave generated by the OSIRIS-REx SRC (clearly visible in Figure 9).

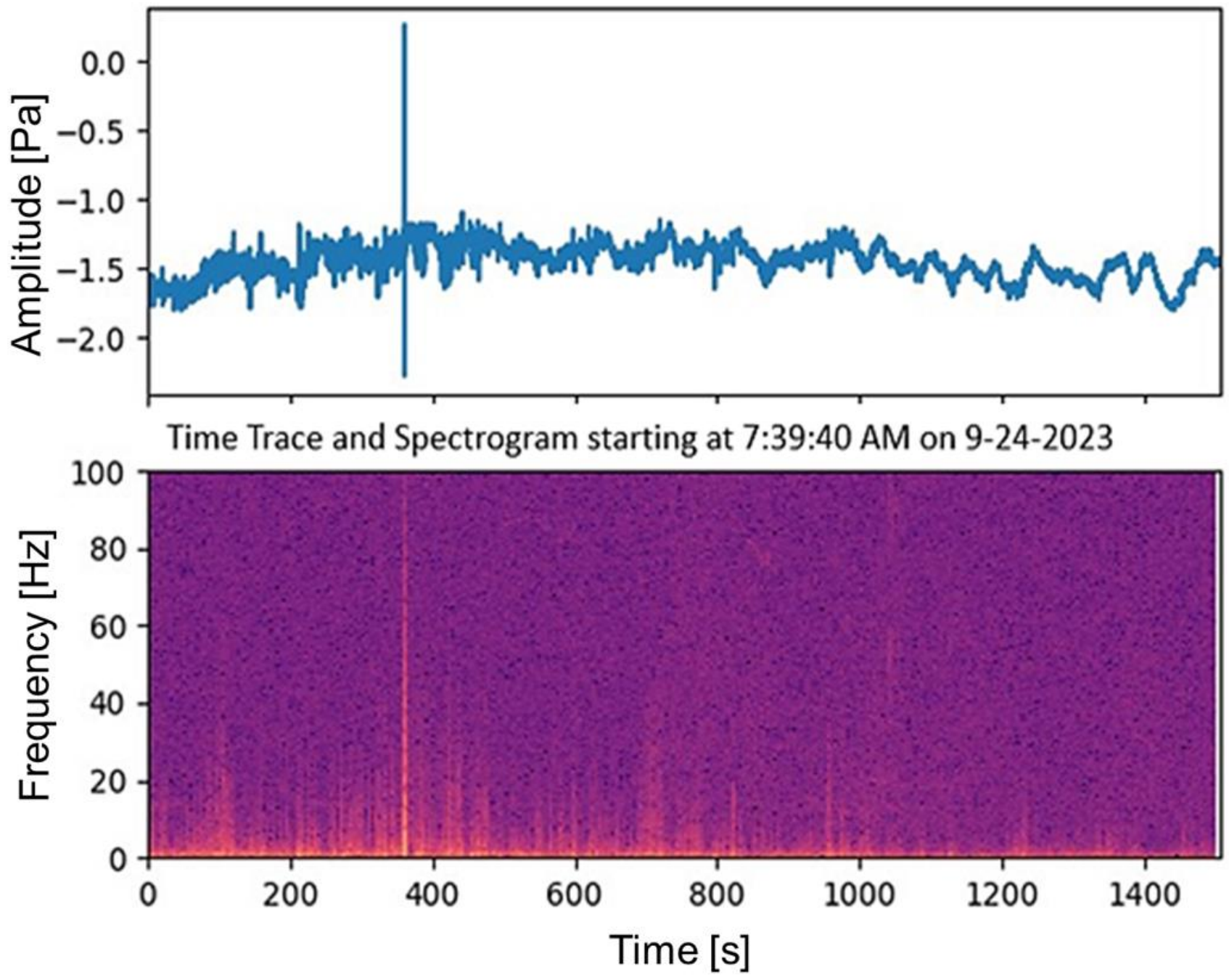

Figure 9: Signal detected at one of the TDA sensors of the large-N array. The strong signal is clearly visible just before the 400-second mark. The upper panel shows the timeseries while the lower panel shows the spectrogram.

Three of the four SMU infrasound sensors made a detection, clearly noticeable even without any data filtering (see Appendix A). KUT's infrasound instruments and microphones also detected the signal generated by the OSIRIS-REx SRC re-entry, with the arrival time around 14:45:59 UTC (see Appendix). The apparent direction of arrival was north-to-south, as expected. JHU recorded a clear sonic boom on all seismic stations, arriving at approximately 14:46 UTC. The lateral offset from the nominal trajectory was less than 2km.

All passive seismic and infrasound sensors deployed by UM recorded the capsule return signal. UM also obtained an extensive dataset of refraction waveforms to be used in developing P and S velocity functions for the sediments of Diamond Valley using body wave arrival times and high frequency surface wave dispersion. The refraction profile data are necessary to develop appropriate Earth models at the site. These velocity models are important for inferring the efficiency of the acoustic to seismic interaction and to understand how secondary seismic waves distort the acoustic source function as seen by a seismic instrument. Seismic signatures can be affected by local P-to-S conversions, creation of Rayleigh waves, and by the absolute values of both the P and S wave velocities in the near surface (e.g., Langston, 2004).

Broadband frequency-slowness analysis of the seismic array data gives an apparent velocity of 2.9 km/s and azimuth of approach of the acoustic N-wave of N02°E. We observed significant differences of both acoustic and seismic signals between the western and eastern infrasound arrays (Figure 10) suggesting differences in atmospheric wave propagation and differences in local seismic site responses. Indeed, we also saw significant differences in the seismic responses between northern and southern stations (Figure 10), suggesting the development of secondary Rayleigh waves.

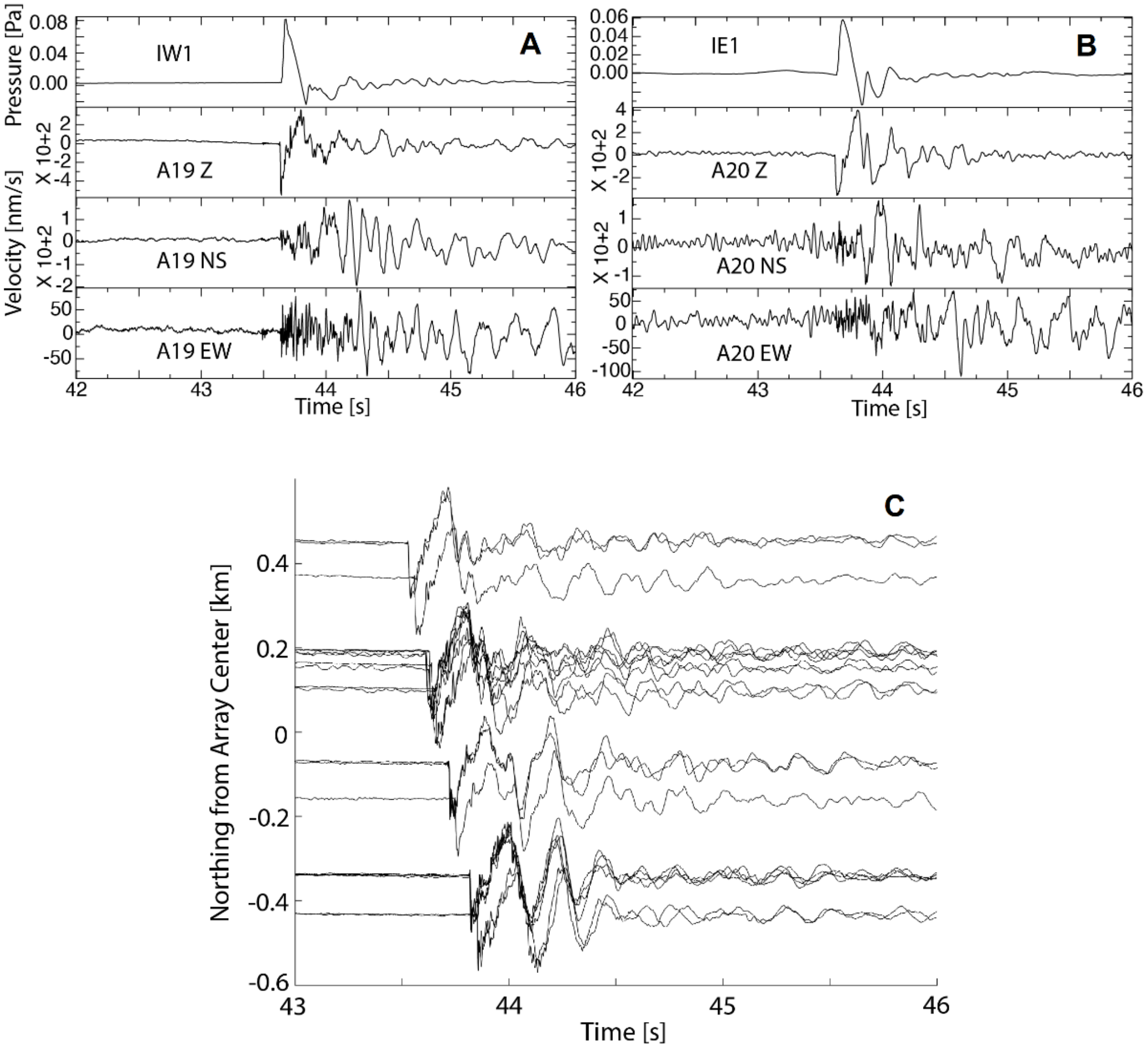

Figure 10: Comparison of pressure and ground velocity data at the central elements of the western (A) and eastern (B) UM seismo-acoustic arrays. Pressure and velocity amplitude values are provisional pending further calibration of the instruments. Time is relative to 14:45:13.818 UTC. (C) Vertical velocity waveforms from the Golay 3x6 array plotted as a function of distance from an azimuth of 0°, i.e., the data have been plotted along a virtual north-south profile with north at the top of the figure. Waveform amplitudes have been normalized. Note the large oscillating secondary arrivals for stations in the south. Time is relative to 14:45:13.818 UTC.

Smart phones running the RedVox infrasound recording app detected strong signals (see Appendix). OSU's sensors at ENV that were operational during the re-entry (14 out of 15 sensors) detected a signal at 14:47:16.5 UTC. All of these sensors were ~58 km from the perpendicular intersection of the OSIRIS-REx SRC trajectory. The received wave was an N-wave with more broadband coherent "rumbling" after the initial arrival. Figure 11 shows the signal recorded by the Chaparral array. The sensors situated south of ENV also captured the signal from the re-entry. The signal first arrived at Gem 092 at 14:46:02.7 UTC and then at Gem 074 at 14:46:17.2 UTC. These two sensors were ~25 km and ~33 km, respectively, from the perpendicular intersection of the OSIRIS-REx SRC trajectory. Both sensors had a dominant N-wave arrival with some broadband coherent "rumbling" after the initial arrival.

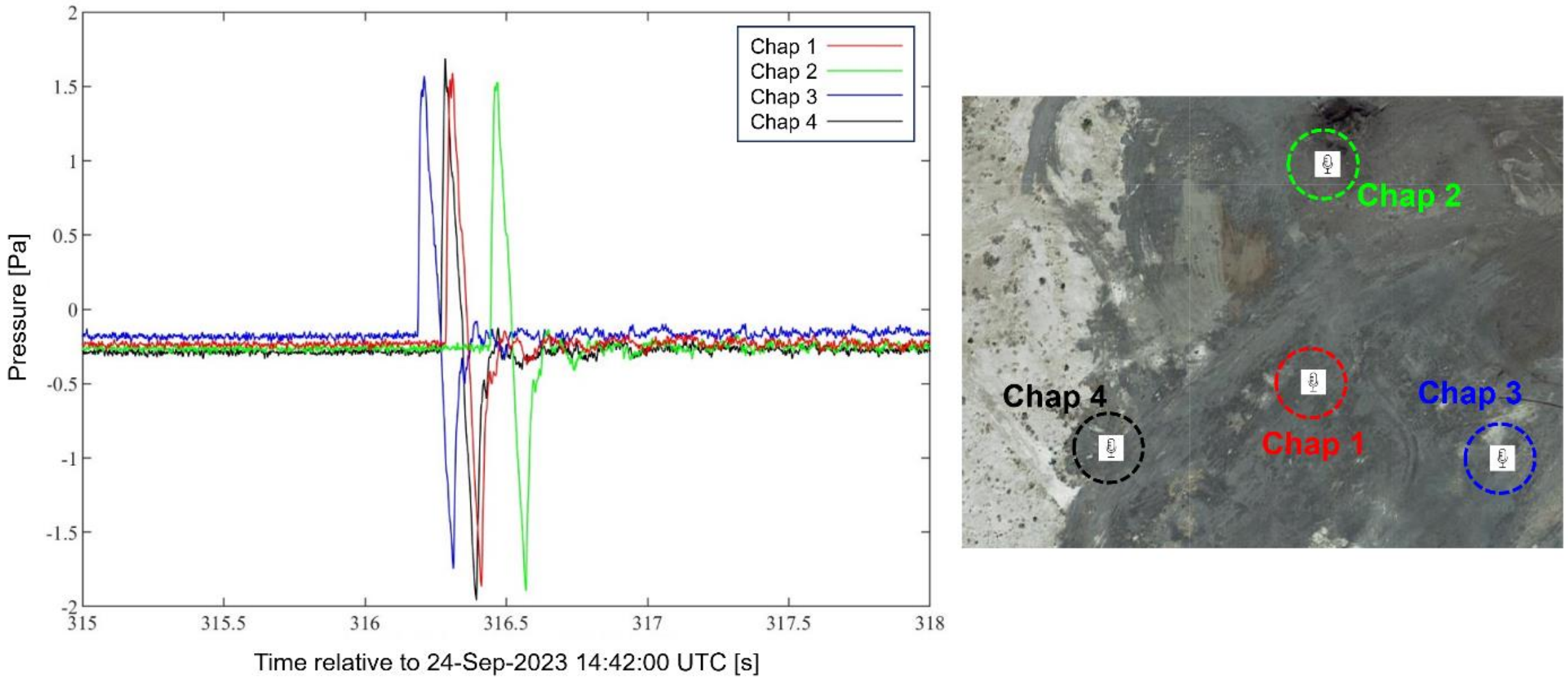

Figure 11: Left panel: Time trace from the four Chaparral Physics model 64s sensors in OSU Array 1 located at the Wendover Airport. Right panel: Satellite image showing the orientation of the sensors (Chap 1-4) with colors matching those in the left panel, which shows the signal was arriving from the south-southeast direction. The width across the map is 175 m.

In the Distal Region, early analysis from the BSU's Idaho stations show possible signals. At this location, we recorded good data from 47 of 66 sensors during the OSIRIS-REx entry. Causes of sensor failure include dead batteries, disturbance by cattle, and theft/loss. The AWE's station at the NNSS detected the signal (Figure 12). All LANL's sensors in the Distal Region performed well. The array near Price, UT did not detect the OSIRIS-REx re-entry. However, the other two arrays (NNSS array (Figure 12) and the St. George, UT array), did capture signals from the re-entry. The stations at NNSS are situated ~260 km from the point of peak heating.

Infrasound signals in the 1-4 Hz passband were observed on two small aperture 4-element arrays at ~420km west-southwest of the OSIRIS-REx landing site (Figure 12) exhibited backazimuths (the direction from which the signal arrived) of ~345° (i.e., from just west of north) consistent with signal generation along the re-entry trajectory. The arrays are separated by a horizontal distance of 1500m and a vertical distance of 300m, with PSDJK (AWE) located on top of a mesa and OREX1 (LANL) located in the base of a steep-sided valley.

Although the signals at each array are qualitatively similar in terms of duration and waveform variation, beamforming results at OREX1 have a higher resolution likely due to the larger array aperture of 160 m (compared to 100 m for PSDJK). Despite the close proximity of the arrays there are significant differences in the temporal variations of background noise amplitudes. On the mesa, PSDJK exhibits low noise in the period prior to the OSIRIS-REx signal, such that a persistent high-frequency low-amplitude acoustic source can be observed towards the west. During and after the OSIRIS-REx signal the noise amplitude at PSDJK increases, obscuring the low-amplitude signal. The opposite is observed at OREX1 in the valley; here, high amplitude noise prior to the OSIRIS-REx signal obscures the persistent source to the west. During and after the OSIRIS-REx signal the noise amplitude drops allowing the persistent source to be observed. This indicates how wind generated noise at an array can be highly localized, and the impact it has upon signal detection.

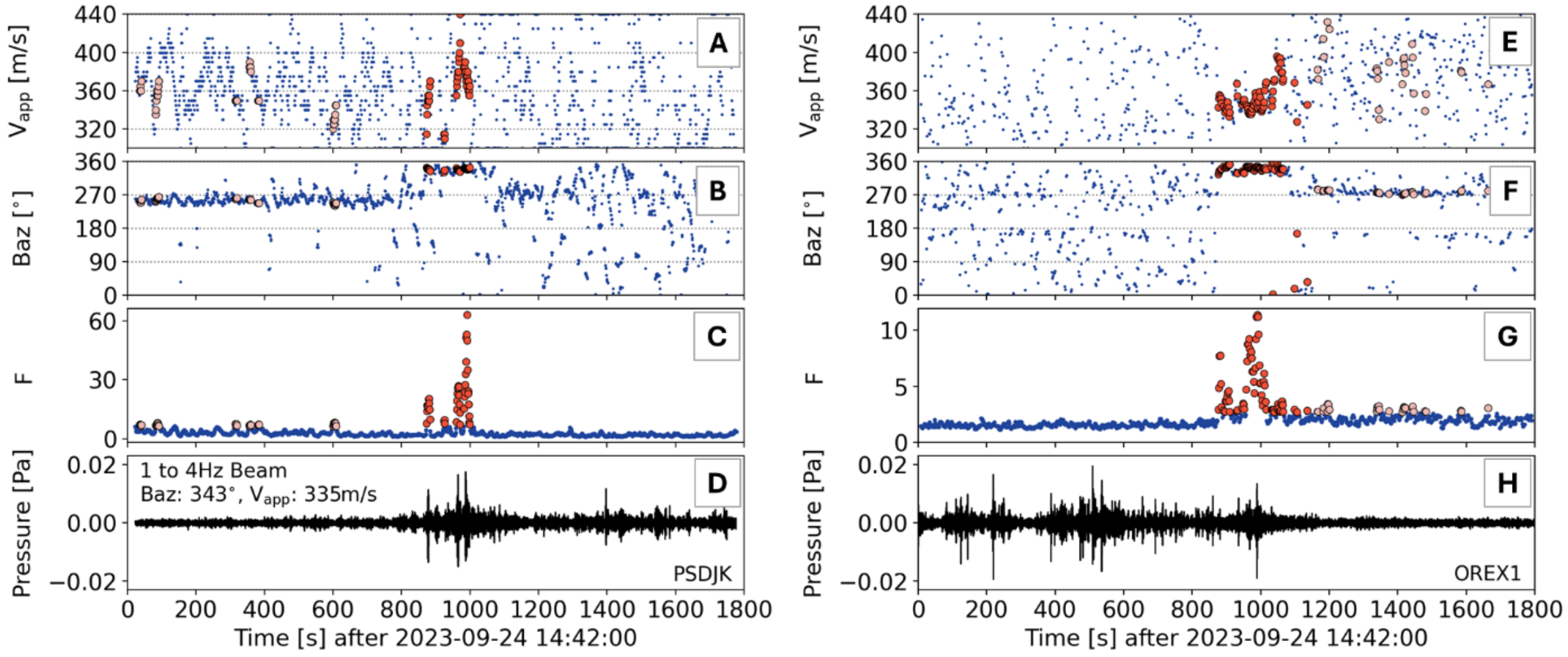

Figure 12: Left: AWE's PSDJK array. Right: LANL's OREX1 array. Panels top to bottom show values of apparent velocity ($V_{app}$), backazimuth (Baz) and F-statistic (F) corresponding to the beam direction that exhibits the highest signal coherence across consecutive 4 s long windows (overlapping by 50%). Orange dots represent time periods for which there is a >95% probability that the window contains a signal with a signal-to-noise ratio greater than, or equal to, four. The lowermost panel shows the 1 to 4 Hz beam for a backazimuth of approximately 345°. Note that the OSIRIS-REx signal is seen above a persistent low signal-to-noise ratio signal arriving from a backazimuth of approximately 260°.

### 7.3 Balloon-borne Infrasound

The high-altitude meteorological balloons and heliotropes traveled furthest west (relative to other balloons) due to their relatively low ascent rates (Figure 13) and were over halfway to the capsule's altitude at the time of overflight (Figure 13). The zero pressure balloons were about 50 km from the launch site at the time of the overflight. The cloudskimmers and low altitude meterological balloon were still quite close to the launch site because they were released just before the overflight. Ground winds were generally low, allowing for the successful inflation and launch of the balloons. While a strong tropospheric jet was present, the winds were not rapid enough to carry the balloons into the termination zone before the OSIRIS-REx SRC re-entered. The balloon data are being analyzed and will be published in the near future.

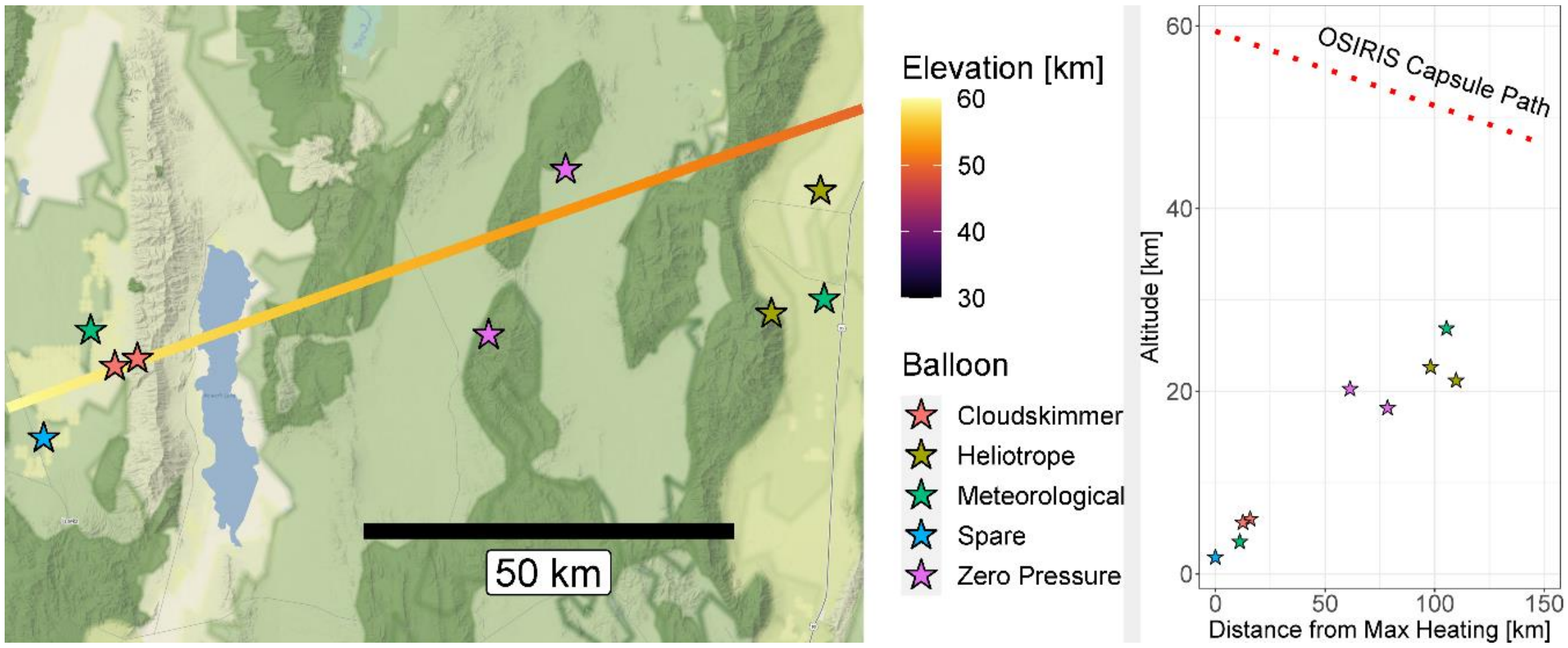

Figure 13: Left: Locations of OSIRIS-REx balloons compared to OSIRIS-REx flight path and altitude. Map imagery from Stadia Maps. Right: Altitude of balloon payloads at the time of OSIRIS signal arrival compared to the altitude of the OSIRIS-REx capsule.

### 7.4 GPS

LANL GPS team observed several signatures originating from the incoming OSIRIS-REx (SRC) (Figure 14). Several faint signatures were detected in the TEC along lines of sight from our GPS ground units through the ionospheric peak density altitudes, arriving more than 10 minutes after the re-entry as expected (having traversed 300 km from maximum shock-wave altitudes to the ionospheric peak) (Figure 14, left panel). Estimated ionospheric signatures were traveling at speeds expected in the thermosphere (greater than 800 m/s), much faster than ground speeds (343 m/s). We were also able to observe signatures (possibly moving at faster speeds) in the refined GPS scintillation measurements. In addition, data from 12 publicly available GPS ground stations within 300 km of Eureka (NV) were used to approximate the location of the final descent after the parachute deployment stage (Figure 14, right panel). The immediate results of the observational campaign are encouraging. The LANL GPS team will continue to examine the data from the OSIRIS-REx capsule return and refine their analysis methodologies.

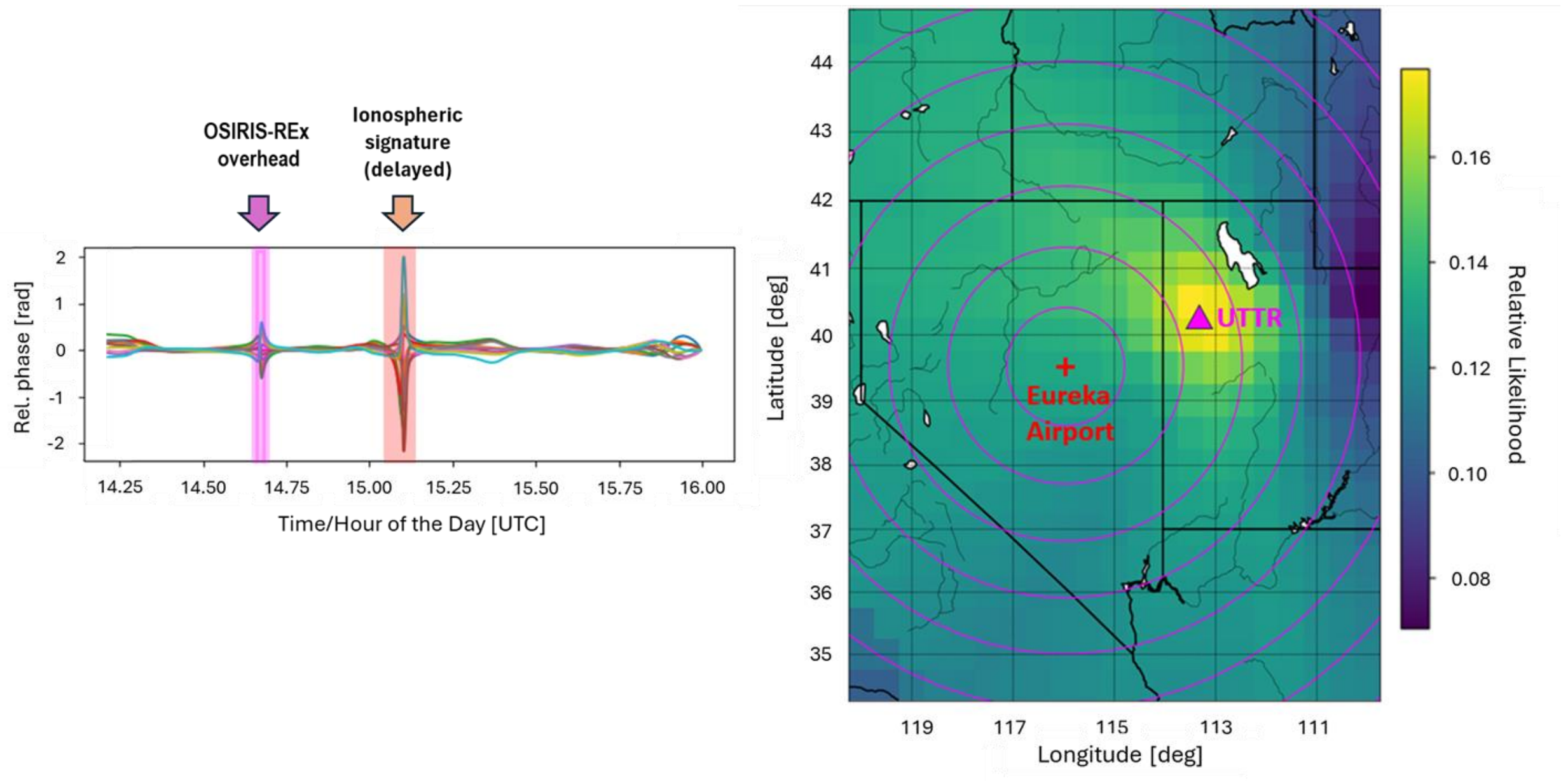

Figure 14: Left: Relative LANL GPS filtered precision TEC measurements, identifying 10-minute acoustic signatures. Right: Most likely acoustic source location identified from 10-minute TEC waves derived from public GNSS data.

## 8. Conclusions

NASA's OSIRIS-REx SRC returned to Earth on September 24, 2023, delivering precious cargo consisting of physical samples of the asteroid Bennu. This was the first asteroid sample return mission for the USA. Considering that SRCs come from interplanetary space at hypervelocity, they can serve as ideal analogues for studying meteor phenomena. The most recent re-entry over the USA was in 2006 with the return of NASA's Stardust mission. The OSIRIS-REx SRC's hypersonic flight through the atmosphere provided an exceptionally rare opportunity to carry out geophysical observations of a well-characterized source with known parameters, including timing and trajectory.

A large team of researchers from 16 institutions gathered to perform a coordinated geophysical observational campaign at strategically and carefully selected locations that were projected to provide robust and high-fidelity data. Over 450 ground-based sensors including infrasound, seismic, DAS, and GPS were deployed. Moreover, several balloons carrying infrasound sensors were launched with the aim of capturing the signals at high altitude. This was the largest geophysical observational campaign of a re-entry ever performed, collecting a wealth of valuable data that is expected to promote scientific inquiry for many years to come.

The observational campaign was highly successful, with detections on nearly all instruments, near and far. Here, we present our early results collectively, noting that more focused studies will be disseminated in due course. Data collected during this effort will be eventually made openly available.

Here we summarize our preliminary findings and conclusions:

i. Infrasound and Seismic: The campaign to record the OSIRIS-REx capsule return seismo-acoustic signals was remarkably successful. Nearly all passive instrumentation at the variety of vantage points recorded the signal generated by the re-entry. Most instruments performed well, including arrays, single stations, Large N-array, and smart phones. A diverse range of instruments recorded the signal at distances from beneath the nominal trajectory to several hundreds of kilometers from the nominal trajectory. The seismic array design proved to be able to accurately determine the wave characteristics of the incoming acoustic signal, and we observed interesting seismo-acoustic interactions at our infrasound subarrays. Our refraction work is also an important element of the experiment since it will yield baseline information on near-surface P and S wave velocities that are important constraints on the nature of the acoustic-seismic interaction.

i. DAS: As expected, within each site (Eureka Airport and Newark Valley), detections sweep across the instruments from North to South, with instruments closer to the trajectory recording arrivals before instruments further from the trajectory. This event marks the first time that DAS recorded a re-entry event. In Newark Valley in particular, the arrivals were clear across much of the 7.5 km of deployed fiber, despite the simple placement of cable directly onto the ground (no trenching). This is particularly promising in light of rapid deployment observation campaigns, where trenching may be logistically prohibitive.

ii. Balloons: A diverse set of balloon-borne acoustic stations were fielded during the OSIRIS-REx observation campaign, with the hope that at least one of them would succeed. Fortunately, every airborne sensor recorded data during the overflight. While careful planning and experienced launch crews played a major role in this achievement, the weather played a critical part as well.

iii. GPS: The GPS observational campaign was successful. Several faint signatures generated by the OSIRIS-REx SRC during re-entry were detected in the total electron content (TEC) and L-band scintillation (directly). Additionally, we were able to infer the location of the final descent after the parachute using data from a dozen publicly available GPS ground stations.

This largest to-date observational campaign of a hypersonic re-entry with a multitude of geophysical instruments provided valuable insight and data collection that can serve as a blueprint not only for terrestrial applications but also for future space mission planning. Future campaigns should attempt to capture the onset of the shock wave as the object transitions into the continuum flow regime. Having radiosondes launched in various locations along the ground track to collect atmospheric data up to 40 km altitude would be of immense value. Moreover, seismic instruments extending over a longer region would further help constrain the source. Because we prioritized the West Region where the likelihood of capturing the signal would be the greatest, we did not have any dedicated seismic sensors further east. Future balloon campaigns should focus on broadening the horizontal range between the sensors and the reentering object, as well as deployments at a wider array of azimuths.

**Acknowledgements:**

The authors thank the two anonymous reviewers for their valuable and supportive feedback. The authors gratefully acknowledge the Board of Eureka County Commissioners (Chairman Rich McKay, Vice Chair Michael Sharkozy, Commissioner Marty Plaskett, and Deputy Clerk Jackie Berg) and the Eureka Municipal Airport personnel (Jayme Halpin). The authors thank Sergeant Adam Zehr (Nevada Highway Patrol) and Eureka County Deputies Jason Flanagan and Allison Flanagan for their assistance during the OSIRIS-REx SRC re-entry. The authors also extend heartfelt thanks to the people of town of Eureka for their hospitality.

We are grateful for the Bureau of Land Management Ely District Office for granting us permission to deploy sensors in eastern Nevada as well as assisting us with coordination. We are also grateful for the Bureau of Land Management Salt Lake Field Office for granting us permission to deploy sensors in western Utah.

E. A. Silber thanks M. Moreau (NASA) for sharing the EDL data.

**SNL:** This article has been authored by an employee of National Technology & Engineering Solutions of Sandia, LLC under Contract No. DE-NA0003525 with the U.S. Department of Energy (DOE). The employee owns all right, title and interest in and to the article and is solely responsible for its contents. The United States Government retains and the publisher, by accepting the article for publication, acknowledges that the United States Government retains a non-exclusive, paid-up, irrevocable, world-wide license to publish or reproduce the published form of this article or allow others to do so, for United States Government purposes. The DOE will provide public access to these results of federally sponsored research in accordance with the DOE Public Access Plan https://www.energy.gov/downloads/doe-public-access-plan. This paper describes objective technical results and analysis. Any subjective views or opinions that might be expressed in the paper do not necessarily represent the views of the U.S. Department of Energy or the United States Government. This work was supported by the Nuclear Arms Control Technology (NACT) program at the Defense Threat Reduction Agency (DTRA).

**LANL:** This research was supported by Los Alamos National Laboratory (LANL) through the Laboratory Directed Research and Development (LDRD) program, under project number 20220188DR and under the LANL Center for Space and Earth Science (CSES) LDRD project number 20240477CR-SES. We also thank the Eureka County Commission and Eureka Airport for access at the airport site, and the Bristlecone BLM Field Office for guidance in complying with casual use requirements at the Newark Valley site. We are grateful for the loan of a field truck by the MPA-Q Division at LANL. Funding for Elisa A. McGhee was provided by the Pat Tillman Foundation Scholarship and the Colorado State University Vice President for Research Graduate Fellowship Program. Colors used in plots of seismic, infrasound, and DAS data are from Diana (2023). All the deployed instruments belong to LANL, with the exception of the OptoDAS which was rented from Alcatel and one personal cell phone. Loïc Viens was partly supported by the Chick Keller Fellowship from the Center for Space and Earth Science (CSES) at LANL. All tools and software used for measurement and analyses were LANL owned/operated, except the personal cellphone.

AWE and LANL acknowledge the assistance and expertise provided by our MSTS colleagues in deploying and maintaining the infrasound arrays at NNSS.

**JPL:** Contributions from JPL authors were performed at the Jet Propulsion Laboratory, California Institute of Technology, under a contract with the National Aeronautics and Space Administration (80NM0018D0004). The authors acknowledge funding from the NASA Planetary Science and Technology through Analog Research (PSTAR) program.

**INL:** INL is operated for the U.S. Department of Energy by Battelle Energy Alliance under DOE contract DE-AC07-05-ID14517. This work was supported by the Department of Energy National Nuclear Security Administration under award DE-NA0003921 (Consortium for Enabling Technologies and Innovation).

**OSU:** This work was funded, in part, by Gordon and Betty Moore Foundation under grant GBMF11559 (doi.org/10.37807/GBMF11559). The OSU team would also like to thank the Wendover Airport for providing access to the airport site.

**UM:** We would like to thank the Damele family of Eureka, NV, for allowing access to the eastern portion of the experiment through their farm which is adjacent to the airport. The authors acknowledge the use of the "Seismic Analysis Code" (SAC) (Goldstein et al., 2003) in preparing this paper. Travel funding was provided by the University of Memphis Research and Innovation Office.

**BSU:** Data collection was funded by US Forest Service award 23-JV-11111135-067 and NSF award EAR-2122188. We thank Z. Cram and V. Porter for help arranging fieldwork in Reynolds Creek Experimental Watershed.

**RedVox:** This work was supported by the Air Force Research Laboratory under Agreement FA8650-18-C-1669.

**UH:** This work was supported in part by the Department of Energy National Nuclear Security Administration under Award Numbers DE-NA0003920 (MTV), DE-NA0003921 (ETI). This report was prepared as an account of work sponsored by agencies of the United States Government. Neither the United States Government nor any agency thereof, nor any of their employees, makes any warranty, express or implied, or assumes any legal liability or responsibility for the accuracy, completeness, or usefulness of any information, apparatus, product, or process disclosed, or represents that its use would not infringe privately owned rights. Reference herein to any specific commercial product, process, or service by trade name, trademark, manufacturer, or otherwise does not necessarily constitute or imply its endorsement, recommendation, or favoring by the United States Government or any agency thereof. The views and opinions of authors expressed herein do not necessarily state or reflect those of the United States Government or any agency thereof. The United States Government is authorized to reproduce and distribute reprints for Governmental purposes notwithstanding any copyright notation thereon.

**Data availability:**

Tables with the locations of the instruments are in Appendices. Tables A1 and A2 are also available in machine readable format. Data from some components of this observational campaign will be made openly available in the future. UM data collected from this experiment will be openly available from the Seismological Facility for the Advancement of Geoscience (SAGE) of the EarthScope Consortium in September 2025. Currently, the data are forming part of a PhD

thesis at the University of Memphis. LANL/GPS observational data may be released at a later date. SMU data will be made available in the near future.

## Appendix A

### Los Alamos National Laboratory (LANL)

### Deployment Photos (DAS)

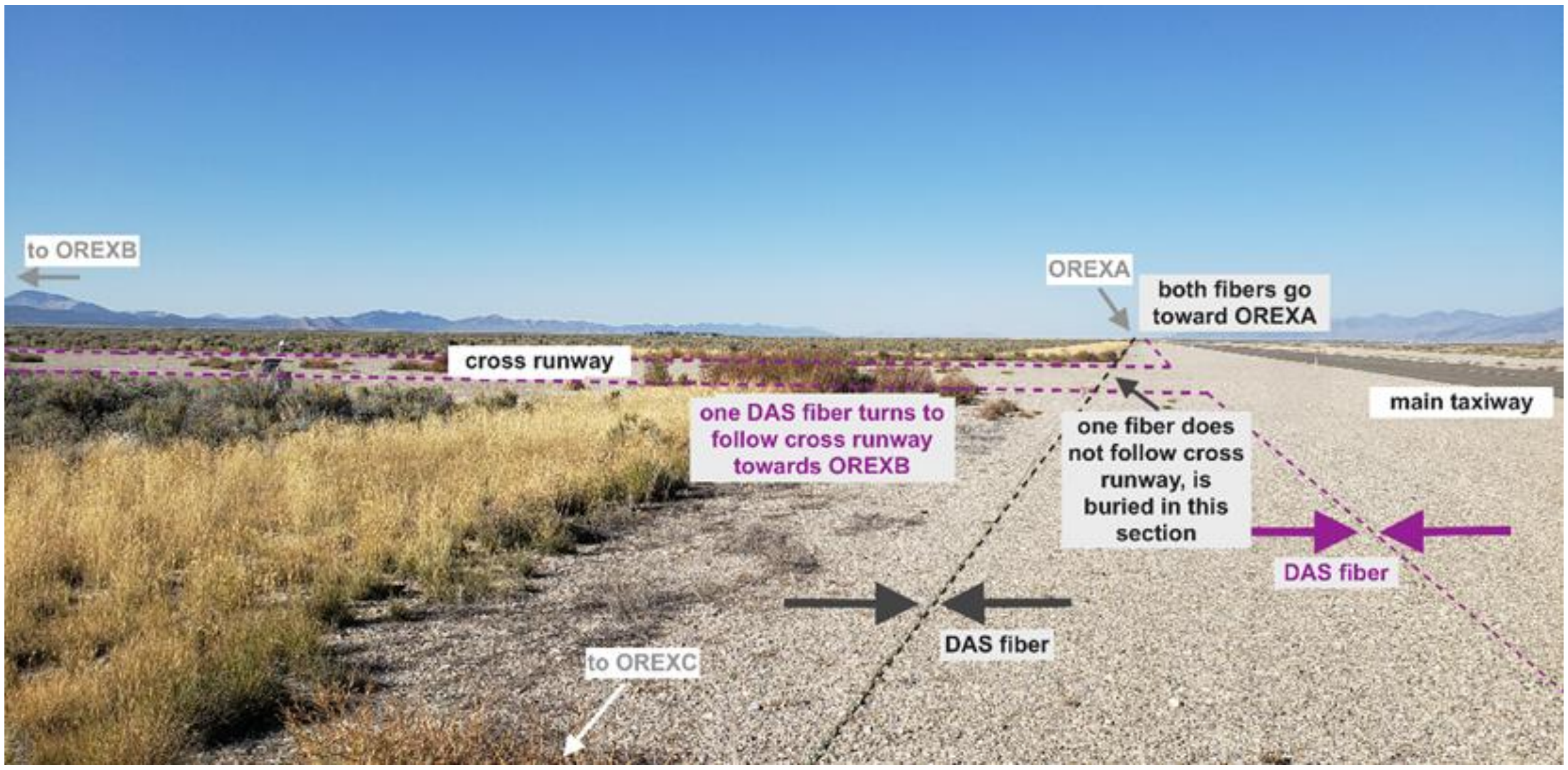

Figure S1: DAS fiber deployment at the Eureka Airport (photo credit: C. Carr).

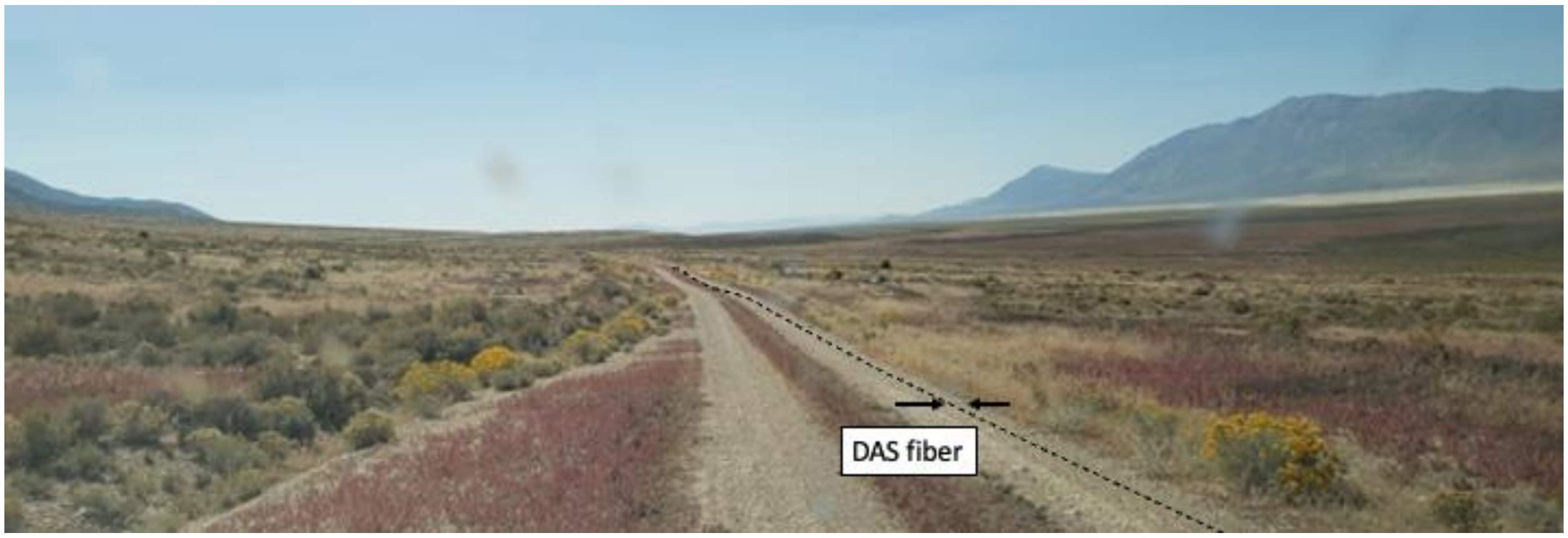

Figure S2: DAS fiber deployment in Newark Valley (photo credit: C. Carr).

## Deployment Photos (GPS)

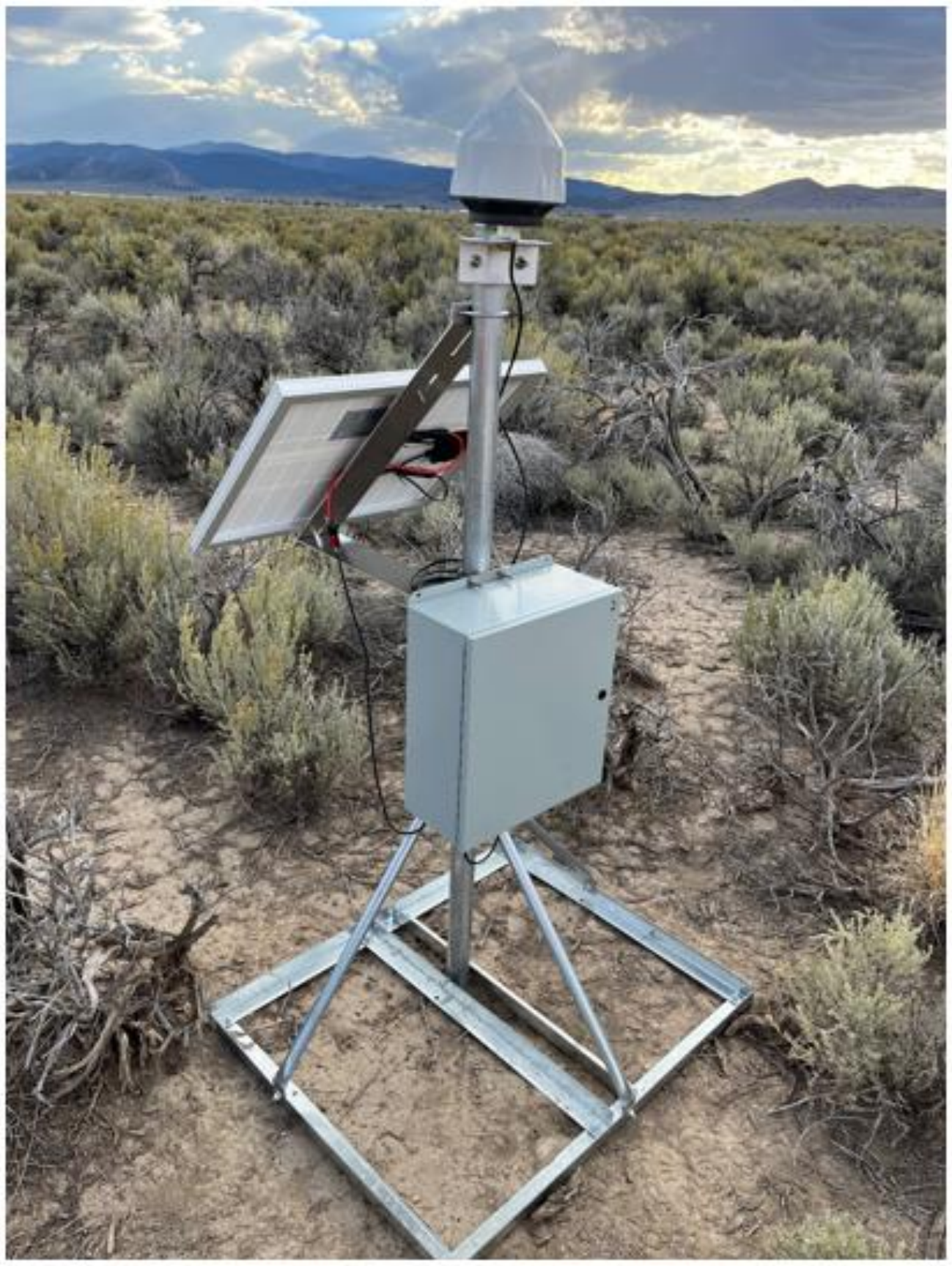

Figure S3: GPS Ground Station: Rex-2 (photo credit: R. Hasser).

## Visual and Audio Observations in Newark Valley

In preparation for observing the re-entry on the morning of 24 September 2023, in Newark Valley, two team members stood at the trailer containing the DAS interrogator unit to monitor operation during the overflight. One team member sat about 50m to the north-northwest of the trailer and one team member laid on the ground about 60m to the north of the trailer. The team members maintained a quiet observation time from 07:35 AM to 08:10 AM local (14:35 – 15:10 UTC 24 Sept 23) in sunny weather. While conditions were calm earlier in the morning, as the sun rose in the valley, the wind increased such that during the quiet observation time observers described the wind as breezy. During the quiet observation time, team members observed several airplanes, bird noise, and various wind noises. At the airport, the two team members were outside the hangar and noted the weather was sunny and slightly windy. The airport observers noted that traffic noise from the nearby highway was present during the re-entry time frame.

One team member out of six likely saw the SRC. The team member (Carr) lying on the ground in Newark Valley observed a very bright whitish-yellow streaking flash. They estimated afterwards the streak covered about 5 degrees, starting nearly vertically overhead and moving roughly toward the NE. They estimated the time as 07:43:06 AM local (14:43:06 UTC) based on counting seconds since the last check of their GPS watch; they did not look at the watch at the time to maintain visual observation overhead. The other three team members in Newark Valley did not see the capsule. Team members at the airport reported they had no visual observations of the capsule.

Most team members in Newark Valley and at Eureka Airport heard a sound, though perceptions of the sound and direction of sound origin varied. Four observers in Newark Valley heard a double boom, Carr recorded the time as 07:45:52 AM local (14:45:52 UTC), other observers recorded the time to the minute as 07:45 AM local (14:45 UTC). Newark Valley observers perceived the sound as coming from the east (two observers), southwest (one observer), northwest (one observer). The four Newark observers agreed the sound was distinct and unmistakable given the quiet conditions but could have been missed if a loud conversation had been happening. At the airport, one team member heard a faint "pop" sound (time not recorded).

Carr recorded a video (duration 35 minutes, 20 seconds) with their personal cell phone, starting just after 07:35:00 AM local time (14:35 UTC) on 24 September 2023. Timing was determined by their personal GPS-enabled smart watch, and the video timing is within a second but slightly behind the time as recorded by the watch at the start of the video. Timing resolution is limited by the watch and phone. A clip from the video is included in a .tar.gz package, the clip begins at 07:45:40 AM local (14:45:40 UTC) and ends at 07:46:00 AM local (14:46:00 UTC). A double boom is audible about 11 seconds into the clip, corresponding to 07:45:51 AM local (14:45:51 UTC). This is consistent with a written observation by Carr of an audible double boom at 07:45:52 AM local (14:45:52 UTC) based on the watch. The sample return capsule is not visible in the recording.

## Signal Detection Metadata

Table S1: Detection times for LANL seismometers, infrasound sensors, the DAS spool near OREXF, human observers, and a cell phone. We manually picked arrivals on unfiltered data because the SNR is so large. For infrasound detection, we chose the corner at the start of the increase in pressure of the incoming N-wave. For the seismic records, we pick the corresponding corner at the start of the rise towards the first high SNR peak amplitude on the vertical channel. For the DAS detection, we pick the corresponding corner at the start of the rise toward the maximum peak in strain. Ground distance (last column) represents the distance measured along a perpendicular back azimuth to the closest ground path of the nominal trajectory.

| Instrument or Observer | Signal notes | Detection time (local) on 24 Sept 2023 | Detection time (UTC) on 24 Sept 2023 | Latitude (°N) | Longitude (°E) | Distance to trajectory |
|---|---|---|---|---|---|---|
| OREXA - infrasound sensor | N wave | 7:45:57.440 | 14:45:57.440 | 39.6109883 | -116.002932 | 5.6 km |
| OREXA - seismometer | impulsive arrival with coda | 7:45:57.480 | 14:45:57.480 | 39.6109883 | -116.002932 | 5.6 km |
| OREXB - infrasound sensor | N wave | 7:45:57.520 | 14:45:57.520 | 39.60899 | -116.011737 | 5.6 km |
| OREXB - seismometer | impulsive arrival with coda | 7:45:57.560 | 14:45:57.560 | 39.60899 | -116.011737 | 5.6 km |
| OREXC - infrasound sensor | N wave | 7:45:57.720 | 14:45:57.720 | 39.6040433 | -116.004643 | 6.3 km |
| OREX C - seismometer | impulsive arrival with coda | 7:45:57.755 | 14:45:57.755 | 39.6040433 | -116.004643 | 6.3 km |
| OREXD - infrasound sensor | N wave | 7:45:51.595 | 14:45:51.595 | 39.7372017 | -115.674093 | 1.8 km |
| OREXD - seismometer | impulsive arrival with coda | 7:45:51.625 | 14:45:51.625 | 39.7372017 | -115.674093 | 1.8 km |
| OREXE - infrasound sensor | N wave | 7:45:52.485 | 14:45:52.485 | 39.7043 | -115.676033 | 5.2 km |
| OREXE - seismometer | impulsive arrival with coda | 7:45:52.520 | 14:45:52.520 | 39.7043 | -115.676033 | 5.2 km |
| OREXF - infrasound sensor | N wave | 7:45:53.255 | 14:45:53.255 | 39.6858783 | -115.676975 | 7.1 km |
| OREXF - seismometer | impulsive arrival with coda | 7:45:53.290 | 14:45:53.290 | 39.6858783 | -115.676975 | 7.1 km |
| DAS spool near OREXF | peak in strain | 7:45:53.352 | 14:45:53.352 | 39.6875 | -115.67696 | 7.1 km |
| Chris Carr | visual bright streak | 7:43:06 (estimated) | 14:43:06 (estimated) | 39.7516 | 115.6736 | 0.3 km |
| Chris Carr | audible double boom | 7:45:52 | 14:45:52 | 39.7516 | 115.6736 | 0.3 km |
| Chris Carr cell phone | audible double boom | 7:45:51 | 14:45:51 | 39.7516 | 115.6736 | 0.3 km |
| Carly Donahue | audible double boom | 7:45 | 14:45 | 39.7511 | -115.6734 | 0.4 km |
| Luke Beardslee | audible double boom | 7:45 | 14:45 | 39.7511 | -115.6734 | 0.4 km |
| Lisa Danielson | audible double boom | 7:45 | 14:45 | 39.7515 | -115.6737 | 0.4 km |
| Loïc Viens | audible faint pop | not recorded | not recorded | 39.600158 | -116.0058 | 6.6 km |

## Johns Hopkins University (JHU)

### Deployment Photos

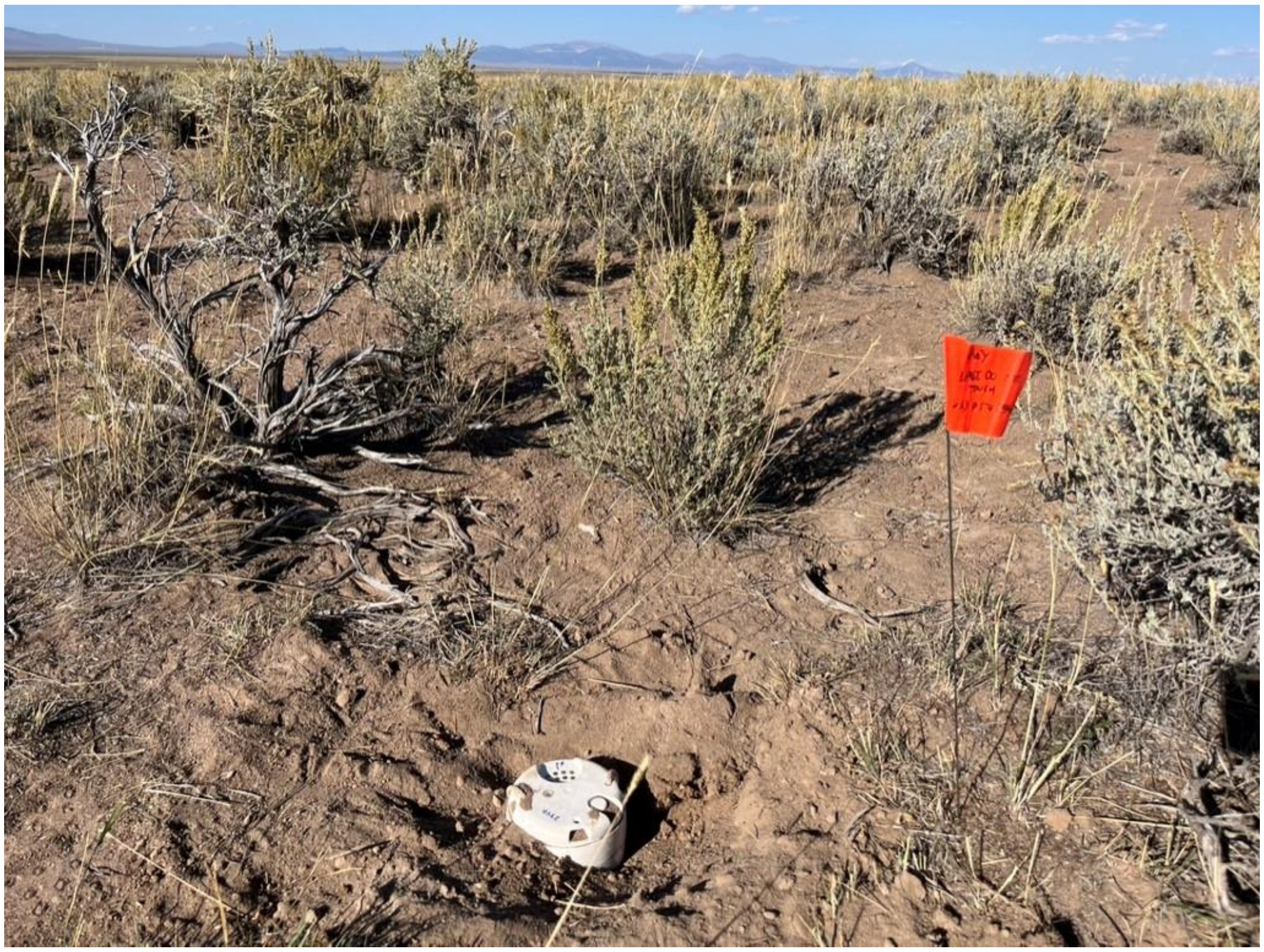

Figure S4: One of the seismic nodes deployed by JHU at the Beans Flat Rest Area (photo credit: B. Fernando).

## Kochi University of Technology (KUT)

### Signal Detection

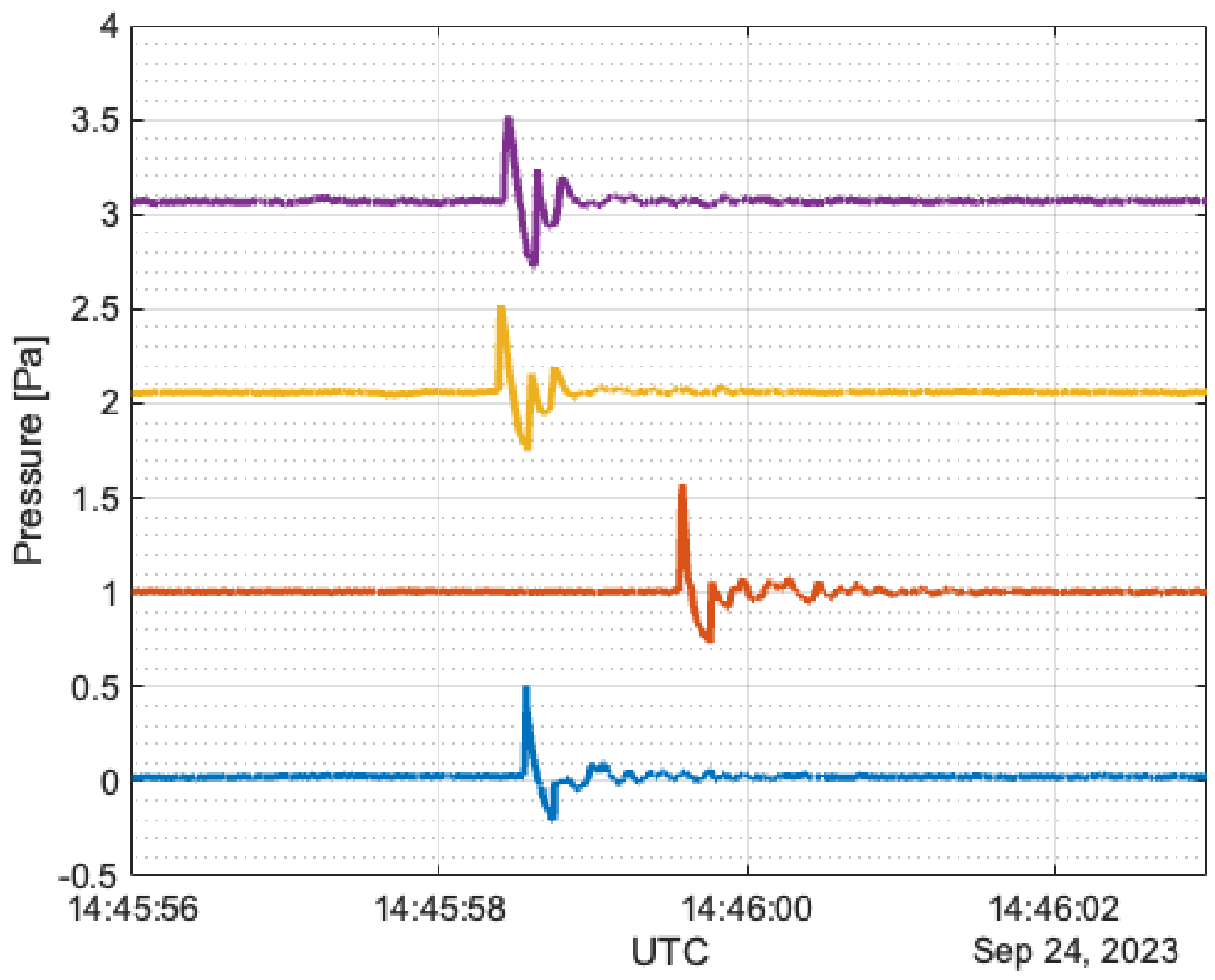

Figure S5: The signals recorded by the INF04 sensors deployed at EUE by KUT.

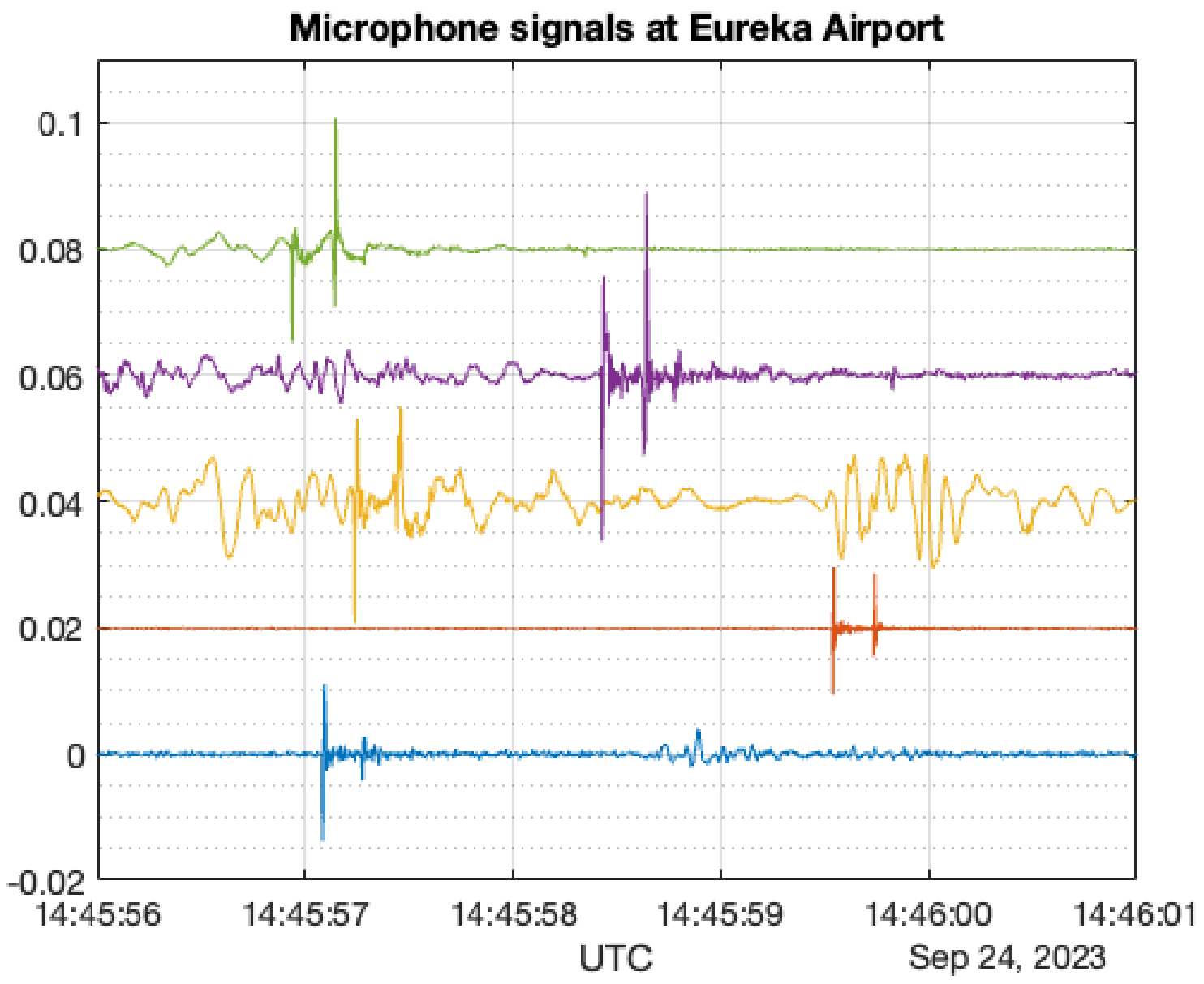

Figure S6: The signals recorded by microphones at EUE.

## Oklahoma State University (OSU)

### Deployment Photos

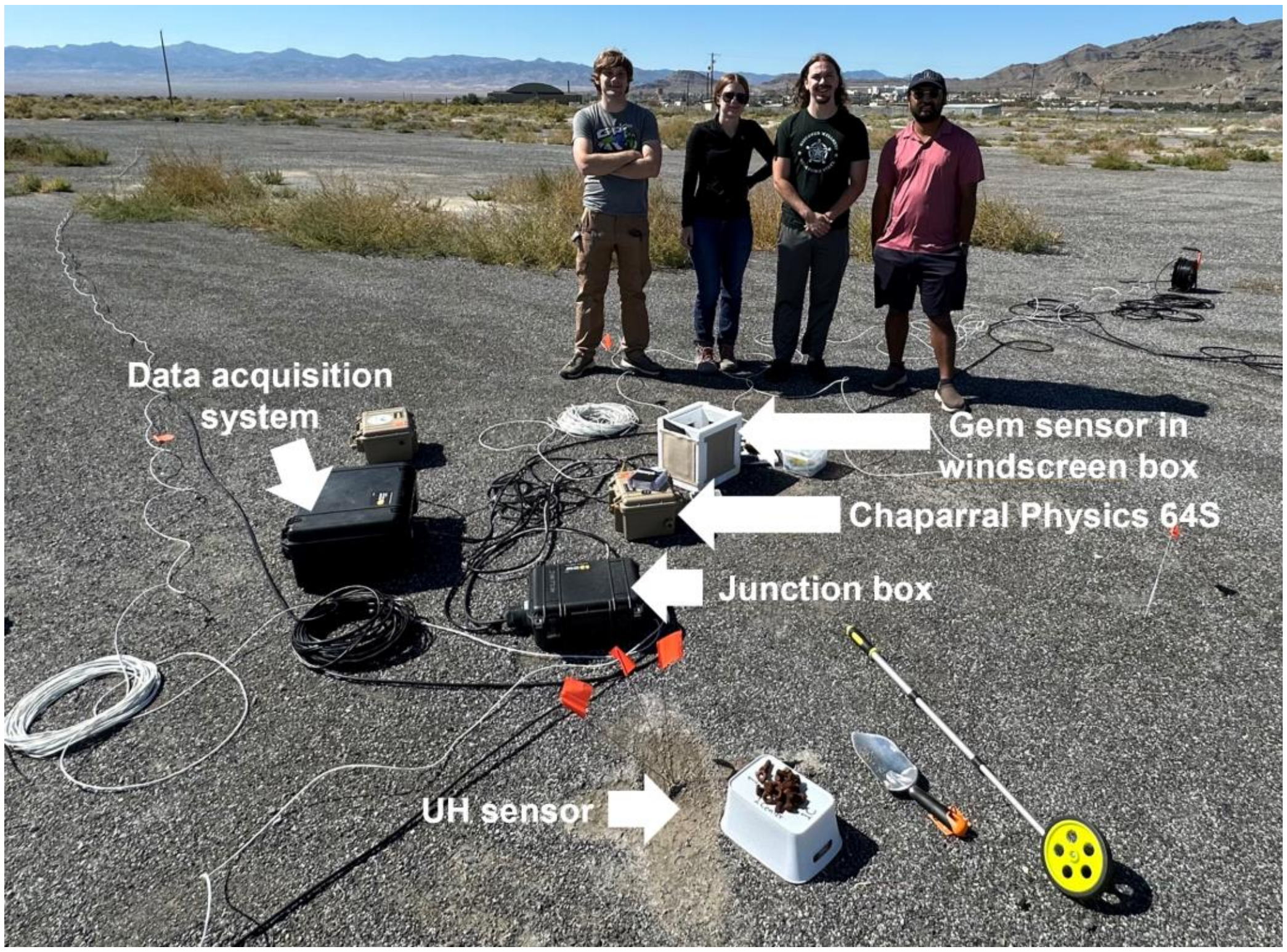

Figure S7: OSU Team (L to R: Douglas Fox, Kate Spillman, Trevor Wilson, and Real KC) at the Wendover Airport near the central location where four different sensors were co-located (Chaparral Physics 64s, GEM, WERD ISSM23, and RedVox deployed by the University of Hawaii). Photo credit: M. Garcés.

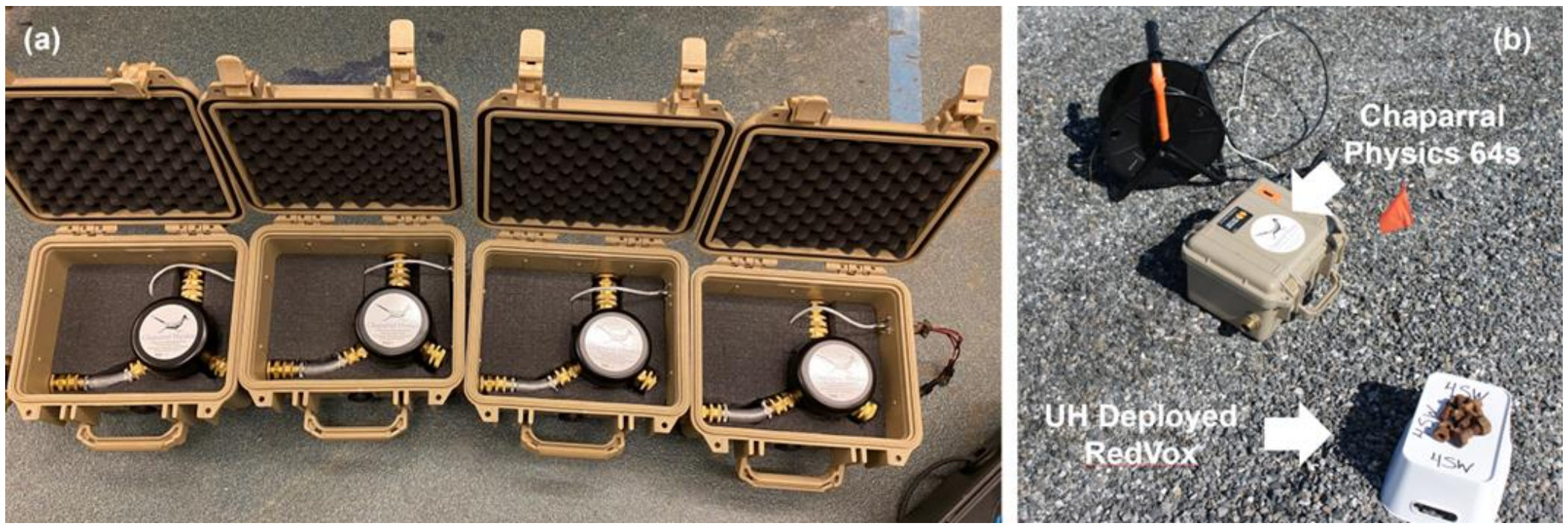

Figure S8: (a) Photo of the four Chaparral Physics 64s sensors in their cases before the OSIRIS-REx SRC re-entry deployment. (b) One of the Chaparral Physics sensors deployed near a RedVox sensor deployed by UH (photo credit: B. Elbing).

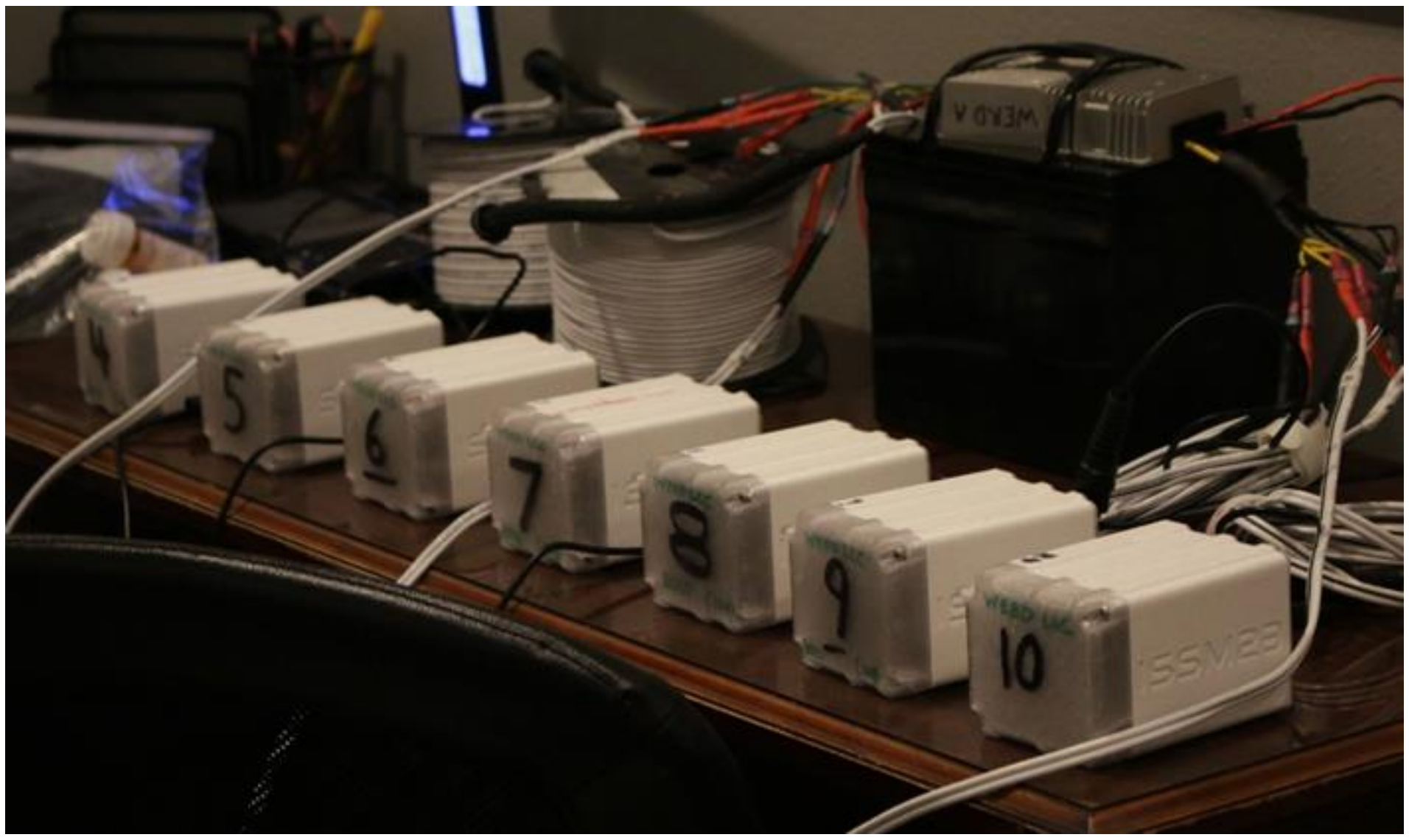

Figure S9: Picture of the seven WERD ISSM23 sensors being tested the night before the deployment for the OSIRIS-REx SRC re-entry deployment (photo credit: B. Elbing).

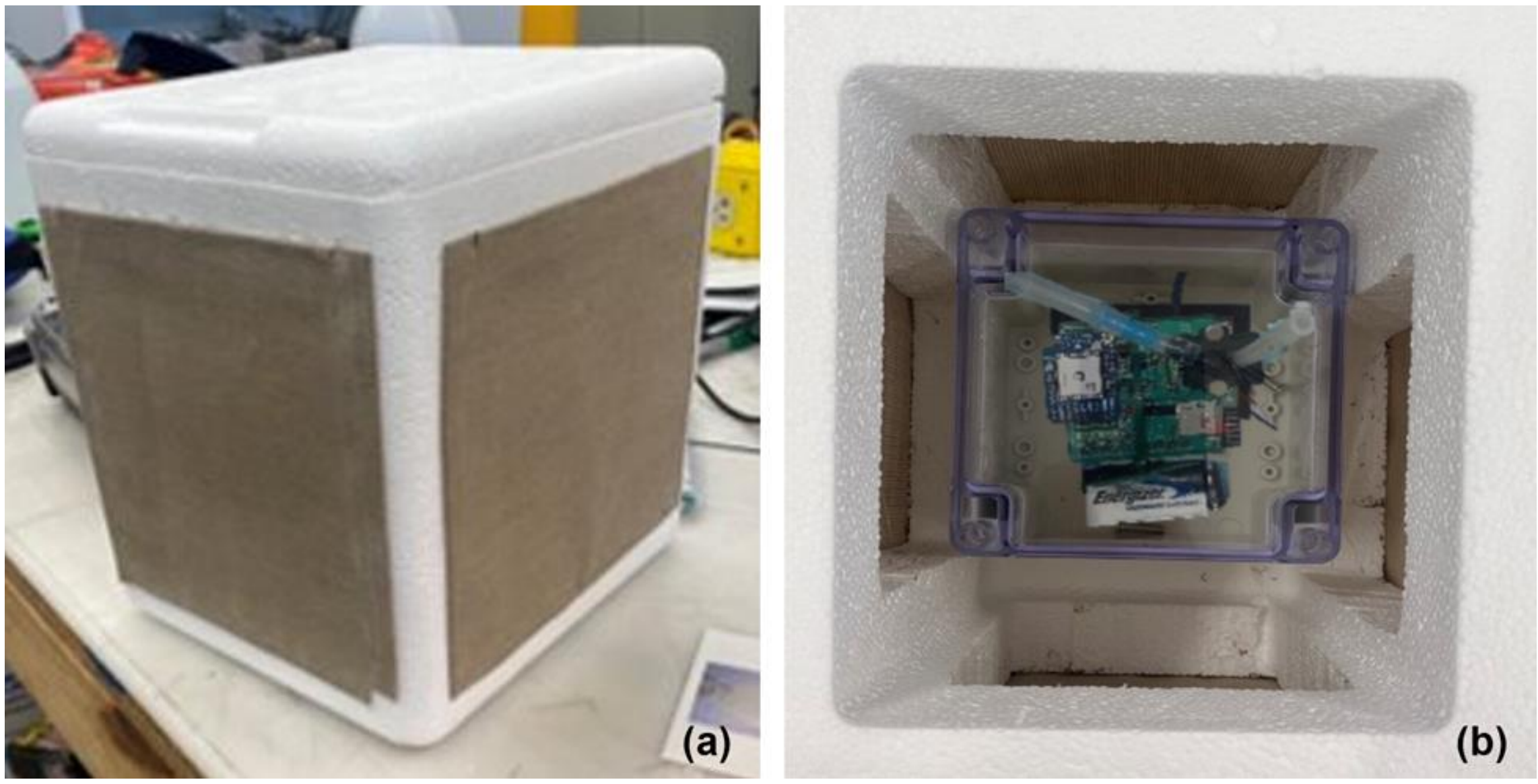

Figure S10: (a) Picture of one of the windscreen boxes (Swaim et al., 2023) that housed a GEM sensor. (b) View from above with the cover removed showing a single GEM sensor held within an enclosure and positioned within the windscreen box (photo credit: B. Elbing).

## Signal Detection

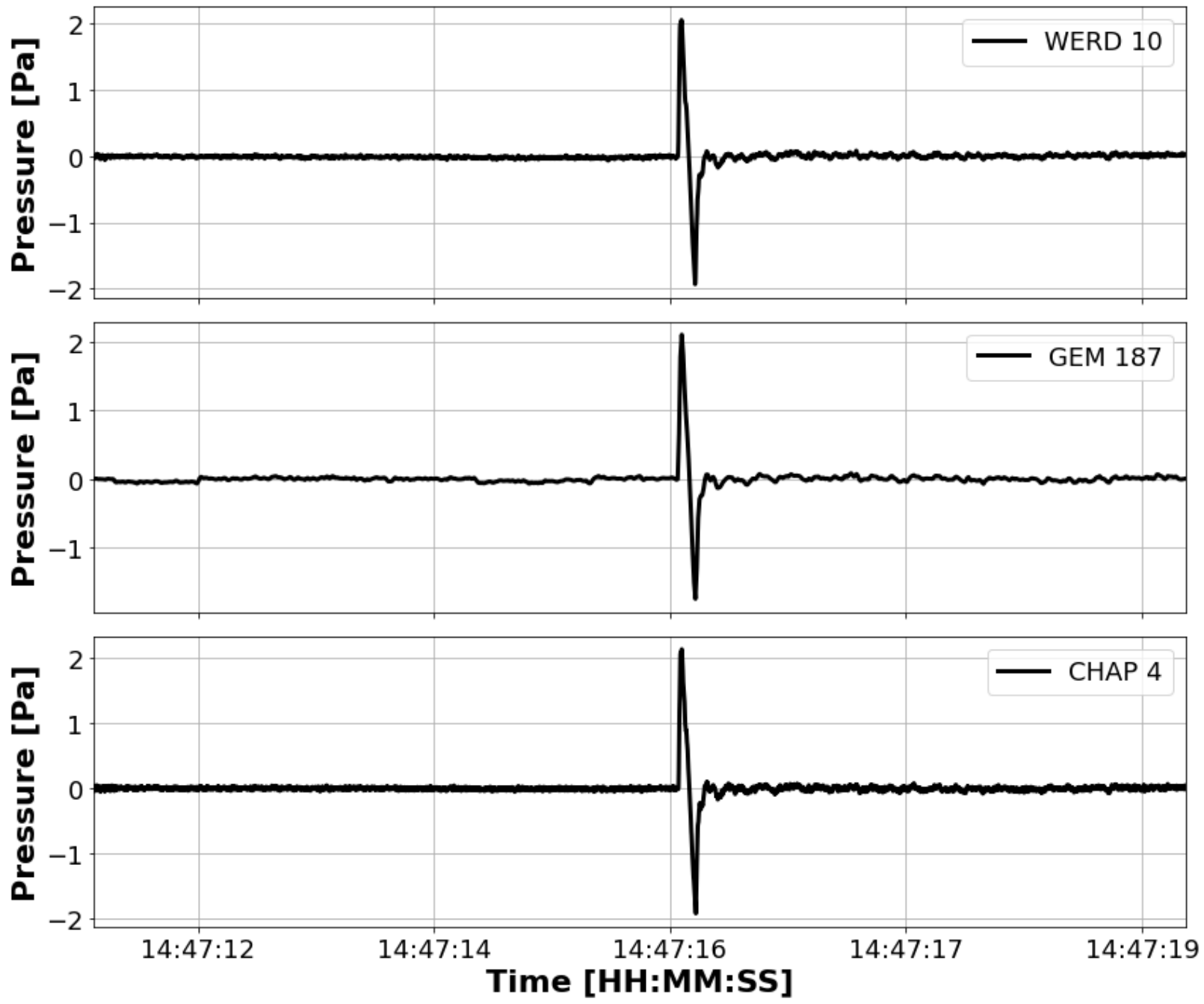

Figure S11: Signal detection at the West Wendover Airport, UT. The figure shows a comparison between the three sensor types used.

## South Methodist University (SMU)

## Signal Detection

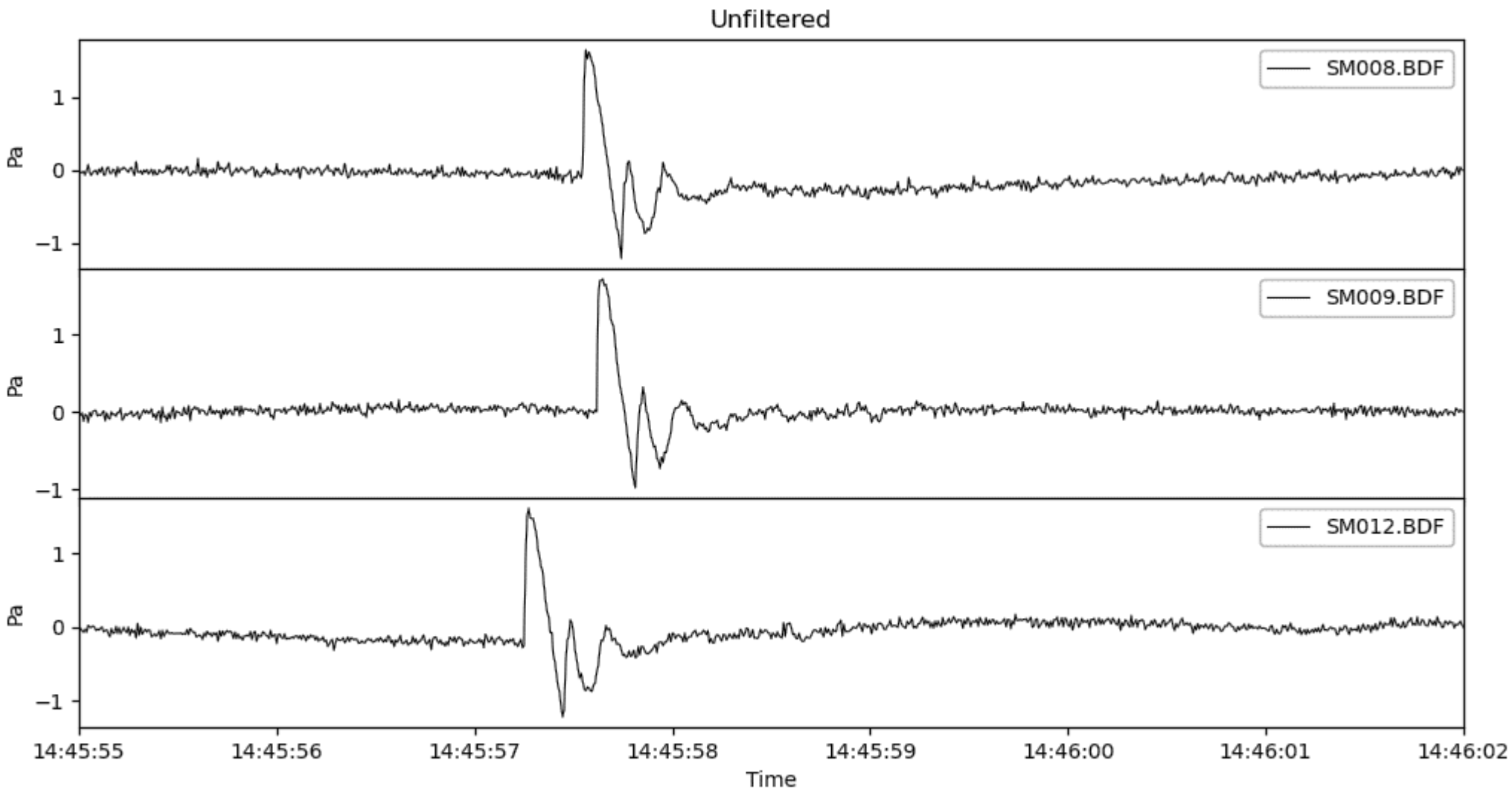

Figure S12: Signal detection at the Eureka Airport, NV. The timeseries shown are unfiltered. Time is shown in UTC.

## University of Hawaii (UH)

### Signal Detections

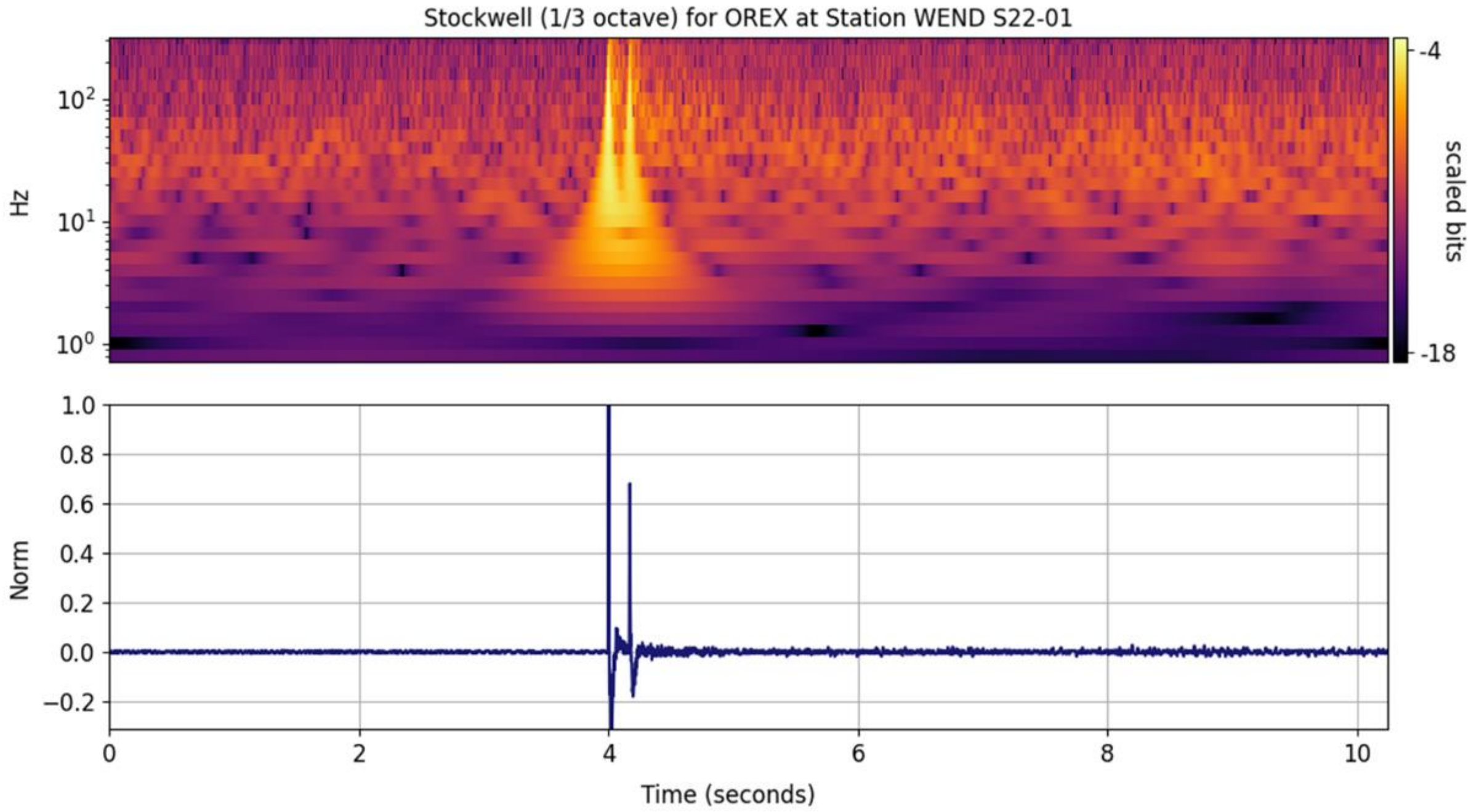

Figure S13: Signal detection at West Wendover Airport, UT. The time series in the lower panel shows the smartphone microphone equivalent high-pass filter response of the N-wave, the time between the two distinct peaks is the N wave duration. The upper panel shows the multiresolution time-frequency representation of the signal using a Stockwell transform (Garces, 2023), and showing the lower-frequency components of the N-wave. All channels of the Wendover array showed nearly identical waveforms time shifted by their arrival time.

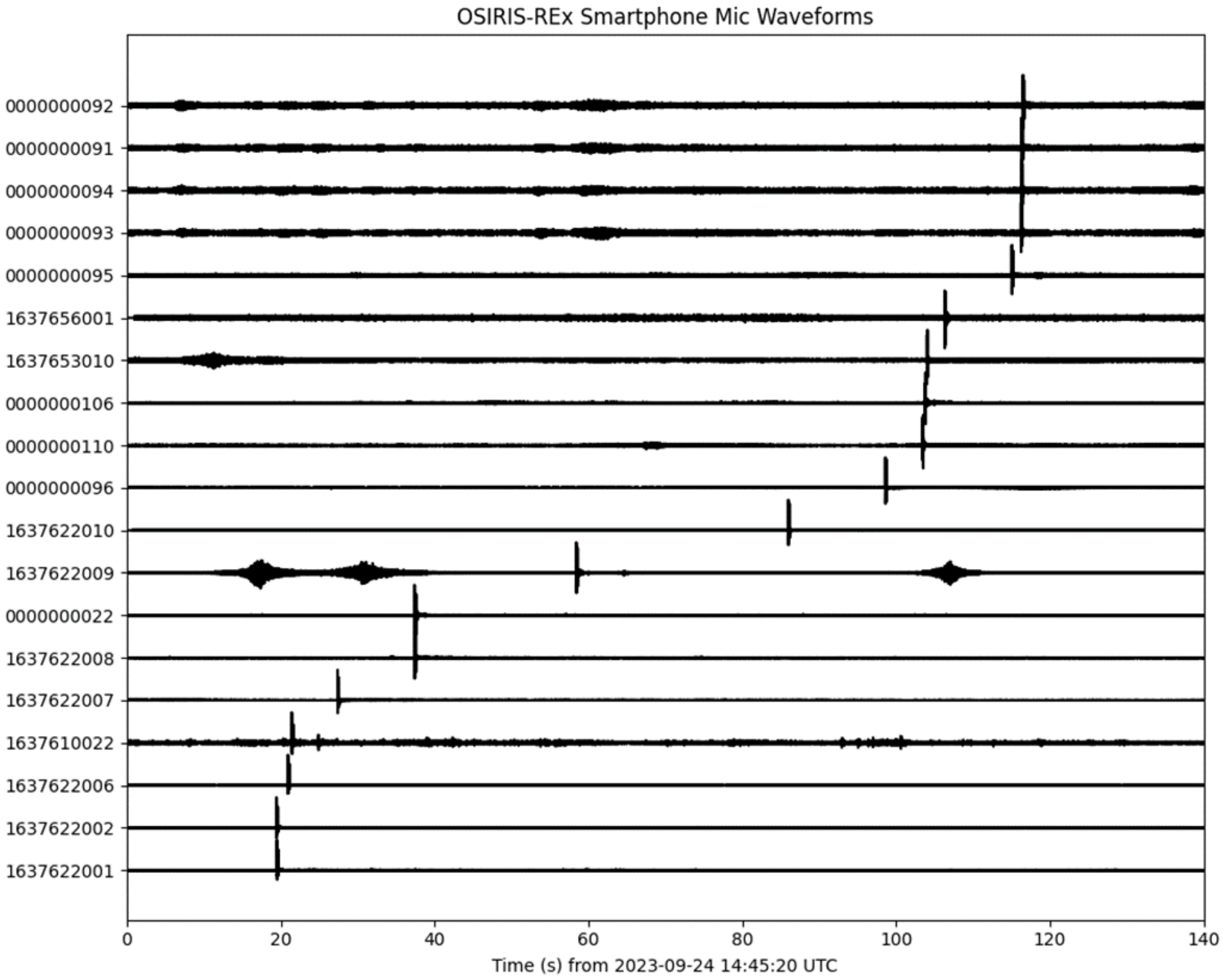

Figure S14: Detections made by smartphones, ordered by arrival time. The arrival waveforms and spectra are similar to those in Figure S13, but the N wave duration depends on the source height and speed. The timing of the arrivals corresponds to the time of closest approach of the source plus the time it would take to reach the station. This arrival pattern is only possible from hypersonic and supersonic sources.

## University of Memphis (UM)

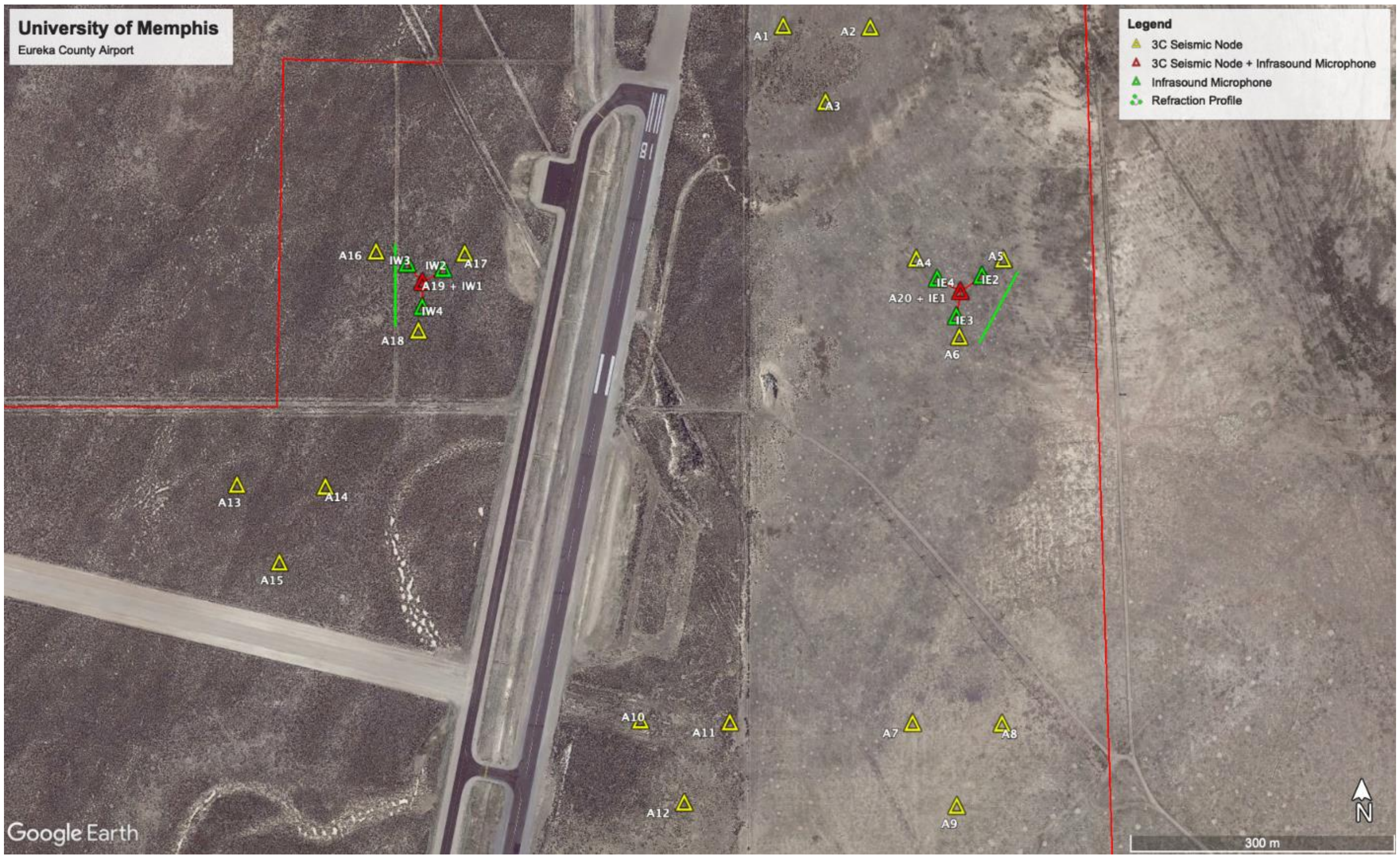

Figure S15: University of Memphis seismo-acoustic array experiment. Yellow triangles show the location of the 3 component seismic nodes in the Golay 3x6 array configuration (A1 through A18). Red triangles show the center infrasound instrument and co-located 3 component seismic node for the western and eastern infrasound arrays (A19 + IW1, A20 + IE1). Green triangles show locations for infrasound microphones. Green lines show locations for the P and SH refraction profiles near the western and eastern infrasound arrays. The red lines show the boundary of the airport property.

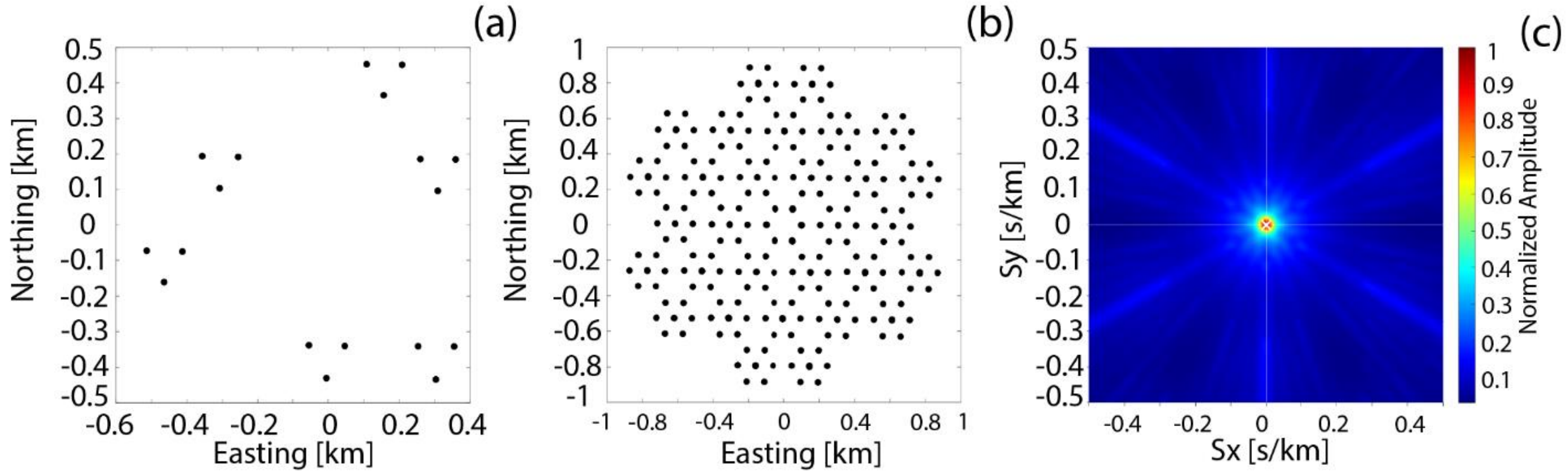

Figure S16: Geometry and response of the Golay 3x6 Array. (a) shows the OSIRIS-REx array design. (b) is the co-array which consists of distances and azimuths between all pairs of stations of the array. (c) is the broadband array response for a vertically incident plane wave for the frequency band 0.25 to 35Hz.

## Appendix B

Table A1: Infrasound instrument installation data. This table is available in machine readable format.

| Institution | Instrument field name | Instrument type | Sampling rate [Hz] | Location | Lat (N) [deg] | Lon (E) [deg] | Comments |
|---|---|---|---|---|---|---|---|
| SNL | C1 | Gem | 100 | Utah | 39.92841 | -113.99949 | Single sensor |
| SNL | C2 | Gem | 100 | Utah | 39.97127 | -113.97749 | Single sensor |
| SNL | C3 | Gem | 100 | Utah | 40.01614 | -113.97312 | Single sensor |
| SNL | C4 | Gem | 100 | Utah | 40.006113 | -113.9766 | Single sensor |
| SNL | C5 | Gem | 100 | Utah | 40.10518 | -113.97008 | Single sensor |
| SNL | C6 | Gem | 100 | Utah | 40.1518 | -113.98394 | Single sensor |
| SNL | C7 | Gem | 100 | Utah | 40.17377 | -113.99698 | Single sensor |
| SNL | C8 | Gem | 100 | Utah | 40.19454 | -113.98697 | Single sensor |
| SNL | C9 | Gem | 100 | Utah | 40.21743 | -113.99005 | Single sensor |
| SNL | C10 | Gem | 100 | Utah | 40.24075 | -113.99157 | Single sensor |
| SNL | C11 | Gem | 100 | Utah | 40.28813 | -113.98843 | Single sensor |
| SNL | A1 | Gem | 100 | Eureka Airport | 39.37657 | -115.82061 | Single sensor |
| SNL | A2 | Gem | 100 | Eureka Airport | 39.41789 | -115.81251 | Single sensor |
| SNL | A3 | Gem | 100 | Eureka Airport | 39.45742 | -115.80031 | Single sensor |
| SNL | A4 | Gem | 100 | Eureka Airport | 39.50356 | -115.78451 | Single sensor |
| SNL | A5 | Gem | 100 | Eureka Airport | 39.54794 | -115.77042 | Single sensor |
| SNL | A6 | Gem | 100 | Eureka Airport | 39.59715 | -115.75859 | Single sensor |
| SNL | A7 | Gem | 100 | Eureka Airport | 39.64012 | -115.77885 | Single sensor |
| SNL | A8 | Gem | 100 | Eureka Airport | 39.681 | -115.7779 | Single sensor |
| SNL | A9 | Gem | 100 | Eureka Airport | 39.72717 | -115.76785 | Single sensor |
| SNL | A10 | Gem | 100 | Eureka Airport | 39.74884 | -115.77333 | Single sensor |
| SNL | A11 | Gem | 100 | Eureka Airport | 39.77546 | -115.76523 | Single sensor |
| SNL | A12 | Gem | 100 | Eureka Airport | 39.81548 | -115.75109 | Single sensor |
| SNL | A13 | Gem | 100 | Eureka Airport | 39.86349 | -115.73202 | Single sensor |
| SNL | A14 | Gem | 100 | Eureka Airport | 39.91697 | -115.73846 | Single sensor |
| SNL | A15 | Gem | 100 | Eureka Airport | 39.96471 | -115.74647 | Single sensor |
| SNL | A16 | Gem | 100 | Eureka Airport | 40.01179 | -115.76422 | Single sensor |
| SNL | A17 | Gem | 100 | Eureka Airport | 40.05231 | -115.77848 | Single sensor |
| SNL | A18 | Gem | 100 | Eureka Airport | 40.09891 | -115.7832 | Single sensor |
| SNL | A19 | Gem | 100 | Eureka Airport | 40.14403 | -115.75413 | Single sensor |
| SNL | A20 | Gem | 100 | Eureka Airport | 40.19088 | -115.74343 | Single sensor |
| SNL | A21 | Gem | 100 | Eureka Airport | 40.23178 | -115.70692 | Single sensor |
| SNL | A22 | Gem | 100 | Eureka Airport | 40.26937 | -115.68173 | Single sensor |
| SNL | T1 | Gem | 100 | | 39.54136 | -116.38015 | Single sensor |
| SNL | T2 | Gem | 100 | | 39.55026 | -116.33675 | Single sensor |
| SNL | T3 | Gem | 100 | | 39.58732 | -116.20592 | Single sensor |
| SNL | T4 | Gem | 100 | | 39.60226 | -116.1448 | Single sensor |
| SNL | T5 | Gem | 100 | | 39.63018 | -116.04795 | Single sensor |
| SNL | T6 | Gem | 100 | | 39.67194 | -115.90301 | Single sensor |
| SNL | T7 | Gem | 100 | | 39.71323 | -115.7629 | Single sensor |
| SNL | T8 | Gem | 100 | | 39.74373 | -115.6735 | Single sensor |
| SNL | T10 | Gem | 100 | | 40.04763 | -114.64058 | Single sensor |
| SNL | T11 | Gem | 100 | | 40.06358 | -114.58548 | Single sensor |
| SNL | T12 | Gem | 100 | | 40.07789 | -114.53723 | Single sensor |
| SNL | T13 | Gem | 100 | | 40.09932 | -114.54961 | Single sensor |
| SNL | T14 | Gem | 100 | | 40.18212 | -114.01282 | Single sensor |
| SNL | HA1-W | Hyperion | 100 | | 39.61691 | -115.99818 | Array |
| SNL | HA1-E | Hyperion | 100 | | 39.61691 | -115.99764 | Array |
| SNL | HA1-N | Hyperion | 100 | | 39.60727 | -115.99791 | Array |
| SNL | HA1-C | Hyperion | 100 | | 39.61705 | -115.99794 | Array |
| SNL | HA2-E | Hyperion | 100 | | 39.63933 | -115.78173 | Array |
| SNL | HA2-W | Ultra Light | 100 | | 39.63933 | -115.78227 | Array |
| SNL | HA2-C | Ultra Light | 100 | | 39.63945 | -115.78200 | Array |
| SNL | HA2-N | Hyperion | 100 | | 39.63969 | -115.78200 | Array |
| SNL | HA3-W | Hyperion | 100 | | 39.54036 | -115.77476 | Array |
| SNL | HA3-C | Hyperion | 100 | | 39.54048 | -115.77450 | Array |
| SNL | HA3-E | Hyperion | 100 | | 39.54036 | -115.77423 | Array |
| SNL | HA3-N | Hyperion | 100 | | 39.54072 | -115.7745 | Array |
| LANL | OREXA | Hyperion 3000 | 200 | | 39.6109883 | -116.002932 | Single sensor |
| LANL | OREXB | Hyperion 3000 | 200 | | 39.60899 | -116.011737 | Single sensor |
| LANL | OREXC | Hyperion 3000 | 200 | | 39.6040433 | -116.004643 | Single sensor |
| LANL | OREXD | Hyperion 3000 | 200 | | 39.7372017 | -115.674093 | Single sensor |
| LANL | OREXE | Hyperion 3000 | 200 | | 39.7043 | -115.676033 | Single sensor |
| LANL | OREXF | Hyperion 3000 | 200 | | 39.6858783 | -115.676975 | Single sensor |
| LANL | OREX1, e1 | Hyperion IFS-3000 | 100 | Price, UT | 39.4751582 | -110.7433235 | Array |
| LANL | OREX1, e2 | Hyperion IFS-3000 | 100 | Price, UT | 39.475359 | -110.7451219 | Array |
| LANL | OREX1, e3 | Hyperion IFS-3000 | 100 | Price, UT | 39.4739251 | -110.7449485 | Array |
| LANL | OREX1, e4 | Hyperion IFS-3000 | 100 | Price, UT | 39.474677 | -110.7442959 | Array |
| LANL | OREX3, e1 | Hyperion IFS-3000 | 100 | St George, UT | 37.0151557 | -113.616179 | Array |
| LANL | OREX3, e2 | Hyperion IFS-3000 | 100 | St George, UT | 37.0149832 | -113.6170342 | Array |
| LANL | OREX3, e3 | Hyperion IFS-3000 | 100 | St George, UT | 37.0155959 | -113.6171397 | Array |
| LANL | OREX3, e4 | Hyperion IFS-3000 | 100 | St George, UT | 37.0162369 | -113.6170218 | Array |
| LANL | OREX3, e5 | Hyperion IFS-3000 | 100 | St George, UT | 37.0160341 | -113.6162266 | Array |
| LANL | OREX3, e6 | Hyperion IFS-3000 | 100 | St George, UT | 37.0155697 | -113.6161785 | Array |
| LANL | OREX2, e1 | Hyperion IFS-3000 | 500 | NNSS, NV | 37.224998 | -116.149168 | Array |
| LANL | OREX2, e2 | Hyperion IFS-3000 | 500 | NNSS, NV | 37.223915 | -116.149249 | Array |
| LANL | OREX2, e3 | Hyperion IFS-3000 | 500 | NNSS, NV | 37.223589 | -116.148195 | Array |
| LANL | OREX2, e4 | Hyperion IFS-3000 | 500 | NNSS, NV | 37.224247 | -116.1488 | Array |
| UM | IW1 | VLF Designs IAM-1 | 1000 | Eureka Airport | 39.61159286 | -116.004319 | Array |
| UM | IW2 | VLF Designs IAM-1 | 1000 | Eureka Airport | 39.61172807 | -116.0040416 | Array |
| UM | IW3 | VLF Designs IAM-1 | 1000 | Eureka Airport | 39.61178112 | -116.0045212 | Array |
| UM | IW4 | VLF Designs IAM-1 | 1000 | Eureka Airport | 39.61133569 | -116.0043187 | Array |
| UM | IE1 | VLF Designs IAM-1 | 1000 | Eureka Airport | 39.61150004 | -115.997155 | Array |
| UM | IE2 | VLF Designs IAM-1 | 1000 | Eureka Airport | 39.61165862 | -115.9968662 | Array |
| UM | IE3 | VLF Designs IAM-1 | 1000 | Eureka Airport | 39.61124095 | -115.9972107 | Array |
| UM | IE4 | VLF Designs IAM-1 | 1000 | Eureka Airport | 39.61162776 | -115.9974629 | Array |
| KUT | | microphone | | Eureka Airport | 39.6166 | -115.9986 | Single sensor |
| KUT | | microphone | | Eureka Airport | 39.6165 | -115.9974 | Single sensor |
| KUT | | INF04 | 100 | Eureka Airport | 39.6175 | -115.9974 | Array |
| KUT | | INF04 | 100 | Eureka Airport | 39.6175 | -115.9986 | Array |
| KUT | | INF04 | 100 | Eureka Airport | 39.6138 | -116.0046 | Array |
| KUT | | microphone | | Eureka Airport | 39.6138 | -116.0046 | |
| KUT | | INF04 | 100 | Eureka Airport | 39.5893 | -116.0049 | Array |
| KUT | | microphone | | Eureka Airport | 39.5893 | -116.0049 | |
| BSU | JDSA1 | GEM, Infrasound Loggers 1.01 | 100 | Southwest Idaho | 43.12185907 | -116.7856059 | Array |
| BSU | JDSA2 | GEM, Infrasound Loggers 1.01 | 100 | Southwest Idaho | 43.12192087 | -116.7855215 | Array |
| BSU | JDSA3 | GEM, Infrasound Loggers 1.01 | 100 | Southwest Idaho | 43.12178317 | -116.7855284 | Array |
| BSU | JDSA4 | GEM, Infrasound Loggers 1.01 | 100 | Southwest Idaho | 43.12181649 | -116.7856906 | Array |
| BSU | JDSB1 | GEM, Infrasound Loggers 1.01 | 100 | Idaho | 43.12157218 | -116.7879959 | Array |
| BSU | JDSB2 | GEM, Infrasound Loggers 1.01 | 100 | Idaho | 43.12160936 | -116.7879871 | Array |
| BSU | JDSB3 | GEM, Infrasound Loggers 1.01 | 100 | Idaho | 43.12155341 | -116.7879649 | Array |
| BSU | JDSB4 | GEM, Infrasound Loggers 1.01 | 100 | Idaho | 43.12156295 | -116.788055 | Array |
| BSU | JDNB1 | GEM, Infrasound Loggers 1.03 | 100 | Idaho | 43.12543365 | -116.7875822 | Array |
| BSU | JDNB2 | GEM, Infrasound Loggers 1.04 | 100 | Idaho | 43.12547297 | -116.7875298 | Array |
| BSU | JDNB3 | GEM, Infrasound Loggers 1.05 | 100 | Idaho | 43.12541687 | -116.7875265 | Array |
| BSU | JDNB4 | GEM, Infrasound Loggers 1.06 | 100 | Idaho | 43.12543144 | -116.7876213 | Array |
| BSU | TOP01 | GEM, Infrasound Loggers 1.08 | 100 | Idaho | 43.12550998 | -116.8013974 | Array |
| BSU | TOP02 | GEM, Infrasound Loggers 1.09 | 100 | Idaho | 43.12560459 | -116.8015603 | Array |
| BSU | TOP03 | GEM, Infrasoun | 100 | Idaho | 43.12572902 | -116.8017619 | Array |
| | | d Loggers 1.10 | | | | | |
| BSU | TOP04 | GEM, Infrasound Loggers 1.11 | 100 | Idaho | 43.12584698 | -116.8019344 | Array |
| BSU | TOP05 | GEM, Infrasound Loggers 1.12 | 100 | Idaho | 43.12596533 | -116.8020799 | Array |
| BSU | TOP06 | GEM, Infrasound Loggers 1.13 | 100 | Idaho | 43.12606252 | -116.8021671 | Array |
| BSU | TOP07 | GEM, Infrasound Loggers 1.14 | 100 | Idaho | 43.12620928 | -116.8023396 | Array |
| BSU | TOP08 | GEM, Infrasound Loggers 1.15 | 100 | Idaho | 43.12634974 | -116.8024834 | Array |
| BSU | TOP09 | GEM, Infrasound Loggers 1.16 | 100 | Idaho | 43.12643686 | -116.8026145 | Array |
| BSU | TOP10 | GEM, Infrasound Loggers 1.17 | 100 | Idaho | 43.12655189 | -116.8027244 | Array |
| BSU | TOP11 | GEM, Infrasound Loggers 1.18 | 100 | Idaho | 43.12666905 | -116.8028433 | Array |
| BSU | TOP12 | GEM, Infrasound Loggers 1.19 | 100 | Idaho | 43.12677948 | -116.8030524 | Array |
| BSU | TOP13 | GEM, Infrasound Loggers 1.20 | 100 | Idaho | 43.12674003 | -116.803228 | Array |
| BSU | TOP14 | GEM, Infrasound Loggers 1.21 | 100 | Idaho | 43.1266241 | -116.8033681 | Array |
| BSU | TOP15 | GEM, Infrasound Loggers 1.22 | 100 | Idaho | 43.12663831 | -116.8035886 | Array |
| BSU | TOP16 | GEM, Infrasound Loggers 1.23 | 100 | Idaho | 43.1265124 | -116.8036435 | Array |
| BSU | TOP17 | GEM, Infrasound Loggers 1.24 | 100 | Idaho | 43.1264339 | -116.8038132 | Array |
| BSU | TOP18 | GEM, Infrasound Loggers 1.25 | 100 | Idaho | 43.12634576 | -116.8039767 | Array |
| BSU | TOP19 | GEM, Infrasound Loggers 1.26 | 100 | Idaho | 43.12624488 | -116.804096 | Array |
| BSU | TOP20 | GEM, Infrasound Loggers 1.27 | 100 | Idaho | 43.12617945 | -116.8042248 | Array |
| BSU | TOP21 | GEM, Infrasound Loggers 1.28 | 100 | Idaho | 43.1260073 | -116.8041484 | Array |
| BSU | TOP22 | GEM, Infrasound Loggers 1.29 | 100 | Idaho | 43.12583556 | -116.8039579 | Array |
| BSU | TOP23 | GEM, Infrasound Loggers 1.30 | 100 | Idaho | 43.12569345 | -116.8037969 | Array |
| BSU | TOP24 | GEM, Infrasound Loggers 1.31 | 100 | Idaho | 43.12556656 | -116.8036705 | Array |
| BSU | TOP25 | GEM, Infrasound Loggers 1.32 | 100 | Idaho | 43.12546093 | -116.8035615 | Array |
| BSU | TOP26 | GEM, Infrasound Loggers 1.33 | 100 | Idaho | 43.1252966 | -116.8034484 | Array |
| BSU | TOP27 | GEM, Infrasound Loggers 1.34 | 100 | Idaho | 43.1251737 | -116.8032903 | Array |
| BSU | TOP28 | GEM, Infrasound Loggers 1.35 | 100 | Idaho | 43.12508251 | -116.8030776 | Array |
| BSU | TOP29 | GEM, Infrasound Loggers 1.36 | 100 | Idaho | 43.12494683 | -116.8029506 | Array |
| BSU | TOP30 | GEM, Infrasoun | 100 | Idaho | 43.12481786 | -116.8028627 | Array |
| | | d Loggers 1.37 | | | | | |
| BSU | TOP31 | GEM, Infrasound Loggers 1.38 | 100 | Idaho | 43.12469091 | -116.8027794 | Array |
| BSU | TOP32 | GEM, Infrasound Loggers 1.39 | 100 | Idaho | 43.12472395 | -116.802606 | Array |
| BSU | TOP33 | GEM, Infrasound Loggers 1.40 | 100 | Idaho | 43.12478858 | -116.8024612 | Array |
| BSU | TOP34 | GEM, Infrasound Loggers 1.41 | 100 | Idaho | 43.12490261 | -116.8023562 | Array |
| BSU | TOP35 | GEM, Infrasound Loggers 1.42 | 100 | Idaho | 43.12498012 | -116.8022125 | Array |
| BSU | TOP36 | GEM, Infrasound Loggers 1.43 | 100 | Idaho | 43.12508385 | -116.8020925 | Array |
| BSU | TOP37 | GEM, Infrasound Loggers 1.44 | 100 | Idaho | 43.1251528 | -116.8019492 | Array |
| BSU | TOP38 | GEM, Infrasound Loggers 1.45 | 100 | Idaho | 43.12523129 | -116.8018297 | Array |
| BSU | TOP39 | GEM, Infrasound Loggers 1.46 | 100 | Idaho | 43.12535378 | -116.8016928 | Array |
| BSU | TOP40 | GEM, Infrasound Loggers 1.47 | 100 | Idaho | 43.12539279 | -116.8015533 | Array |
| BSU | TOP41 | GEM, Infrasound Loggers 1.48 | 100 | Idaho | 43.12550101 | -116.8019804 | Array |
| BSU | TOP42 | GEM, Infrasound Loggers 1.49 | 100 | Idaho | 43.12544889 | -116.8022824 | Array |
| BSU | TOP43 | GEM, Infrasound Loggers 1.50 | 100 | Idaho | 43.12556576 | -116.8025817 | Array |
| BSU | TOP44 | GEM, Infrasound Loggers 1.51 | 100 | Idaho | 43.12561661 | -116.8028346 | Array |
| TDA | AA 1 | TDA sensor | 200 | Eureka Airport | 39.61752711 | -115.9986176 | Large N-array |
| TDA | AA 2 | TDA sensor | 200 | Eureka Airport | 39.61752834 | -115.9986619 | Large N-array |
| TDA | AA 3 | TDA sensor | 200 | Eureka Airport | 39.61754657 | -115.998621 | Large N-array |
| TDA | AA 4 | TDA sensor | 200 | Eureka Airport | 39.61752664 | -115.9985874 | Large N-array |
| TDA | AA 5 | TDA sensor | 200 | Eureka Airport | 39.61749151 | -115.9986127 | Large N-array |
| TDA | AB 1 | TDA sensor | 200 | Eureka Airport | 39.61752118 | -115.9984251 | Large N-array |
| TDA | AB 2 | TDA sensor | 200 | Eureka Airport | 39.6175235 | -115.9984637 | Large N-array |
| TDA | AB 3 | TDA sensor | 200 | Eureka Airport | 39.61754496 | -115.9984252 | Large N-array |
| TDA | AB 4 | TDA sensor | 200 | Eureka Airport | 39.61752115 | -115.9983805 | Large N-array |
| TDA | AB 5 | TDA sensor | 200 | Eureka Airport | 39.61748155 | -115.9984185 | Large N-array |
| TDA | AC 1 | TDA sensor | 200 | Eureka Airport | 39.61751905 | -115.9982339 | Large N-array |
| TDA | AC 2 | TDA sensor | 200 | Eureka Airport | 39.61753585 | -115.9982558 | Large N-array |
| TDA | AC 3 | TDA sensor | 200 | Eureka Airport | 39.61753085 | -115.9982101 | Large N-array |
| TDA | AC 4 | TDA sensor | 200 | Eureka Airport | 39.61750178 | -115.9982132 | Large N-array |
| TDA | AC 5 | TDA sensor | 200 | Eureka Airport | 39.61749968 | -115.9982592 | Large N-array |
| TDA | AD 1 | TDA sensor | 200 | Eureka Airport | 39.61752679 | -115.9980451 | Large N-array |
| TDA | AD 2 | TDA sensor | 200 | Eureka Airport | 39.61753938 | -115.9980647 | Large N-array |
| TDA | AD 3 | TDA sensor | 200 | Eureka Airport | 39.61754099 | -115.9980263 | Large N-array |
| TDA | AD 4 | TDA sensor | 200 | Eureka Airport | 39.61750802 | -115.9980294 | Large N-array |
| TDA | AD 5 | TDA sensor | 200 | Eureka Airport | 39.61750673 | -115.9980685 | Large N-array |
| TDA | AE 1 | TDA sensor | 200 | Eureka Airport | 39.61752578 | -115.9978596 | Large N-array |
| TDA | AE 2 | TDA sensor | 200 | Eureka Airport | 39.61754488 | -115.9978823 | Large N-array |
| TDA | AE 3 | TDA sensor | 200 | Eureka Airport | 39.61754334 | -115.997839 | Large N-array |
| TDA | AE 4 | TDA sensor | 200 | Eureka Airport | 39.61750814 | -115.9978419 | Large N-array |
| TDA | AE 5 | TDA sensor | 200 | Eureka Airport | 39.61750734 | -115.9978785 | Large N-array |
| TDA | AG 1 | TDA sensor | 200 | Eureka Airport | 39.61750459 | -115.9973942 | Large N-array |
| TDA | AG 2 | TDA sensor | 200 | Eureka Airport | 39.61752665 | -115.9974294 | Large N-array |
| TDA | AG 3 | TDA sensor | 200 | Eureka Airport | 39.61752726 | -115.9973741 | Large N-array |
| TDA | AG 4 | TDA sensor | 200 | Eureka Airport | 39.61748733 | -115.9973661 | Large N-array |
| TDA | AG 5 | TDA sensor | 200 | Eureka Airport | 39.61747382 | -115.9974161 | Large N-array |
| TDA | AO 1 | TDA sensor | 200 | Eureka Airport | 39.61729741 | -115.9973552 | Large N-array |
| TDA | AO 2 | TDA sensor | 200 | Eureka Airport | 39.61732845 | -115.9973682 | Large N-array |
| TDA | AO 3 | TDA sensor | 200 | Eureka Airport | 39.61731556 | -115.9973271 | Large N-array |
| TDA | AO 4 | TDA sensor | 200 | Eureka Airport | 39.61727182 | -115.9973251 | Large N-array |
| TDA | AO 5 | TDA sensor | 200 | Eureka Airport | 39.61728508 | -115.9973932 | Large N-array |
| TDA | AL 1 | TDA sensor | 200 | Eureka Airport | 39.61731577 | -115.9980295 | Large N-array |
| TDA | AL 2 | TDA sensor | 200 | Eureka Airport | 39.61733268 | -115.9980542 | Large N-array |
| TDA | AL 3 | TDA sensor | 200 | Eureka Airport | 39.61733605 | -115.998002 | Large N-array |
| TDA | AL 4 | TDA sensor | 200 | Eureka Airport | 39.61729622 | -115.9980052 | Large N-array |
| TDA | AL 5 | TDA sensor | 200 | Eureka Airport | 39.61729946 | -115.9980556 | Large N-array |
| TDA | AH 1 | TDA sensor | 200 | Eureka Airport | 39.61731637 | -115.9985973 | Large N-array |
| TDA | AH 2 | TDA sensor | 200 | Eureka Airport | 39.61730556 | -115.9986386 | Large N-array |
| TDA | AH 3 | TDA sensor | 200 | Eureka Airport | 39.61735896 | -115.9985988 | Large N-array |
| TDA | AH 4 | TDA sensor | 200 | Eureka Airport | 39.61732078 | -115.9985483 | Large N-array |
| TDA | AH 5 | TDA sensor | 200 | Eureka Airport | 39.61727528 | -115.9985985 | Large N-array |
| TDA | AP 1 | TDA sensor | 200 | Eureka Airport | 39.6171509 | -115.9985694 | Large N-array |
| TDA | AP 2 | TDA sensor | 200 | Eureka Airport | 39.61712029 | -115.9985639 | Large N-array |
| TDA | AP 3 | TDA sensor | 200 | Eureka Airport | 39.61714777 | -115.9986103 | Large N-array |
| TDA | AP 4 | TDA sensor | 200 | Eureka Airport | 39.61718634 | -115.9985563 | Large N-array |
| TDA | AP 5 | TDA sensor | 200 | Eureka Airport | 39.61715099 | -115.9985163 | Large N-array |
| TDA | AT 1 | TDA sensor | 200 | Eureka Airport | 39.61713803 | -115.9980301 | Large N-array |
| TDA | AT 2 | TDA sensor | 200 | Eureka Airport | 39.61711037 | -115.9980552 | Large N-array |
| TDA | AT 3 | TDA sensor | 200 | Eureka Airport | 39.61715787 | -115.9980779 | Large N-array |
| TDA | AT 4 | TDA sensor | 200 | Eureka Airport | 39.61717115 | -115.9979966 | Large N-array |
| TDA | AT 5 | TDA sensor | 200 | Eureka Airport | 39.6171214 | -115.9979858 | Large N-array |
| TDA | AW 1 | TDA sensor | 200 | Eureka Airport | 39.61711683 | -115.9973592 | Large N-array |
| TDA | AW 2 | TDA sensor | 200 | Eureka Airport | 39.61713963 | -115.9973922 | Large N-array |
| TDA | AW 3 | TDA sensor | 200 | Eureka Airport | 39.61713069 | -115.9973274 | Large N-array |
| TDA | AW 4 | TDA sensor | 200 | Eureka Airport | 39.61709491 | -115.9973306 | Large N-array |
| TDA | AW 5 | TDA sensor | 200 | Eureka Airport | 39.61709719 | -115.997395 | Large N-array |
| TDA | BE 1 | TDA sensor | 200 | Eureka Airport | 39.61689621 | -115.9979975 | Large N-array |
| TDA | BE 2 | TDA sensor | 200 | Eureka Airport | 39.61692138 | -115.9979683 | Large N-array |
| TDA | BE 3 | TDA sensor | 200 | Eureka Airport | 39.61687534 | -115.9979646 | Large N-array |
| TDA | BE 4 | TDA sensor | 200 | Eureka Airport | 39.61687088 | -115.998023 | Large N-array |
| TDA | BE 5 | TDA sensor | 200 | Eureka Airport | 39.6169141 | -115.9980322 | Large N-array |
| TDA | BB 1 | TDA sensor | 200 | Eureka Airport | 39.61696885 | -115.998569 | Large N-array |
| TDA | BB 2 | TDA sensor | 200 | Eureka Airport | 39.61697081 | -115.9985362 | Large N-array |
| TDA | BB 3 | TDA sensor | 200 | Eureka Airport | 39.61693557 | -115.9985649 | Large N-array |
| TDA | BB 4 | TDA sensor | 200 | Eureka Airport | 39.61696674 | -115.9986111 | Large N-array |
| TDA | BB 5 | TDA sensor | 200 | Eureka Airport | 39.61700185 | -115.9985761 | Large N-array |
| TDA | BI 1 | TDA sensor | 200 | Eureka Airport | 39.61678249 | -115.9985538 | Large N-array |
| TDA | BI 2 | TDA sensor | 200 | Eureka Airport | 39.61678141 | -115.9985191 | Large N-array |
| TDA | BI 3 | TDA sensor | 200 | Eureka Airport | 39.61675684 | -115.9985499 | Large N-array |
| TDA | BI 4 | TDA sensor | 200 | Eureka Airport | 39.61678143 | -115.9986053 | Large N-array |
| TDA | BI 5 | TDA sensor | 200 | Eureka Airport | 39.61681847 | -115.9985575 | Large N-array |
| TDA | BL 1 | TDA sensor | 200 | Eureka Airport | 39.61669171 | -115.9979882 | Large N-array |
| TDA | BL 2 | TDA sensor | 200 | Eureka Airport | 39.61667753 | -115.9979595 | Large N-array |
| TDA | BL 3 | TDA sensor | 200 | Eureka Airport | 39.61667106 | -115.9980087 | Large N-array |
| TDA | BL 4 | TDA sensor | 200 | Eureka Airport | 39.61670666 | -115.9980145 | Large N-array |
| TDA | BL 5 | TDA sensor | 200 | Eureka Airport | 39.61671435 | -115.9979715 | Large N-array |
| TDA | BO 1 | TDA sensor | 200 | Eureka Airport | 39.61673257 | -115.9973895 | Large N-array |
| TDA | BO 2 | TDA sensor | 200 | Eureka Airport | 39.61671555 | -115.9973527 | Large N-array |
| TDA | BO 3 | TDA sensor | 200 | Eureka Airport | 39.61670968 | -115.9974107 | Large N-array |
| TDA | BO 4 | TDA sensor | 200 | Eureka Airport | 39.61675925 | -115.997413 | Large N-array |
| TDA | BO 5 | TDA sensor | 200 | Eureka Airport | 39.61675637 | -115.9973539 | Large N-array |
| TDA | BS 1 | TDA sensor | 200 | Eureka Airport | 39.61650194 | -115.9979802 | Large N-array |
| TDA | BS 2 | TDA sensor | 200 | Eureka Airport | 39.61648103 | -115.9980106 | Large N-array |
| TDA | BS 3 | TDA sensor | 200 | Eureka Airport | 39.61652107 | -115.9980084 | Large N-array |
| TDA | BS 4 | TDA sensor | 200 | Eureka Airport | 39.61652053 | -115.9979524 | Large N-array |
| TDA | BR 1 | TDA sensor | 200 | Eureka Airport | 39.61652715 | -115.9981568 | Large N-array |
| TDA | BR 2 | TDA sensor | 200 | Eureka Airport | 39.61649997 | -115.9981276 | Large N-array |
| TDA | BR 3 | TDA sensor | 200 | Eureka Airport | 39.61650461 | -115.998182 | Large N-array |
| TDA | BR 4 | TDA sensor | 200 | Eureka Airport | 39.61654842 | -115.9981822 | Large N-array |
| TDA | BR 5 | TDA sensor | 200 | Eureka Airport | 39.61654919 | -115.9981286 | Large N-array |
| TDA | BQ 1 | TDA sensor | 200 | Eureka Airport | 39.61653333 | -115.9983295 | Large N-array |
| TDA | BQ 2 | TDA sensor | 200 | Eureka Airport | 39.61649627 | -115.9983174 | Large N-array |
| TDA | BQ 3 | TDA sensor | 200 | Eureka Airport | 39.61652912 | -115.9983727 | Large N-array |
| TDA | BQ 4 | TDA sensor | 200 | Eureka Airport | 39.61656311 | -115.998334 | Large N-array |
| TDA | BQ 5 | TDA sensor | 200 | Eureka Airport | 39.61653614 | -115.9982894 | Large N-array |
| TDA | BP 1 | TDA sensor | 200 | Eureka Airport | 39.61654933 | -115.998579 | Large N-array |
| TDA | BP 2 | TDA sensor | 200 | Eureka Airport | 39.61654556 | -115.9985301 | Large N-array |
| TDA | BP 3 | TDA sensor | 200 | Eureka Airport | 39.61651148 | -115.9985766 | Large N-array |
| TDA | BP 4 | TDA sensor | 200 | Eureka Airport | 39.61655246 | -115.9986164 | Large N-array |
| TDA | BP 5 | TDA sensor | 200 | Eureka Airport | 39.6165886 | -115.9985852 | Large N-array |
| TDA | AX 1 | TDA sensor | 200 | Eureka Airport | 39.61646076 | -115.9973911 | Large N-array |
| TDA | AX 2 | TDA sensor | 200 | Eureka Airport | 39.61648133 | -115.9974156 | Large N-array |
| TDA | AX 3 | TDA sensor | 200 | Eureka Airport | 39.61648015 | -115.9973568 | Large N-array |
| TDA | AX 4 | TDA sensor | 200 | Eureka Airport | 39.61643304 | -115.9973639 | Large N-array |
| TDA | AX 5 | TDA sensor | 200 | Eureka Airport | 39.61643632 | -115.997415 | Large N-array |
| TDA | BT 1 | TDA sensor | 200 | Eureka Airport | 39.61645917 | -115.9976089 | Large N-array |
| TDA | BT 2 | TDA sensor | 200 | Eureka Airport | 39.61644781 | -115.9975688 | Large N-array |
| TDA | BT 3 | TDA sensor | 200 | Eureka Airport | 39.61644094 | -115.9976325 | Large N-array |
| TDA | BT 4 | TDA sensor | 200 | Eureka Airport | 39.61647984 | -115.9976417 | Large N-array |
| TDA | BT 5 | TDA sensor | 200 | Eureka Airport | 39.61648358 | -115.9975813 | Large N-array |
| UH | redvox_1173028730 | | 800 | Eureka, NV | 39.60085476 | -116.0061335 | Single station |
| UH | redvox_0000000022 | | 800 | Nevada | 39.61727241 | -115.9979189 | Single station |
| UH | redvox_1637622001 | | 800 | Nevada | 40.06032154 | -114.525342 | Single station |
| UH | redvox_1637622002 | | 800 | Nevada | 40.06941936 | -114.5299165 | Single station |
| UH | redvox_1637622006 | | 800 | Nevada | 40.1184172 | -114.5319345 | Single station |
| UH | redvox_1637622007 | | 800 | Nevada | 40.22897117 | -114.4351343 | Single station |
| UH | redvox_1637622008 | | 800 | Nevada | 40.35166394 | -114.2376673 | Single station |
| UH | redvox_1637622009 | | 800 | Nevada | 40.47726143 | -114.155499 | Single station |
| UH | redvox_1637622010 | | 800 | Nevada | 40.59260421 | -114.1392218 | Single station |
| UH | redvox_0000000096 | | 800 | Nevada | 40.6452582 | -114.1235271 | Single station |
| UH | redvox_0000000095 | | 800 | Nevada | 40.71097088 | -114.0889773 | Single station |
| UH | redvox_0000000094 | | 800 | West Wendover Airport | 40.7278619 | -114.0218836 | Array |
| UH | redvox_0000000091 | | 800 | West Wendover Airport | 40.72804373 | -114.0211986 | Array |
| UH | redvox_0000000093 | | 800 | West Wendover Airport | 40.72782152 | -114.0205354 | Array |
| UH | redvox_0000000092 | | 800 | West Wendover Airport | 40.728606 | -114.0211872 | Array |
| UH | redvox_1637653010 | | 800 | Nevada | 40.7124639 | -113.1244734 | Single station |
| UH | redvox_0000000110 | | 800 | Clive, UT | 40.70939688 | -113.1214597 | Single station |
| UH | redvox_0000000106 | | 800 | Clive, UT | 40.7088681 | -113.1167137 | Single station |
| UH | redvox_1637656001 | | 800 | Clive, UT | 40.7165156 | -113.1127525 | Single station |
| UH | redvox_0000000103 | | 800 | Dugway, UT | 40.25711486 | -112.7404369 | Single station |
| UH | redvox_0000000104 | | 800 | Dugway, UT | 40.25731765 | -112.7404951 | Single station |
| UH | redvox_1637610021 | | 800 | Nevada | 39.69948222 | -115.8906106 | Single station |
| UH | redvox_1637610022 | | 800 | Eureka, NV | 39.7076942 | -115.8617927 | Single station |
| INL | redvox_1637622022 | Samsung S22 | 800 | Idaho Falls, ID | 43.65981128 | -111.844751 | Single station |
| INL | redvox_1637622023 | Samsung S22 | 800 | Idaho Falls, ID | 43.49978502 | -112.0490684 | Single station |
| INL | redvox_1637622024 | Samsung S22 | 800 | Idaho Falls, ID | 43.49979985 | -112.0491303 | Single station |
| INL | redvox_1637622025 | Samsung S22 | 800 | Idaho Falls, ID | 43.50714182 | -111.9710928 | Single station |
| INL | redvox_1637622026 | Samsung S22 | 800 | Idaho Falls, ID | 43.48783296 | -112.0784373 | Single station |
| INL | redvox_1637622029 | Samsung S22 | 800 | Idaho Falls, ID | 43.49990583 | -112.0491253 | Single station |
| INL | redvox_1637622030 | Samsung S22 | 800 | Idaho Falls, ID | 43.48782929 | -112.0784318 | Single station |
| SMU | SN10 | Sapphire | | Eureka Airport | 39.61727 | -116.00037 | Array |
| SMU | SN12 | Sapphire | | Eureka Airport | 39.61704 | -116.0004 | Array |
| SMU | SN09 | Sapphire | | Eureka Airport | 39.61691 | -116.00064 | Array |
| SMU | SN08 | Sapphire | | Eureka Airport | 39.61691 | -116.0001 | Array |
| OSU | Loc04 - Center | Chaparral Physics, 64S | 1000 | West Wendover Airport | 40.7280 | -114.0212 | Array |
| OSU | Loc03 | Chaparral Physics, 64S | 1000 | West Wendover Airport | 40.7286 | -114.0211 | Array |
| OSU | Loc02 | Chaparral Physics, 64S | 1000 | West Wendover Airport | 40.7278 | -114.0205 | Array |
| OSU | Loc01 | Chaparral Physics, 64S | 1000 | West Wendover Airport | 40.7279 | -114.0219 | Array |
| OSU | WERD 10 - center | WERD, ISSM 23 | 400 | West Wendover Airport | 40.7280 | -114.0212 | Single sensor |
| OSU | WERD 9 | WERD, ISSM 23 | 400 | West Wendover Airport | 40.7282 | -114.0210 | Single sensor |
| OSU | WERD 7 | WERD, ISSM 23 | 400 | West Wendover Airport | 40.7278 | -114.0213 | Single sensor |
| OSU | WERD 5 | WERD, ISSM 23 | 400 | West Wendover Airport | 40.7281 | -114.0214 | Single sensor |
| OSU | WERD 3 | WERD, ISSM 23 | 400 | West Wendover Airport | 40.7281 | -114.0218 | Single sensor |
| OSU | WERD 6 | WERD, ISSM 23 | 400 | West Wendover Airport | 40.7276 | -114.0219 | Single sensor |
| OSU | WERD 4 | WERD, ISSM 23 | 400 | West Wendover Airport | 40.7278 | -114.0222 | Single sensor |
| OSU | GEM 185 | GEM, 1.01 Flight version | 100 | West Wendover Airport | 40.7277 | -114.0208 | Single sensor |
| OSU | GEM 186 | GEM, 1.01 Flight version | 100 | West Wendover Airport | 40.7281 | -114.0208 | Single sensor |
| OSU | GEM 187 - center | GEM, 1.01 Flight version | 100 | West Wendover Airport | 40.7280 | -114.0212 | Single sensor |
| OSU | GEM 074 | GEM, 1.01 Flight version | 100 | West Wendover Airport | 40.4924 | -114.0425 | Single sensor |
| OSU | GEM 092 | GEM, 1.01 Flight version | 100 | West Wendover Airport | 40.4254 | -114.0155 | Single sensor |

## Appendix C

Table A2: Seismic instrument installation data. This table is available in machine readable format.

| Institution | Instrument field name | Latitude (N) [deg] | Longitude (E) [deg] | Comments |
|---|---|---|---|---|
| SNL | EA-S1 | 39.61727 | -116.00026 | |
| SNL | EA-S2 | 39.61551 | -115.99598 | |
| SNL | EA-S3 | 39.61203 | -116.00454 | |
| SNL | EA-S4 | 39.61013 | -116.00019 | |
| SNL | EA-S5 | 39.60808 | -116.01050 | |
| SNL | EA-S6 | 39.60755 | -116.00766 | |
| SNL | EA-S7 | 39.60702 | -116.00482 | |
| SNL | EA-S8 | 39.60623 | -116.00057 | |
| SNL | EA-S9 | 39.60286 | -116.00646 | |
| SNL | EA-S10 | 39.59980 | -116.00269 | |
| SNL | EA-S11 | 39.59782 | -116.00742 | |
| SNL | EA-S12 | 39.59347 | -116.00436 | |
| SNL | HA1-SNC | 39.61702 | -115.99792 | Co-located with infrasound |
| SNL | HA1-SN | 39.61727 | -115.99791 | Co-located with infrasound |
| SNL | HA2 | 39.63969 | -115.78261 | Co-located with infrasound |
| SNL | HA2 | 39.63945 | -115.78200 | Co-located with infrasound |
| SNL | HA3 | 39.54050 | -115.77453 | Co-located with infrasound |
| SNL | HA3 | 39.54073 | -115.77453 | Co-located with infrasound |
| LANL | OREXA | 39.61099 | -116.00293 | Co-located with infrasound |
| LANL | OREXB | 39.60899 | -116.01174 | Co-located with infrasound |
| LANL | OREXC | 39.60404 | -116.00464 | Co-located with infrasound |
| LANL | OREXD | 39.73720 | -115.67409 | Co-located with infrasound |
| LANL | OREXE | 39.70430 | -115.67603 | Co-located with infrasound |
| LANL | OREXF | 39.68588 | -115.67698 | Co-located with infrasound |
| UM | A1 | 39.61424 | -115.99951 | |
| UM | A2 | 39.61422 | -115.99834 | |
| UM | A3 | 39.61345 | -115.99894 | |
| UM | A4 | 39.61183 | -115.99774 | |
| UM | A5 | 39.61182 | -115.99658 | |
| UM | A6 | 39.61103 | -115.99717 | |
| UM | A7 | 39.60709 | -115.99781 | |
| UM | A8 | 39.60709 | -115.99663 | |
| UM | A9 | 39.60625 | -115.99723 | |
| UM | A10 | 39.60712 | -116.00141 | |
| UM | A11 | 39.60710 | -116.00023 | |
| UM | A12 | 39.60629 | -116.00083 | |
| UM | A13 | 39.60951 | -116.00676 | |
| UM | A14 | 39.60950 | -116.00559 | |
| UM | A15 | 39.60872 | -116.00619 | |
| UM | A16 | 39.61191 | -116.00494 | |
| UM | A17 | 39.61189 | -116.00376 | |
| UM | A18 | 39.61109 | -116.00437 | |
| UM | A19 | 39.61159 | -116.00432 | |
| UM | A20 | 39.61150 | -115.99716 | |
| UM | Refraction West | 39.61113563 | -116.00468 | South end of line |
| UM | Refraction East | 39.61096743 | -115.99691 | SouthWest end of line |
| JHU | S1 | 39.49957 | -116.50945 | Array centre |
| JHU | S2 | 39.49919 | -116.51052 | Uprange 1 |
| JHU | S3 | 39.49883 | -116.51157 | Uprange 2 |
| JHU | S4 | 39.49844 | -116.51262 | Uprange 3 |
| JHU | S5 | 39.49790 | -116.50855 | South 1 |
| JHU | S6 | 39.50036 | -116.50994 | North 1 |
| JHU | S7 | 39.50117 | -116.51045 | North 2 |
| JHU | S8 | 39.49991 | -116.50838 | Downrange 1 |
| JHU | S9 | 39.50027 | -116.50731 | Downrange 2 |
| JHU | S10 | 39.50060 | -116.50623 | Downrange 3 |
| JHU | S11 | 39.49874 | -116.50920 | South 2 |

## Appendix D

Table A3: GPS instrument installation data.

| Institution | Instrument field name | Latitude (N) [deg] | Longitude (E) [deg] | Elevation [m] |
|---|---|---|---|---|
| LANL | West Runway | 39.608664 | -116.011496 | 1811.8 |
| LANL | Mid-North Runway | 39.607281 | -116.004027 | 1810.4 |
| LANL | Mid-South Runway | 39.603620 | -116.004756 | 1812.0 |
| LANL | South Runway | 39.595354 | -116.007202 | 1814.4 |
| LANL | North Runway | 39.612798 | -116.002877 | 1810.3 |